\PassOptionsToPackage{dvipsnames}{xcolor}
\documentclass[11pt]{Latex/Classes/PhDthesisPSnPDF}

\usepackage{lmodern}
\usepackage{enumitem}
\usepackage{csquotes}
\usepackage{blindtext}         
\usepackage{setspace}

\title{ \Large{  Topics on Foundations of Physics:}\par \Large{From the quantum to the (semi) classical, gravity, thermodynamics, and (or beyond) our possible detections.}}
\author{Ricardo Muciño Gómez}       
\degree{Doctor en Ciencias (Física)}
\director{Dr. Elias Okon Gurvich}

\degreedate{March 2026}
\lugar{Ciudad de México}

\portadatrue

\programa{Posgrado en Ciencias Físicas}
\institucion{Instituto de Ciencias Nucleares}
\campo{FÍSICA DE ALTAS ENERGÍAS, FÍSICA NUCLEAR, GRAVITACIÓN Y FÍSICA MATEMÁTICA}

\posgradotrue

\ctutoruno{Dr. Daniel Sudarsky Saionz}
\ctutordos{Dr. Yuri Bonder Grimberg}

\comitetrue

\begin{document}
\maketitle

							


\begin{acknowledgements}
\addcontentsline{toc}{chapter}{\protect\numberline{}Acknowledgements}

La realización de todo aquí no habría sido posible sin mi familia y sin muchos entes más con los que me he cruzado y viajado a través de este tiempo.

No hay manera de agradecer lo suficiente aquí primero a mi madre Rosa, a mi padre Adrián y a mi hermano Adrián, con quienes viví mi ingreso al doctorado en medio de una pandemia. Fueron y serán siempre el hogar donde aprendí a sembrar las primeras semillas, de ``los cantos y las flores'', dirían los \textit{tlamatinime} \cite{portilla}. Seguiré regresando y me seguirán enseñando.

Segundo, pero en cronología y no en importancia, agradezco a Carolina, mi \textit{yollotl}, porque me enseñó a construir, y construimos un hogar, temporal, donde sembramos mas semillas que se hicieron selva, de nuevos cantos e impredecibles flores, como la de Tigridia, fugaz e imborrable. 

Tercero, agradezco a otro constructor, Jorge, que reencontré en la selva y que me apoyó y cuidó sin saber que lo necesitaba. Igualmente tengo qué mencionar a personas que me apoyaron cantidades que no imaginan ni ellos, como mi tío Óscar y mi tía Chucha, Liss, mi gran amigo Toño con quien siempre podré dicutir amenamente por horas donde sea que nos encontremos,  a Xime con quien sé que cuento aunque a veces sobran las palabras, así como Ayan y Andrés. Mis amigos de la vida, Vianey e Isra. Y los amigos que formé en el tiempo del posgrado, como los ``filósofos de la física'' Rodrigo, Juan y Jorge, o los ``físicos'' como Johas. 

A Elias, le agradezco desde la licenciatura haber sido un gran mentor, a su manera pero tal vez justo el que yo necesitaba. Gracias a el mi pasión por entender el mundo nunca se tambaleó, sé que mi vida sería otra si el no me hubiera señalado, hace casi una década, la ``nueva'' forma de ver la física que a ciegas yo buscaba. Asimismo quiero agradecer enormemente a Daniel, un gran científico al que admiro y me honra haber podido aprender y trabajar con él desde la maestría. Hay  muchos profesores más a quienes les debo uun agradecimiento, sin embargo no puedo no mencionar a Yuri, quien siempre ha estado ahí.

Also, I really want to thank my friend Siddhant Das for his enormous support since we met some years ago, for his advice, his \textit{time}, the pleasant talks and his food. His vision has inspired me and also motivated part of this thesis.  

Por otro lado, quiero agradecer también al ``último río'' y a la comunidad de Atlitic que lo mantiene vivo, porque ascender a él en bici todos éstos años era como entrar al \textit{Omeyocan}.

Finalmente, agradezco enormemente a Gabriel, Benito, Miguel y Pablo por leer, comentar y corregir ésta tesis. Su revisión ayudó a mejorar considerablemente este trabajo.
\end{acknowledgements}

\begin{dedication}
	
A Ometeotl...

Mirar al abismo,\\
es como mirar al cielo,\\
se encuentran nuestros ojos,\\
perdidos en el dzonot,\\
y sentimos que...\\
     ...como Nezahualcóyotl\\
¡por fin lo comprende nuestro corazón!\\
Escuchamos un canto,\\
contemplamos una flor, \\
¡Ojalá no se marchiten!\\

\vspace{6mm}

[...]consult experience first and then with reasoning show why such experience is bound to operate in such a way. And this is the true rule by which those who analyse the effects of nature must proceed: and although
nature begins with the cause and ends with the experience, we must follow the opposite course, namely, begin with the experience, and by means of it investigate the cause.

---Leonardo da Vinci

It is the theory which decides what can be observed.

---Albert Einstein

\end{dedication}
                  


\begin{abstracts}        
\addcontentsline{toc}{chapter}{\protect\numberline{}Abstract}

Part I.1: Newton constructed a universe composed of absolute space and time where material bodies followed his dynamical laws. His theory presupposes that, among the real objective properties of bodies, lies the absolute space a body covers during an absolute amount of time, that is, its \textit{absolute velocity}. However, in our universe, it seems that we simply cannot detect such a property. Thus, it emerges the theoretical question of whether or not the Newtonian universe, \textit{by itself}, possesses this same characteristic, i.e., the non-detectability of absolute velocities. It is standardly thought it does; that from Newton's theory standpoint alone, it follows that absolute velocities are, in principle, \textit{undetectable}, just as it appears to be the case in our world. Here it will be shown that the standard arguments, starting from the one given by Newton himself, crucially end begging the question and, thus, contrary to the appearances in our universe, that absolute velocities are not proved to be, in principle, undetectable in a Newtonian universe.

Part I.2: The simplest, best-developed hidden-variable alternative to standard quantum mechanics is the de Broglie-Bohm pilot-wave theory. According to it, a complete characterization of the universe is given by the position of the $N$ particles in it and its universal wave function. The wave function is assumed to always evolve according to the Schrödinger equation and the $N$-positions are assumed to be defined at all times evolving due to the non-local deterministic ``guiding equation''. A complete \textit{empirical agreement} with the standard measurement formalism of quantum mechanics, including the collapse postulate and the probabilistic Born rule, is then said to emerge from the pilot-wave postulates, without providing a special role to measurements or observers. Two key elements behind the proof of this complete agreement are the so-called \textit{absolute uncertainty}, and the \textit{POVM theorem}. These results allegedly convey a ``naturally emerging, irreducible limitation on the possibility of obtaining knowledge within pilot-wave theory'' and that the statistical descriptions of outcomes in all possible measurements can be given by POVMs, respectively. Here, It will be argued that the derivation of absolute uncertainty and the POVM theorem both rely upon the \textit{unjustified} assumption that ``information is always configurationally grounded''. I shall explain in detail why there is no, in principle, theoretical reason behind such an assumption and explore some far-reaching possible empirical consequences of having let go of it.
\newpage

Part II:	The universally observed time-asymmetrical evolution towards a (macro-)state of \textit{equilibrium} (or `maximum entropy') shown by macroscopic systems has resisted a general, coherent, and objective explanation from our most fundamental physics. Boltzmann's classical approaches, from the `H-theorem' to ergodicity, although insightful and influential, made a variety of unjustified assumptions and applied only to very specific systems. The advent of quantum phenomena makes it indispensable to tackle such a reductive project in quantum terms. If the world is quantum, our explanations of its fundamental objective properties should also be. However, almost every modern approach to the subject of `thermalization' treats the properties of the wavefunction and its behavior considering only the unitary evolution part of the standard quantum formalism. Notwithstanding, the macroscopic objects we observe to equilibrate are never in superpositions of macro-states. The fundamental problems of the quantum formalism have not been taken into account. Here, some of the above issues will be alleviated putting forward an approach motivated by Albert's `semi-classical' suggestion to employ the GRW theory and its stochastic perturbations in a classical phase space. Although the proposal in this manuscript will rest in the underlying dynamics of generic collapse models, it will be argued that the equilibration process is, after all, just a particular classical behavior followed by general macroscopic quantum systems. 

Part III: Although non-relativistic quantum collapse theories have been well-recognized as resolutions to the fundamental problems of standard quantum mechanics, their relativistic extension has been unrecognized, criticized or misunderstood with influential literature apparently showing a variety of conflicts in combining both: the collapse of the wavefunction and the lessons of relativity. These alleged conflicts include: the impossibility of a relativistic-invariant model, the possibility of sending signals faster than light  \cite{sorkin1993impossible}, the meaning of indeterminism in a four dimensional block universe \cite{Putnam1967-PUTTAP,Rietdijk}, or even further problems when coupled to gravity. Here we will expel how relativistic theories work, reviewing how they avoid the alleged issues. Moreover, although for different reasons, semiclassical gravity has been criticized as self-inconsistent or empirically nonviable, here we manage to put forward a fully self-consistent, empirically viable semiclassical gravity framework, in which the expectation value of a quantum field, evolving via a relativistic objective collapse dynamics, couples to a wholly classical Einstein tensor. We present the general framework, a concrete example, and briefly explore possible empirical consequences of our model to Gravitationally-Induced-Entanglement experiments. 

\end{abstracts}


%
\setcounter{secnumdepth}{3} 
\setcounter{tocdepth}{3}    

\tableofcontents            


\mainmatter

\def\baselinestretch{1}                   


\chapter*{Introduction}
\addcontentsline{toc}{chapter}{\protect\numberline{}Introduction}	  		
\label{chapter:intro}


The work leading to this thesis focuses on assessing and extending quantum theories in order to explore and test their implications across various regimes---including thermodynamics, semiclassical and quantum gravity scenarios, and the \textit{in principle} detectable predictions of such theories. The general motivation stems from a basic desire to understand the world form its very \textit{foundations}. For instance, how can we bridge the gap between what we observe or `perceive' and the fundamental quantum nature in our theories. This work started, in particular, from a search to better understanding of the \textit{nature of time} according to our physical theories and the common perception that it invariably `flows' to the future, or, in other words, why do we observe distinct natural processes evolving asymmetrically in time?. These motivations lead to three different, but successful, lines of research.

The first PhD project (appearing as \hyperlink{PartII}{Part II} in this thesis) was directed towards a clearer explanation of the thermodynamic arrow of time or the \textit{approach to equilibrium}. Boltzmann's proposals were important attempts but, of course, they lacked generality and did not consider the quantum nature of matter. von Neumann's quantum ergodic theorem tried to fill that gap, but he only considered the unitary evolution of quantum mechanics and left the fundamental problems of standard quantum mechanics (QM) untouched \cite{goldstein2010normal,von_Neumann_2010}. Thus, motivated by Albert's suggestion \cite{Albert2021}, I analyzed the possible explanations that quantum collapse theories could offer for the thermodynamic equilibration, given its inherent \textit{asymmetrical} dynamics. Although it was found that Albert's expectations could not really be fulfilled by collapse theories, it was managed to propose a clearer explanation of equilibration that, in a way, backed up Botlzmann's classical arguments as approximately true. Our proposal rests on the results of \cite{bassi2010long}, showing that a macroscopic quantum system described by objective collapse models will show three well-distinguished \textit{temporal regimes}\footnote{The first two regimes are: the \textit{collapse regime}, where the system has not collapsed and is well described by the Schrödinger equation; and after collapse, the \textit{classical regime} approximately well-described by classical trajectories.}. In brief, we argue that what collapse models can do is to give us a clear explanation of how and when macroscopic systems acquire the classical statistical mechanics properties of the classical regime---which permits us to describe them as typically following trajectories in phase-space towards an equilibrium macrostate.

It became interesting to explore the different emerging pictures of time that relativistic extensions of `realist' quantum theories could provide. In the pilot-wave case, it is striking that although a privileged foliation (providing an absolute notion of simultaneity) is needed in the most natural extensions, it is frequently argued that such structure is \textit{empirically inaccessible} \cite{PhysRevA.53.2062}. These claims were assessed in this work and were found to rely on an argument from non-relativistic pilot-wave theory: namely, the argument for the \textit{empirical agreement} of pilot-wave theory with standard QM and the \textit{inaccessibility} of exact trajectories \cite{Maudlin1996,durr1992quantum}. Analyzing another argument involving empirical constraints within a completely different theoretical framework, it became noticeable that both lines of reasoning shared a common deficiency: they presupposed the very conclusion they sought to establish concerning \textit{possible detections}. The other theory under consideration was Newtonian mechanics, in which a circularity was identified in standard arguments claiming that, although absolute velocities are real, they are nonetheless undetectable \cite{manero2025Newton,Roberts2008}. This analysis could have several implications, both for pilot-wave theory and for our general understanding of physical theories, which are explored in this thesis. If our general assessment is right, it could be argued that---from a natural and broad characterization of detection---whenever a physical theory postulates an entity or property in its ontology, it is at least in principle detectable according to the theory, provided we do not introduce ad hoc postulates that presuppose its non-detectability and thereby beg the question. This second PhD project appears as \hyperlink{PartI}{Part I} in this thesis\footnote{There are already two preprints in collaboration with Jorge Manero: one concerns the Newtonian case and another concerns pilot-wave theory \cite{manero2025Newton,manero2025Bohm}.}.

The third and final PhD project (appearing as \hyperlink{PartIII}{Part III} in this thesis) was the following. In the relativistic extensions of collapse theories interesting pictures of time also emerge, which I wanted to carefully study. For example, they suggest a way out of the \textit{problem of time} in quantum gravity \cite{okon2014benefits}. However, although \textit{fundamentally covariant} extensions have already been proposed, some objections and misunderstandings appear to persist, and more are thought to arise if we include gravity. For instance, relativistic collapse theories are still viewed by some as incompatible with relativity, inconsistent with natural notions of temporal 'becoming' or 'unfolding', or even at odds with indeterminism.

Thus, after gaining a clearer understanding of how relativistic collapse theories operate, it was proposed a novel, self-consistent, empirically viable, semiclassical gravity framework incorporating collapses. Semiclassical gravity itself has been subject to scrutiny over the years, facing criticisms of inconsistency, empirical inadequacy, and even for possibly allowing superluminal signaling. Our framework is able to evade these accusations relying on a relativistic collapse dynamics for the quantum state, and a proper coupling of the classical Einstein tensor to the `local beables', i.e., the ontology of the collapse theory. This ontology is provided by postulating the existence in spacetime of an \textit{energy-momentum density} $\mathcal{T}_{ab}(x)$: the expectation value of the energy-momentum tensor operator, calculated in a state over the past null cone of $x$.
That is,
$$
	G_{ab}(x)=\mathcal{T}_{ab}(x)\equiv \langle \psi |  \hat{T}_{ab} | \psi \rangle_{\partial J^-(x)},
$$
with $\partial J^-(x)$ the past null cone of $x$; importantly, such a prescription is explicitly Lorentz-invariant.
We provided an example of the functioning of our framework in a scenario with superpositions of masses, and we also explored some of the potential empirical implications of our approach for well-known Gravitationally-Induced Entanglement experiments such as \cite{Bose2017} and \cite{MarlettoVedral2017}, including possible observational signatures that could distinguish our model from both standard semiclassical theories and full quantum gravity scenarios. This work thus offers a concrete and testable path forward in the ongoing effort
to reconcile the quantum and gravitational domains\footnote{A paper concerning these matters has already been published in \textit{Physical Review D} in collaboration with Prof. Daniel Sudarsky and Martin Wiedemann \cite{mucino2025fully}.}.

\chapter*{Introducción}
\addcontentsline{toc}{chapter}{\protect\numberline{}Introducción}	

El trabajo que dio lugar a esta tesis se centra en evaluar y ampliar las teorías cuánticas con el fin de explorar y poner a prueba sus implicaciones en diversos regímenes---incluyendo escenarios termodinámicos, semiclásicos y de gravedad cuántica, así como las predicciones \textit{en principio} detectables de dichas teorías. La motivación general surge de un deseo primordial de comprender el mundo físico desde sus mismos \textit{fundamentos}. Por ejemplo, ¿cómo podemos tender un puente entre lo que observamos o ``percibimos'' y la naturaleza cuántica fundamental en nuestras teorías? Este trabajo comenzó, en particular, con la búsqueda de una mejor comprensión de la naturaleza del tiempo según nuestras teorías físicas y de la percepción común de que éste fluye invariablemente hacia el futuro o, dicho de otro modo, por qué observamos que ciertos procesos naturales evolucionan de manera asimétrica en el tiempo. Estas motivaciones dieron lugar a tres líneas de investigación distintas, pero exitosas, que se exponen brevemente a continuación.

El primer proyecto de doctorado (apareciendo como \hyperlink{PartII}{Parte II} en esta tesis) estuvo dirigido hacia una explicación más clara de la flecha termodinámica del tiempo o el \textit{acercamiento al equilibrio}. Las propuestas de Boltzmann fueron intentos importantes, pero, por supuesto, carecían de generalidad y no consideraban la naturaleza cuántica de la materia. El teorema ergódico cuántico de von Neumann intentó llenar ese vacío, pero únicamente consideró la evolución unitaria de la mecánica cuántica y dejó intactos los problemas fundamentales de la mecánica cuántica estándar (QM) \cite{goldstein2010normal} \cite{von_Neumann_2010}. Así, motivados por la sugerencia de Albert \cite{Albert2021}, se analizaron las posibles explicaciones que las teorías de colapso cuántico podrían ofrecer para la termalización, dada su dinámica inherentemente \textit{asimétrica}. Aunque se encontró que las expectativas de Albert no podían realmente cumplirse mediante teorías de colapso, se logró proponer una explicación más clara de la termalización que, de cierta manera, respaldaba los argumentos clásicos de Boltzmann como aproximadamente válidos. 

Nuestra propuesta se apoya en los resultados de \cite{bassi2010long}, que muestran que un sistema cuántico macroscópico descrito por modelos de colapso objetivo presentará tres \textit{regímenes temporales} bien diferenciados\footnote{Los dos primeros regímenes son: el \textit{régimen de colapso}, donde el sistema no se ha colapsado y está bien descrito por la ecuación de Schrödinger; y, después del colapso, el \textit{régimen clásico}, aproximadamente bien descrito por trayectorias clásicas.}. En resumen, argumentamos que lo que los modelos de colapso pueden hacer es darnos una explicación clara de cómo y cuándo los sistemas macroscópicos adquieren las propiedades de la mecánica estadística clásica propias del régimen clásico---lo cual nos permite describirlos como sistemas que típicamente siguen trayectorias en el espacio de fases hacia un macroestado de equilibrio.

Se volvió interesante explorar también las diferentes imágenes emergentes del tiempo que las extensiones relativistas de las teorías cuánticas ``realistas'' podrían proporcionar. En el caso de la teoría de onda-piloto, resultaba impresionante que, aunque en las extensiones más naturales se requiere una foliación privilegiada (que proporciona una noción absoluta de simultaneidad), con frecuencia se argumenta que dicha estructura es \textit{empíricamente inaccesible} \cite{PhysRevA.53.2062}. Estas afirmaciones fueron evaluadas en este trabajo, y se encontró que se basaban en un argumento propuesto dentro de la teoría de onda-piloto no-relativista. Este es el argumento de la \textit{coincidencia empírica} de la teoría de onda-piloto con la mecánica cuántica estándar y de la \textit{inaccesibilidad} a las trayectorias exactas \cite{Maudlin1996,durr1992quantum}. Al analizar otro argumento que involucraba restricciones empíricas dentro de un marco teórico completamente distinto, se hizo evidente que ambas líneas de razonamiento compartían una misma deficiencia: presuponían la propia conclusión que buscaban establecer con respecto a las \textit{posibles detecciones}. La otra teoría considerada fue la mecánica newtoniana, en la cual se identificó una circularidad en los argumentos estándar que afirman que, aunque las velocidades absolutas son reales, sin embargo son indetectables \cite{manero2025Newton,Roberts2008}. 

Este análisis podría tener varias implicaciones empíricas que serán exploradas al final del Capítulo \ref{Chapter:Bohm} en el caso de la teoría de ondas piloto. Si nuestro análisis es correcto, podría sostenerse que---a partir de una caracterización natural y general de detección---siempre que una teoría física postule una entidad o propiedad en su ontología, ésta podría ser, al menos en principio, detectable según la teoría, mientras no se agreguen postulados ad hoc que presupongan su no-detectabilidad y, con ello, incurran en petición de principio. Este segundo proyecto de doctorado aparece como \hyperlink{PartI}{Parte I} en esta tesis\footnote{Ya se encuentran terminados dos preprints en colaboración con Jorge Manero: uno sobre el caso newtoniano y otro sobre la teoría de ondas piloto \cite{manero2025Newton,manero2025Bohm}}.

El tercer y último proyecto de doctorado (apareciendo como \hyperlink{PartIII}{Parte III} en esta tesis) fue el siguiente. En las extensiones relativistas de las teorías de colapso también emergen interesantes imágenes del tiempo, que quise estudiar cuidadosamente. Por ejemplo, sugieren una posible salida al \textit{problema del tiempo} en gravedad cuántica \cite{okon2014benefits}. Sin embargo, aunque ya se han propuesto extensiones \textit{fundamentalmente covariantes}, algunas objeciones y malentendidos parecen persistir, y se piensa que más podrían surgir si incluimos la gravedad. Por ejemplo, todavía se considera que las teorías de colapso relativistas son incompatibles con la relatividad, inconsistentes con nociones naturales de ``devenir'' o ``despliegue'' temporal, o incluso que están en desacuerdo con el indeterminismo.

Así, después de obtener una comprensión más clara de cómo operan las teorías de colapso relativistas, se propuso un marco novedoso, autoconsistente y empíricamente viable, de gravedad semiclasica que incorpora colapsos. La gravedad semiclasica en sí ha estado sujeta a escrutinio a lo largo de los años, enfrentando críticas de inconsistencia, inadecuación empírica e incluso por la posibilidad de permitir señales superlumínicas. Nuestro marco logra evadir estas acusaciones al apoyarse en una dinámica relativista de colapso para el estado cuántico y en un acoplamiento adecuado del tensor clásico de Einstein a los `local beables', es decir, a la ontología de la teoría de colapso. Esta ontología se provee postulando la existencia en el espacio-tiempo de una \textit{densidad de energía-momento} $\mathcal{T}_{ab}(x)$: el valor esperado renormalizado del operador tensor de energía-momento, calculado en un estado sobre el cono de luz pasado de $x$.
Es decir,
$$
G_{ab}(x)=\mathcal{T}_{ab}(x)\equiv \langle \psi |  \hat{T}_{ab} | \psi \rangle_{\partial J^-(x)},
$$
con $\partial J^-(x)$ el cono de luz pasado de $x$; lo cual, de manera importante, constituye una prescripción explícitamente invariante de Lorentz.

Proporcionamos un ejemplo del funcionamiento de nuestro marco en un escenario esféricamente simétrico con superposición de masas. Además, exploramos algunas de las posibles implicaciones empíricas de nuestro enfoque para experimentos bien conocidos de Entrelazamiento-inducido-Gravitacionalmente, como \cite{Bose2017} y \cite{MarlettoVedral2017}, incluyendo posibles firmas observacionales que podrían distinguir nuestro modelo tanto de las teorías semiclasicas estándar como de los escenarios completos de gravedad cuántica. Este trabajo, por tanto, ofrece un camino concreto y comprobable en el esfuerzo actual por reconciliar los dominios cuántico y gravitacional\footnote{Un artículo sobre este tema ya ha sido publicado en \textit{Physical Review D} en colaboración con el Prof. Daniel Sudarsky y Martin Wiedemann \cite{mucino2025fully}}.            
\part*{\hypertarget{PartI}{Part I}: \\ On Possible detections within Physical theories}
\addcontentsline{toc}{part}{\protect\numberline{}Part I: On Possible detections within Physical theories}\label{PartI}

\chapter{On the Detection of Absolute Velocity in a Newtonian Universe}\label{ch:Newton}

\section{Introduction}\label{section:introNewton}

The famous `Principia' was published by Newton in 1687, probably the most influential book written throughout the history of science. Newton tried to accurately describe our universe with the universe he outlined in this book. In his construction, he presupposed the existence of an absolute space and absolute time, as the arena where the material bodies exist and follow his dynamical laws. Newton's theory implies that, among the real, objective properties of bodies, lies the absolute space a body covers during an absolute amount of time, that is, its absolute velocity. However, in \textit{our} actual universe, it seems that we simply cannot have any means to detect such an absolute velocity. Thus, the following, purely \textit{theoretical} question emerges: Is it the case that the \textit{hypothetical} Newtonian universe possesses this same characteristic---namely the non-detectability of absolute velocities---by itself?

Although it is nothing new, much less controversial, to think that the Newtonian universe does not adequately describe our actual universe, it is almost never questioned the concordance between both universes regarding the above characteristic. That is, it is standardly thought that, from Newton's theory standpoint, and nothing more, it indeed follows that absolute velocities---although real---are, in principle, undetectable. Just as it appears to be the case in our world, i.e., we do not seem to observe absolute velocities. In fact, it is strongly believed that our actual world does not even possess such an objective absolute property of matter and that this will also be the case in a further more fundamental physical theory (even if it is still unknown) given that such theory should account for the success of special and general relativity where, of course, there are no absolute velocities.

Newton himself gave an influential argument to answer the above question in the positive. But, another highly influential argument in the literature, arguing for the non-detectability of absolute velocities, appeared in a much more recent work by Roberts \cite{Roberts2008}. It should be clear, however, that in order to assess the detectability of a given property in a hypothetical universe, one must at least provide a minimal conception of what constitutes a detection. This, in a few words, is what differs from the answers given by Newton and Roberts, among some other subtleties. And, more importantly, to carry out such assessment, one should not employ empirical facts of our universe to conclude something about the  possibilities regarding `detections' in another different universe.

In this chapter, it shall be explained how all the proposed answers arguing for the non-detectability of absolute velocity in a Newtonian universe arise from essentially the same reasoning, in the sense that all of them end up begging the question, one way or another. Furthermore, I will present an analysis that---at least potentially—could allow for the detection of absolute velocities in a Newtonian universe in the same way as relational velocities are thought to be detectable. To this end, I will first provide a very broad and explicit characterization of what is meant by ``detection''. I will adopt a natural but minimal definition, such that adding further complexity to it would not alter our conclusions. This chapter is based on the paper \cite{manero2025Newton}, submitted for publication in collaboration with my advisor, Elias Okon, and our colleague Jorge Manero. 

Now, one might think that this all sounds fine but still wonder about the point of discussing these questions, given that no one seriously believes our universe is Newtonian. Such a universe has already been shown to be an incorrect description of reality. While we all agree on this, the analysis developed in this work will allow us to compare the spirit of these flawed answers---offered by Newton and Roberts—with the spirit of widely accepted arguments in modern quantum theories regarding the non-detectability of certain real properties. These results are crucial for evaluating the alleged empirical adequacy of such quantum theories. In Chapter \ref{Chapter:Bohm}, I will thoroughly analyze these arguments in the context of pilot-wave theory. This work also stems from a collaboration with Elias Okon and Jorge Manero, resulting in the preprint \cite{manero2025Bohm}. There, we show—as I will explicitly detail in Chapter \ref{Chapter:Bohm}—that the same type of mistake found in the Newtonian case is repeated in arguments for the non-detectability of certain real and objective properties within the Pilot-wave case. It becomes clear that this kind of reasoning, whether in a Newtonian universe or in a `Bohmian' one, succeeds only by presupposing the very conclusion it seeks to establish. This highlights the usefulness of the type of conceptual analysis developed in this work for contemporary physics.

And there's more. The lessons drawn from the analysis in this chapter and in Chapter \ref{Chapter:Bohm} suggest an argument relevant to a highly debated topic in the philosophy of physics: the underdetermination of physical theories. That is, the possibility that two competing theoretical descriptions of the world might nonetheless yield, in principle, empirically indistinguishable predictions. The general insights gained from the Newtonian and pilot-wave cases seem to point to the following: whenever a physical theory posits some real and objective entity, structure, or a property of them, it is, at least in principle, detectable within the universe of such a theory. Therefore, if a theory implies the existence of such entities or properties but no corresponding detection has been obtained in our actual world, we may have further grounds to question the theory's validity or to seek new experimental avenues.

Note that common claims appealing to an alleged empirical equivalence---or ``indistinguishability''---among fundamental scientific theories typically begin with the assumption that there exist distinct physical theories (e.g., with different ontologies) that, nevertheless, cannot be empirically distinguished in any way. By the end of the following chapters, if my analysis is accepted, it should sound more convincing for the readers the idea that every instance in which such indiscernible ontologies are claimed ultimately rests on circular reasoning, similar to the flawed arguments for non-detectability exposed in those chapters.

Thus, what appeared to be an innocent and simple analysis on detectability in the context of an old physical theory like that of Newton, uncovers related implications from modern physics to influential discussions in philosophy and metaphysics of science. 

In the following section \ref{Section:NDabsoluteVel} we will assess the influential arguments claiming that absolute velocities are undetectable in a Newtonian universe. We will start by presenting Newton's argument, followed by our assessment. Then, we shall first characterize a general notion of detectability in \ref{section:detectability}, before presenting the more elaborated argument by Roberts in section \ref{Roberts}. We will expose the problems with this argument in \ref{RobertsProblems}. And we will finally state our conclusions in \ref{ConclusionsNewton}.

\section{Assessing the standard non-detectability arguments for absolute velocity}\label{Section:NDabsoluteVel}

Consider a Newtonian universe.  Here material bodies are composed of point particles with mass, which move on absolute space in an absolute amount of time, according to Newton's dynamics. The truth of the assertion that absolute velocities are undetectable in such a Newtonian universe is such a strong and almost unmovable belief, that it populates the mind of almost every physicist or philosopher. Here, we shall show that the arguments in favor of such an assertion, starting with the one offered by Newton himself, are question-begging. 

The arguments in favor of this assertion go like this. First, suppose that all forces are contact forces or forces that only depend on relative positions and velocities of bodies. Then, it can be shown that these relative quantities are invariant under Galilean transformations. Newton, for example, said the following for the case of Galilean boosts under the influence of forces of the sort described above:
\begin{quote}
	``the motions of bodies included in a given space are the same among themselves, whether that space is at rest, or moves uniformly forwards in a right line without any circular motion'' \cite{newton2022sir}. 
\end{quote}
  Finally, from this fact, that relative quantities are Galilean-invariant, and the assertion that \textit{only} the invariant-under-symmetries quantities can be detected, it is concluded that absolute velocities, and all absolute properties, are undetectable.
  
  The assertion concerning the detectability solely of the invariant quantities is commonly supported by making reference to an influential argument,  due to Galileo himself, about the symmetries of our world. Galileo convinced us with his thought experiments in a boat---and the actual experiments confirmed---that \textit{we} would not be able to detect absolute (variant) properties.

However, these arguments, within a Newtonian universe, are flawed in two distinct ways. To begin with, the restriction to only consider
contact forces, or forces that only depend on relative quantities, is somewhat ad hoc. In such a hypothetical universe, absolute properties are real and objective, so there is simple no a priori reason to impose that such properties cannot influence forces. And, of course, without such a restriction, it is immediate to see that absolute velocities can be detected. But our stronger point is that, even if such a restriction on forces is implemented, there is, still, another crucial deficiency in the arguments claiming that absolute velocities are undetectable as, in order to prove that, one must, essentially, assume what one is trying to prove. Let us lay out all this in detail.	

Let us start by emphasizing again that: given that in a Newtonian universe absolute properties are real and objective, there is no a priori reason to establish that such objective properties of bodies cannot influence the forces exerted or felt by them. It could be thought that allowing for absolute properties to affect forces would necessarily break either Newton's third law or the assumption of homogeneity and isotropy of absolute space. However, that is not the case. For example, consider a Newtonian universe populated by point particles, and a universal law similar to Newtonian gravity, describing a pairwise attractive force between the $i-$th and $j-$th particle with magnitude,
\begin{equation}
	F_{ij}=g_{ij}(|v_{ij}^{cm}|)\frac{m_i m_j}{r_{ij}^2},
\end{equation}
where the coupling constant of the
force is a function of the magnitude of the absolute velocity $v_{ij}^{cm}$ of the center of mass of the pair of particles involved, with their respective masses $m_i$ and $m_j$. It can readily be seen that such a force, neither disturbs Newton's third law, as for any pair of particles it is the case that the forces exerted between them are opposite and equal in magnitude, nor that it severs the homogeneity and isotropy of absolute space, as all points and directions are treated equally.

Thus, within a Newtonian universe, it is not a mandatory restriction to only consider forces solely dependent on relative quantities. Or in other words, such a restriction does not follow from the fundamental properties of a Newtonian universe. And without such a restriction, it is certainly not the case that absolute velocities are undetectable. Concerning our own universe, of course, it is the case that Galilean invariance seems to be a more adequate property to describe it. That is, our universe seems to be such that absolute velocities are undetectable \footnote{Although Einstein's relativity theory teaches us that Galilean invariance is also not adequate to describe our universe, in his theory, we have the means to explain why it was approximately adequate to accept it.}. One must not confuse, though, an empirical fact within our universe (e.g., that it approximately respects Galilean invariance), with a theoretical question within another universe (e.g., whether absolute velocities are detectable in a Newtonian universe). This remark concerns the other deficiency mentioned above in the arguments in favor of the non-detectability of absolute velocities in a Newtonian universe.  Empirical facts in our universe can not be used in order to prove that certain properties are, or are not, detectable in another hypothetical universe.

Thus, the important point I want to make is that, even if one artificially stipulates that all forces depend solely on relative quantities, it still does not follow that absolute velocities are undetectable. The issue is that, as mentioned earlier, all one can show with such a stipulation is that all (and only) relative quantities are Galilean invariant. However, to conclude from this that absolute velocities are undetectable, one must assume that only relative quantities are detectable. But, of course, this assumption appears to be precisely what one was trying to prove in the first place!

In order to save the argument, one could: i) claim that we do not need to impose by hand that only relative quantities are detectable, because such a characteristic follows from the principles of a Newtonian universe, or ii) one could claim that such an assumption does not carries a circularity problem because the natural and general characterization of detection allows us to use it without begging the question. We will show below that none of these options succeeds, the argument can not be saved. To see this, we need to say something about what constitutes a measurement or a perception. That, clearly, is a very complicated matter, but for our purposes, it suffices to give a very general and fundamental characterization of a detection, so that it would be a part of any other more detailed characterization.

\section{On the general characteristics of detections}\label{section:detectability}

The most general and fundamental characteristic, that we can think, of a detection, measurement, or perception carried out by a system $M$ on a target system $S$, is the idea that, as a result of an interaction between $M$ and $S$, certain properties of $S$ must end up \textit{correlated} with certain properties of $M$. The result of the detection, is said to be \textit{recorded} in the properties of $M$ participating in the correlation with the properties of $S$ \footnote{Of course, for a complete analysis of measurements, many more things, regarding, e.g., faithfulness of measurements, must be said; but this fundamental feature of measurements is enough for our purposes.}. With this in mind, it follows that determining---within any given theory, classical or quantum---whether certain properties are detectable by particular devices requires examining the interactions that the theory permits and characterizes. In other words, it is the \textit{dynamics} of the theory which naturally decides what kind of devices are able to detect what kind of properties. First, it will be said that:
	\begin{center}
	\textit{For some property of $S$ to be detectable by $M$, the appropriate interaction between $M$ and $S$ must be available by the dynamical laws of the theory. }
\end{center}  
This general characterization, I think, encodes our own common notions of a detection. For instance, the relative position between the needle and the dial in a Geiger counter (the measuring system $M$) is the objective, physical property that encodes the resulting radiation of a sample (the target system $S$). Similarly, when someone (acting now as the measuring system $M$) hears a Geiger counter (the target system $S$) producing audible clicks, that auditory perception becomes encoded in the objective physical state of their brain. Moreover, it should be emphasized that this general characterization---although quite natural---implies that, in principle, all objective physical properties of a system could participate in the encoding of a perception or measurement, which goes against the beliefs of many. Second, considering the above discussion, I will stipulate that: 
\begin{center}
	\textit{It is the dynamics of a given theory which determines what degrees of freedom of $M$ are the ones able to codify, as the record of a detection, certain properties of $S$.} 
\end{center}
This means that, unless there is a fundamental restriction or an ad hoc stipulation within a theory to the contrary, \textit{all} objective, physical properties could, in principle, be involved in the registration of the result. 

It is important to bear in mind that if, within a given theory, one wishes to claim, or stipulate, that the results of all measurements can be recorded in only a certain kind of degrees of freedom, then, we must make sure that the theory possess the appropriate dynamics to \textit{justify} such claim. That is, if we wish to sustain that all sorts of measurements of $M$ over $S$ \textit{can be} recorded \textit{solely} in terms of a certain kind of properties, say the $r-$degrees of freedom, then, one must be able to show that: in all the possible measurements, a correlation between the properties of $S$ and the $r-$degrees of freedom of $M$ can be established. If, on the contrary, there is an objective $v-$property of $S$ such that the given theory does not have the resources to establish a correlation between such property and the $r-$degrees of freedom of $M$, then, we should conclude that the apparatus $M$ is not appropriate to conduct a measurement of $v$. Thus, we would be able to say that the above claim ---that any result of a measurement can be recorded in only one kind of degrees of freedom--- is, actually, \textit{unjustified}.

Having said all this, let us consider a universe populated by particles with $A-$ and $B-$properties, with a given dynamics such that there are no inter-type influences: $A-$properties only interact with $A-$properties, and similarly for the $B-$properties. If one wishes to assert that the results of all measurements can be recorded in terms of, say, the $A-$properties, the dynamics should allow us to prove that the $B-$properties of a set of particles can be correlated with the $A-$properties of another set of particles. But, by construction, this hypothetical universe does no have the appropriate dynamics to do that. Thus, we must conclude that a measurement device that can only record its results in terms of $A-$properties is inappropriate to conduct a measurement of the $B-$properties. One can \textit{impose} that every measurement device should only record results in terms of the $A-$properties, but we must not confuse this imposition, causing an inability to detect $B-$properties, with a \textit{fundamental} unobservability of the $B-$properties. In fact, by construction, both properties are equally real and objective, so there is no means, within the theory itself, to favor one instead of the other, as being responsible for the records.

Let us return now to the case of the detection of absolute velocity in a Newtonian universe, in which all forces only depend on relative quantities. Consider, then, a particular system, $M$, and ask whether it could detect the absolute velocity of a second target system $S$ with which it interacts. To do this, we consider two scenarios, related by an active Galilean boost: The scenario I) where the absolute  velocity of $M$, after the interaction with $S$, is zero; and the scenario II) where $M$ has some non-zero velocity after it interacts with $S$. By assuming that all forces depend only on relative quantities, it is certainly the case that, all relational properties of $M$ are identical in both scenarios\footnote{Moreover, the functional form of the correlation formed through the dynamics between $M$ and $S$ is also the same in both scenarios.}. Therefore, \textit{if} the result of the purported detection performed by $M$ is encoded \textit{solely} in terms of relational properties---invariant under Galiean transformations---of $M$, then $M$ would necessarily record exactly the same result in both scenarios; given that the absolute velocity of $S$ is different in scenarios I and II, under such circumstances, we would be able to say that the absolute velocity of $S$ is undetectable for a device like $M$. However, why would it be the case that the records of \textit{any} measuring device must be entirely encoded in its purely relational properties? 

As was explained above, in general, such an encodement is performed in terms of some subset of the objective, physical properties of the apparatus and, in our example, there is an objective, physical property of $M$ which differs between scenarios I and II, namely the absolute velocity of its center of mass. Moreover, there does not seem to be any fundamental reason, exclusively extracted from the internal standpoint of the Newtonian universe, for which such an objective, physical property could not be included in the subset of properties responsible for encoding the results of some measurement for some apparatus. Of course, it could simply be stipulated ``by hand'', motivated in our own external prejudices, that absolute properties are never involved in detections, but that would amount to impose, not prove, that absolute velocities cannot be measured.

\section{Roberts's influential argument}\label{Roberts}

 A thorough discussion of this issue and an influential argument in favor of the claim that, in a Newtonian universe, absolute velocities are undetectable, is given by Roberts in \cite{Roberts2008}. Before presenting his argument, it is readily acknowledged that, in order to prove such an assertion, one must \textit{assume} that, \textit{for all} measurements, ``results are encoded in physical quantities that are themselves invariant under the dynamical symmetries''. That is, that all pointer variables necessarily are Galilean invariant. However, \cite{Roberts2008} does not see the inclusion of this assumption as endangering its argument for question-begging, as it deems it as following from an important fundamental feature of our world--a feature which physicists are entitled to take for granted. 

The idea is to try to construct an argument to show that the assumption must be true; i.e., to show that all perceptible facts must be preserved under the empirical symmetries of our world. To do so, Roberts starts by noting that it does seem to be the case that we cannot perceive facts that are not preserved by the symmetries. In relation with this, he points out that the results of an empirical measurement must be communicable in a public medium \footnote{In other works in the literature this is related to the stability of measurement outcomes.}. From this, he designs a scenario to show that, if we were to perceive variant facts, then things would be quite strange.

The argument is the following. Consider the world $U_1$, where Sally is able to accurately detect her absolute velocity and communicate the result by sending to Harry a letter (or another type of message composed of purely invariant quantities). Consider now world $U_2$, related to $U_1$ by a Galilean boost. Given that Sally's velocity is affected by the boost, but the message is not, Sally's report would not be accurate in $U_2$. Maybe, Sally could find another medium to send the message employing variant properties. However, if that could be the case, many of the standard means of communication (e.g., letters) would not be available to reliably report a certain type of information (e.g., absolute velocity). Then, Roberts reasons that, either there are things that Sally can perceive, but not communicate, or that she cannot use certain standard media to communicate all information. Then, Roberts alleges, we are not like that. Creatures in those circumstances might have communicative abilities very different from our own, according to Roberts. From that, it is concluded that the assumption required to prove that absolute velocity cannot be measured follows, supposedly, from very general and fundamental truths of our `form of life'. 

\section{The problem with this type of reasoning}\label{RobertsProblems}

Let us now explain why we designate the above line of reasoning as problematic. To begin with, as we mentioned earlier, one must not confuse the empirical fact that our universe seems to (approximately) respect Galilean invariance with the theoretical question of whether absolute velocities are detectable in a Newtonian universe. That is, one should not employ empirical features of our world, which is patently not Newtonian, to try to determine what properties would be detectable in such an hypothetical universe. Particular features about how communication among humans works is simply irrelevant to explore the issue at hand. Concerning the ``strangeness'' uncovered through Sally's attempt to communicate her absolute velocity, this is not really as mysterious as it might seem and, more importantly, has absolutely nothing to do with certain properties being detectable or not. The particular characteristic behind Sally's troubles communicating her result is the fact that, in a Newtonian universe with forces only depending on relative quantities, the center of mass of the universe fully decouples from all relative properties. This, read carefully, is just a dynamical fact that must be considered when using certain kind of devices to measure a certain kind of properties.  But this dynamical fact, is simply innocuous concerning the question of the detectability of absolute properties.

To convince us why this is so, consider again the hypothetical world populated by particles with just two types of properties, $A$ and $B$, such that there are no inter-type interactions. A system of particles would generically have $A-$properties, $B-$properties and even relational $AB-$properties. Suppose now that a system $M$ measures or detects something. If so, the result would end up encoded as a function of such properties; and, until now, it should be clear that there is anything prohibiting the system to detect and record results of, both,
$A-$ and $B-$properties. However, it is also clear that, due to the lack of inter-type interactions, it would be \textit{dynamically impossible} for, say, a measurement of an $A-$property to be encoded in terms of $B-$properties. This, in brief, is a dynamical fact of this universe. A fact that can be used to develop an argument, analogous to Sally's, in which two worlds with different $A-$properties, but identical $B-$properties would demonstrate the
inaccuracy of a certain devices to measure a certain property in most worlds.

Properly speaking it is not that such measurements are inaccurate but that, due to the decoupling of properties, it would simply be impossible to set up a procedure to record one type of property in terms of the other. This is just a natural consequence of the decoupling. The important point, though, is that this inability to record one sort of
property in terms of the other has absolutely nothing to do with the issue of what properties are measurable and which are not. Both $A-$ and $B-$properties are equally real and objective, so there is no reason whatsoever to treat them differently regarding detection, nor to favor one over the other.

Going back to the Newtonian case, one thing is the fact that the center of mass decouples from relative degrees of freedom in such a universe, and another thing is the question of whether absolute velocities are detectable or not. The first fact only tells us that the result of a measurement of absolute velocity cannot end up recorded in terms of relative features. However, it does not imply that only relative properties are detectable; both absolute and relative properties are real and objective and there is simply no way to conclude, out of the fundamental features of a Newtonian universe, that absolute and relative properties would behave differently regarding detection. The upshot of all this is that there is no reason to conclude that in a Newtonian universe absolute velocities would be undetectable and, that the arguments in favor of such an assertion, from Newton to Roberts, are question-begging.

\subsection{Addressing the accusations of being `superfluous' }

Still, some may accept our arguments above, agreeing with us (like Roberts does) that any claim against the detectability of absolute quantities necessarily relies on assuming essentially what one was willing to prove from the start, but insist that the detectability of such a property is somehow nonsense. This is also a very common fallacy, and the ultimate defense against the detectability of absolute velocities. The idea is to resort to the purported redundancy of something like absolute velocity.

It is true that Newton's dynamics can be stated without the need of introducing absolute velocities. A system of particles shows the same relative motions among the particles, independently of the absolute velocity associated with its center of mass. Thus, whether or not absolute velocities are introduced, it is said, it makes no difference on how the system evolves over time; in that sense, they are dispensable from physical explanations or, what amounts to the same thing, their values can be assigned arbitrarily \footnote{For an exhaustive assessment of the redundancy claims over absolute velocity, see \cite{Dasgupta2016}, section 2.1.}. From here, it is concluded that such dynamically `superfluous' properties cannot be real. However, the characteristics of absolute velocities, regarding detection, are no different from that of any other real and objective property, even relational ones. The absolute velocity of a body at any time $t$ is a function of its initial absolute velocity, $V_t=V_t(V_0)$. In order to explain why a particle has a certain absolute velocity at a certain time, the precise value of its initial absolute velocity is crucially relevant, so it cannot be assigned arbitrarily. Moreover, as it is possible to correlate the relative quantities of a system with that of another `measuring' system, it is also possible to do it for absolute quantities. 

To see this, consider a system $S$ at an initial absolute state of motion $X_0^S,V_0^S$, and another system $M$ at an initial absolute state of motion $X_0^M,V_0^M$. Then, even if all forces between them depend on relative quantities, the final state of $M$ will generically depend on the initial state. Think if you wish in the case of the gravitational force: the absolute state of a measuring needle $M$ at a certain time $t$ is correlated to the initial absolute state of a stellar object $S$, i.e.,  $X_t^M(X_0^S,V_0^S, X_0^M,V_0^M)$ and $V_t^M(X_0^S,V_0^S, X_0^M,V_0^M)$. If we fix the ready state of the needle $( X_0^M,V_0^M)$, then, out of the final absolute state of the needle $( X_t^M,V_t^M)$ we can infer, i.e, detect, the initial absolute state of the stellar object $(X_0^S,V_0^S)$.

Given this, one cannot simply say that, based on their actual unobservability in our world, absolute velocities are `superfluous' and conclude, from this, that they are not real. This, again, would be just to beg the question \footnote{Although we and Dasgupta \cite{Dasgupta2016} fully agree on all of this, the author of \cite{Dasgupta2016} still endorses Roberts's argument for the non-detectability of absolute velocities.}. Thus, with the above discussion concerning the actual correlations between the absolute properties of systems, we can now say, not only that the arguments for the non-detectability of absolute velocities are circular and therefore inconclusive; but also describe dynamically how a measuring device actually encodes the result of such a detection.


\section{Conclusions}\label{ConclusionsNewton}

We have seen that, in assessing what can or cannot be detected within the universe associated with a given physical theory, it is irrelevant what can actually be detected in our own universe---even if one expects the former to approximate the latter. The reason is twofold: first, the two universes may not in fact coincide; and second, scientific explanations aim to account for what \textit{could} be observed within the theoretical universe, not to merely stipulate the observations. This basic point has been overlooked in the main arguments asserting the non-detectability of absolute velocities in a Newtonian universe---arguments that, moreover, assume from the outset essentially what they intend to prove.

This purported `in principle' non-detectability is what led, on metaphysical grounds, to the construction of a spacetime---Galilean spacetime---lacking the structure of points in absolute space persisting through time \cite{maudlin2012philosophy}. However, our conclusions suggest that empirical motivations could have been given instead of metaphysical ones. That is, had it been recognized shortly after the publication of Newton's theory that the theory did not forbid the possibility of detecting absolute velocities, it might have been judged empirically inadequate.

The conclusions of this work also resonate with one of the most famous passages in the history of quantum theory: after Heisenberg insisted to build quantum theory by limiting himself only to what is observable, Einstein responded to his colleague with his famous claim that ``\textit{it is the theory which alone decides what is observable''} \cite{heisenberg1971physics}. 

As will be seen in the next chapter, the general reasoning developed here can likewise be applied to evaluate contemporary results in the foundations of quantum mechanics that purportedly constrain the `possibility of acquiring knowledge', leading to far more interesting consequences.



\chapter{On the empirical agreement between Pilot-wave Theory and Standard Quantum Mechanics}\label{Chapter:Bohm}

\section{Introduction}

The de Broglie-Bohm pilot-wave theory is the simplest, best-developed hidden-variable alternative to the standard formalism of quantum mechanics. It was first presented by Louis de Broglie in 1927 and then rediscovered by David Bohm in 1952. It asserts that the universe is composed of $N$ particles and that a complete characterization is provided by the positions and the universal wave function of such particles. The theory postulates the $N$ positions to be defined \textit{at-all-times}, evolving \textit{non-locally} according to the deterministic ``guiding equation'', which depends on the wave function, assumed to \textit{always} satisfy the Schrödinger equation. 

The empirical agreement with the standard measurement formalism of quantum mechanics, which includes collapses and the probabilistic Born rule, is then said to \textit{naturally} emerge from these equations, without having to confer special status to observers or to invoke an artificial distinction between a quantum world and a classical one where the quantum laws cease to apply. Two key results to prove this empirical agreement are the so-called \textit{absolute uncertainty}, and the POVM theorem. The first result allegedly conveys an ``irreducible limitation on the possibility of obtaining knowledge within pilot-wave theory''. And the second result purportedly covers the statistical description of the outcomes of \textit{all} the possible measurements. In particular, absolute uncertainty is said to imply that the wave function exhaust (or limits) all the possible empirical access to the positions of particles; and the POVM theorem, which motivates a `criterion of measurability', is said to show how the statistical predictions for any measurement are modeled by positive operator valued measures (POVMs). 

If these results and its assumptions are accepted as valid, then,  pilot-wave theory would certainly explain away the conceptual obstacles in the standard formalism of quantum mechanics. Moreover, it would provide as well a justification for its indeterministic flavor and empirical success, showing how it all naturally emerges from an underlying theory of particles moving deterministically, and nothing more.

However, as it will clearly be exposed in \S \ref{Section:EmpiricalAgreementBohm}, the derivation of absolute uncertainty and the POVM theorem both crucially depend upon the assumption that ``information is always configurationally grounded''. This assumption means, in particular, that \textit{all} the knowledge an external observer can gather of a system, is necessarily contained,  or recorded, on the actual \textit{instantaneous} particle configuration of the system's environment. I will explain in detail how the common justifications for such an assumption are not well founded and often rely on a circular reasoning. This, of course, calls into question the strength of the two key results, and, in consequence, challenges the validity of the many important claims that depended on these results. In general, this covers the purported empirical agreement between pilot-wave theory and standard quantum mechanics.
 
Furthermore, it will be explored in \S \ref{Section:ConsequencesBohm} the profound consequences of not assuming knowledge, detections, records or perceptions, solely on the instantaneous configurations of particles. Such consequences, for the case of the pilot-wave theory, will range from the possibility of signaling, to the violation of the uncertainty principle, and more. If the analysis in this work is correct and the de Broglie-Bohm theory is left with no resources to justify the, in principle, unobservability of such strong consequences, then, either we should make new experiments to test this novel possible predictions, or the pilot-wave approach is not empirically adequate.

This chapter is organized as follows, in \S \ref{Section:Bohm} the principles of the pilot-wave theory will be outlined, in \S \ref{Section:EmpiricalAgreementBohm} the standard justification of the empirical agreement between pilot-wave and quantum theory is going to be presented, where two key results are central, Absolute Uncertainty and the POVM theorem, outlined in subsections \S \ref{sSection:AbsUnc} and \S \ref{sSection:POVM} respectively. In \S \ref{Section:ConsequencesBohm} the consequences of our analysis to the pilot-wave approach will be explored and the conclusions will be given in \S \ref{ConclusionsBohm}.

	\section{The de Broglie-Bohm Pilot-wave theory}\label{Section:Bohm}

According to the de Broglie-Bohm pilot-wave theory, the complete description of a non-relativistic universe composed of $N$ particles is given by the universal wave function, $\Psi(q,t)$, together with the \emph{actual} configuration of the universe $Q$ determined by the $N$ positions $\textrm{\textbf{Q}}_k(t)$ of the particles
\begin{equation}
	 Q(t)=(\textrm{\textbf{Q}}_1(t),\textrm{\textbf{Q}}_2(t),...,\textrm{\textbf{Q}}_N(t)),
\end{equation}
taken at all times to possess well-defined values. Here, $q(t)=( \textrm{\textbf{q}}_1,...,\textrm{\textbf{q}}_N )$ denotes the \emph{generic} configuration, in contrast to the actual $Q(t)$. The wave function is postulated to always satisfy the Schrödinger equation
\begin{equation}\label{SchrodEq}
	i\hbar\frac{\partial \Psi}{\partial t}= - \sum_{k=1}^{N} \frac{\hbar^2}{2m_k}\nabla^2_k\Psi + V(q)\Psi
\end{equation}
and the particle configuration to evolve according to the guide equation
\begin{equation}\label{GuideEq}
	\frac{d \textrm{\textbf{Q}}_k}{d t} = \frac{\hbar}{m_k} \textrm{Im} \left[  \frac{\nabla_k \Psi(q)}{\Psi}  \right]\bigg|_{q=Q(t)}.
\end{equation} 

This covers the dynamical and ontological aspects in the standard presentations of the theory---a deterministic framework in which particles have well-defined trajectories, as in classical mechanics, but whose motion is, in contrast, influenced nonlocally by the universal quantum state.. However, to make this theory empirically consistent with standard quantum mechanics, more needs to be said---specifically, how the actual positions of de Broglie–Bohm particles relate to the probabilistic predictions of standard quantum mechanics derived from the wave function in generic measurement outcomes.

In order to answer this, it is commonly assumed that the positions at some initial time $t=0$ were randomly distributed according to $\rho_0=|\Psi_0|^2$. It follows from (\ref{SchrodEq}) and (\ref{GuideEq}) that if a density is of the form $\rho=|\Psi|^2$  (known as the \textit{quantum equilibrium} distribution) it will satisfy the continuity equation
\begin{equation}
		\frac{d \rho}{d t} = - \sum_{k=1}^{N} \nabla_k \cdot  \textbf{j}_k, \ \ \ \ \textrm{with}\ \ \ \ \textbf{j}_k= -\frac{\hbar}{m_k} \textrm{Im} \left[ \Psi^* \nabla_k \Psi(q)  \right].
\end{equation} 
This, in turn, implies its \emph{equivariance}, i.e., that if $\rho$ has a $|\Psi|^2$ form at some initial time, it will maintain that functional form at all other times\footnote{The assumption that $\rho$ has a $|\Psi|^2$ form at the \textit{initial} state of the universe is frequently called the `quantum equilibrium hypothesis' \cite{durr1992quantum}.}. From this point, it is claimed that complete empirical agreement with the quantum formalism can be demonstrated, including the derivation of various elements such as the Heisenberg uncertainty principle and the operator formalism used in measurement scenarios where the collapse postulate and the Born rule are typically invoked. Let us now see how this is achieved.

\section{The empirical agreement with standard quantum mechanics}\label{Section:EmpiricalAgreementBohm}

First, we should emphasize that, to assess the empirical agreement between the pilot-wave theory and standard quantum mechanics, it is necessary, but not enough, to show that the configurational distribution $\rho(q)$ of the $N$ particles composing the de Broglie-Bohm universe, has a $|\Psi|^2$ form as suggested by the standard Born rule. To properly cover the assessment, we should also make sure that the universal statistical properties are, in fact, inherited by the configuration of the subsystems, and that all the standard measurement predictions can be explained from there. Note that it is not obvious that in a de Broglie-Bohmn universe we can treat any subsystem of $n < N$ particles with configuration $X=(\textrm{\textbf{X}}_1,...,\textrm{\textbf{X}}_n)$, as possessing a wave function $\psi(x)$ that follows the Schrödinger equation, and defines a guide equation, like $\Psi(q,t)$ in  (\ref{SchrodEq}) and (\ref{GuideEq}). In fact, it can be seen that this does not happen in general (e.g., when the subsystem is part of an entangled system). However, we will see that it is possible to always define a so-called \emph{conditional wave function} for the subsystem, such that the proper statistical properties follow for its configuration, conditioned on the configuration of its complement, or environment. A purported, in principle, limitation of knowledge arises from here, which will be the first key result---known as \emph{absolute uncertainty}---to argue for the empirical equivalence of pilot-wave theory with standard quantum mechanics \cite{durr1992quantum}. 

 Furthermore, we will see that in macroscopic measurement-scenarios, an \emph{effective wave function} can be defined for subsystems.  The second key result, the \emph{POVM theorem}, depends on it.  This result starts from considering an, allegedly, natural generalization of all kinds of measurement situations in the pilot-wave perspective, to derive a statistical formalism for the predictions of any possible outcomes from the statistical behavior of the de Broglie-Bohm configurations, associating to every measurement situation a POVM \cite{durr2004quantum,BeckLazarovici2025}. This, notably, turns out to be equivalent to the Hilbert space operator formalism of standard quantum mechanics, whose statistical predictions rely on a collapse rule for the wave function and the Born rule for the operators associated to the observables. This formalism then motivates a `measurability criterion' that restricts the physical properties that can be measured. 
 
 However, to obtain both results, an assumption on records (or `knowledge') is made, which we will call the \emph{instantaneous configurations-only assumption}. The purported limitations or restrictions to knowledge and measurability arise from it.  After reviewing both results, we shall explain why the assumption is unjustified and then explore the serious consequences of letting it go. First, let us lay out the definitions of conditional and effective wavefunction and their relation.

\subsection{The notion of conditional and effective wave function}

Consider in general a subsystem of particles in a given \textit{actual} configuration $X$. The configuration of the entire de Broglie–Bohm universe can be thought as being split into that of the subsystem and its environment and denoted by
\begin{equation}\label{psisplit}
	Q=(X,Y),
\end{equation}
where $Y$ represents the configuration of the environment\footnote{Think, if you wish, of a target subsystem $\mathcal{X}$ and a measurement apparatus $\mathcal{Y}$. Their mutual interaction could leave a trace, or `record', in $Y$, about the subsystem's state, e.g., about $X$.}. For now, this considerations do not depend on the size of the subsystem, nor on observers. The question of interest is if the subsystem can have its own wave function $\psi(x)$, and if it shows the right statistical implications, as it is assumed in the standard quantum formalism. To achieve this, the notions of effective and conditional wave functions are employed.

The decomposition in (\ref{psisplit}) suggests us to express the universal wave function in terms of the \textit{generic} configurations $x$ and $y$, as
\begin{equation}
	\Psi=\Psi(x,y).
\end{equation} 
In particular, we will say that a subsystem has an \emph{effective wave function} $\psi^{eff}(x)$ at time $t$, if $Y \in \textrm{supp }\Phi$, if the universal wave function satisfies 
\begin{equation}\label{effecpsi}
	\Psi(x,y)=\psi^{eff}(x)\Phi(y)+\Psi^{\perp}(x,y),
\end{equation}
at the same time $t$, and if $\Phi$ and $\Psi^{\perp}$ have \textit{macroscopically} disjoint supports on $y$. Here, macroscopically means that the results, as obtained e.g., by the orientation of pointers, are determined by a function of $y$, whose values in the support of $\Phi$ are macroscopically \textit{distinguishable} from its values in the support of $\Psi^{\perp}$. This notion is motivated by noting that in the standard quantum mechanics formalism, in contrast to the pilot-wave case, we don't have an actual configuration $Y$ at all times, that selects a corresponding single outcome associated to $\Phi$, but, after the measurement, we end up with the linear combination $\Psi_{qm}=\sum_{\alpha}\psi_{\alpha}\otimes\phi_{\alpha}$. And although in the standard formalism every $\phi_{\alpha}$ is a possible end state, just one $\phi_{\alpha_0}=\Phi$ is selected after the measurement by the collapse mechanism, with probabilities given by the Born rule. Thus, defining $\Psi^{\perp}:= \sum_{\alpha\neq \alpha_0}\psi_{\alpha}\otimes\phi_{\alpha}$, we can rewrite $\Psi_{qm}$ as in (\ref{effecpsi}). With this, it can be said that when the standard formalism assigns a wave function to a system, we can also assign it as an effective wave function in pilot-wave theory. 

Given the vagueness on the above the definition concerning what counts as macroscopic, it will be useful only for practical purposes, but otherwise, we can work with the more precise notion of the \emph{conditional wave function} 
\begin{equation}\label{condpsi}
	\psi(x)=\Psi(x,Y).
\end{equation}
Note that $\psi$ here, works as a tool for enconding information about a system independently of its configuration, and depending only on the actual configuration $Y$ of its environment. In  \cite{durr1992quantum} and \cite{durr2004quantum}, it is claimed that the conditional or effective wave functions possess the right empirical statistical regularities to make the de Broglie-Bohm theory agree with the standard formalism regarding `knowledge' and measurements. Let us now review these claims.

\subsection{Absolute uncertainty}\label{sSection:AbsUnc}

 As we saw above, in the de Broglie-Bohm universe we are assuming, at the beginning, for a given initial $\Psi_0$, that the initial actual configuration $Q_0$ was selected randomly according to the quantum equilibrium distribution, that is: $\textrm{\textbf{P}}(dq)=\textrm{\textbf{P}}^{\Psi_0}(dq)=|\Psi_0(q)|^2$. Then, by the equivariance property, the distribution at later times $t$ of the total configuration  of the universe satisfies $\rho_t(q)=|\Psi_t(q)|^2$. Let us consider now, what can be said about the distribution of particles of a subsystem, assuming knowledge of the universal wave function $\Psi_t(q)$.  The conditional probability distribution of the configuration of a subsystem, given the configuration of its environment (associated to the corresponding splitting $q=(x,y)$) is
 \begin{equation}
 	\textrm{\textbf{P}}(X_t \in dx | Y_t)=|\Psi_t(x,Y)|^2 dx.
 \end{equation}
  And from the definition of a conditional wave function (\ref{condpsi}) at time $t$, we can write this as the so-called fundamental conditional probability formula\footnote{Given that the effective wave function and the conditional one coincide when de former can be defined, Eq. (\ref{absoluteuncertainty}) is also true for the effective wave function (\ref{effecpsi}) in such case.} 
 \begin{equation}\label{absoluteuncertainty}
	\textrm{\textbf{P}}(X_t \in dx | Y_t)=|\psi_t(x)|^2 dx.
\end{equation}
If we now further assume that \emph{all} the information about the $x$-system can \emph{only} be registered, or encoded, in the configuration $Y$ of its environment, or, in other more `practical' words, that all detection results can only be \textit{configurationally grounded} at an instant, then, `the wave function $\psi_t$ of a subsystem represents maximal information about its configuration $X_t$'. As said in \cite{durr2004quantum}, the formula (\ref{absoluteuncertainty}) is `the cornerstone' in the standard pilot-wave analysis on the origin of randomness and empirical adequacy. This is the so called \textbf{absolute uncertainty}. Thus, \cite{durr1992quantum} states that, knowing the wave function of a subsystem to be $\psi_t$, it is in principle impossible to have more information, or knowledge about the configuration of the subsystem than what is already contained in $|\psi|^2$; even if the most detailed information of the configuration $Y_t$ of the environment is available, i.e.,
\begin{equation}\label{absoluteuncertainty2}
	\textrm{\textbf{P}}(X_t \in dx | Y_t)=\textrm{\textbf{P}}(X_t \in dx | \psi_t)=|\psi_t(x)|^2 dx,
\end{equation}
provided, again, that we have assumed that the information about any physical properties of a system can only be gathered through \textit{instantaneous} configurational records.
The authors of \cite{durr1992quantum} highlight the `most noteworthy' consequence of absolute uncertainty saying, that:
\begin{quote}
	In a universe governed by Bohmian mechanics there are sharp, precise, and irreducible limitations on the possibility of obtaining knowledge, limitations which can in no way be diminished through technological progress leading to better means of measurement. (Durr et. al \cite{durr1992quantum})
\end{quote}

\subsection{The POVM theorem}\label{sSection:POVM}

The standard quantum formalism is a recipe to get predictions, not just for the positions of particles, but for all kinds of observables in actual measurement situations. The proponents of the de Broglie-Bohm theory have an analysis that it is claimed to explain, within their theory, the empirical success of the general Hilbert space operators formalism.

We will now give a brief outline of their analysis, which starts from the following rough characterization of measurements in the de Broglie-Bohm framework. In a measurement scenario, there is an interaction between a macroscopic measuring apparatus and a `target' system to be measured that can be microscopic. Their interaction is assumed to be, for all practical purposes, isolated from the rest of the world. Thus, an initial universal wave function $\Psi_0(x,y)$ is assigned to them, with $x$ the configuration of the system and $y$ that of the apparatus. Moreover, it is assumed that the conditional wave function of the target system can be prepared (say, with a macroscopic apparatus) initially, such that it is an effective wave function $\psi(x)$. After this, the measurement apparatus is left to interact with the target system, so that it ends encoding information of the target system, as the measurement outcome, in its final configuration. According to \cite{durr2004quantum}:
\begin{quote}
	It is important to bear in mind that regardless of which observable one chooses to measure, the result of the measurement can be assumed to be given configurationally, say by some pointer orientation or by a pattern of ink marks on a piece of paper. 
\end{quote}
Now, concerning the quantum state of the whole system, target+apparatus, the wave function evolves unitarily 
\begin{equation}
	\Psi_0(x,y)\rightarrow\Psi_T(x,y)=\sum_k \alpha_k\psi_k(x,y)
\end{equation}
where the $\psi_k(x,y)$ can be said to be localized in macroscopically disjoint regions of the apparatus configuration space. Then, due to their isolation from the rest of the world, we can say that the total final configuration $Q_T$---which includes the final pointer orientation---is in quantum equilibrium $|\Psi_T|^2$. In other words, the result of the measurement is effectively random. Concretely speaking, the probabilities for the final position of the pointer are given by \cite{BeckLazarovici2025}
\begin{equation}
	p_k=\int_{G_k}|\Psi_T(x,y)|^2dxdy,
\end{equation}
with $G_k$ the macroscopically disjoint regions in the apparatus configuration space, whose associated $y-$supports of $\psi_k(x,y)$ correspond to different, and macroscopically distinguishable, final configurations of the measuring apparatus. 

We can encode the above statistical predictions in POVMs in the following way. First, we will say that, \emph{if} we perform a measurement on a target system with initial effective wave function $\psi$, then, the completed experiment will lead to an outcome $Z$, encoded in, and read  \emph{just} from, the final \emph{instantaneous} configuration $Q_T$, trough a \emph{calibration function} $F$, such that
\begin{equation}
	Z=F(Q_T).
\end{equation}
Second, a probability distribution for $Z$ can be given. This is done with the measure 
\begin{equation}
	\rho_{\psi}^{Z} = \rho_T \circ F^{-1},
\end{equation}
induced from the quantum equilibrium measure $\rho_T$, which we already know to describe the random distribution for $Q_T$. 
And third, given that this can be seen as a quadratic map in the initial wave function $\psi$
\begin{equation}
	\psi \rightarrow 	\rho_{\psi}^{Z},
\end{equation}
the Riesz representation theorem says that it can be associated to a POVM. Thus, within the de Broglie–Bohm theory, any measurement scenario that can be described in this way---which, according to proponents of the theory, includes all possible measurement scenarios—has predictions that can be encoded in some POVM. Note that this scenario concerns not a succession of configurations—that is, a set of different pointer positions at successive times---but only an instantaneous final configuration $Q_T$.

In \cite[section 7.1]{durr2004quantum}, the strength of the POVM theorem is expressed after employing a condition claimed necessary for \emph{measurability}, motivated from the POVM theorem and stating that:
\begin{itemize}
\item[\namedlabel{Meas-Crit}{MCr}: ] A quantity is not measurable when there is a value for it, which is possible
	when the wave function is $\psi_1+\psi_2$, but not possible when the wave function is either $\psi_1$ or $\psi_2$.
\end{itemize}
It happens that, in the de Broglie-Bohm theory, the properties that can be ultimately characterized just by the configurations are the ones that fulfill this criterion. Thus, there is no outcome $Z$, such that it irreducibly arises from final \emph{non-}configurational degrees of freedom, e.g. $Z=F(\dot{Q}_T)$\footnote{Note that since every wave function $\psi$ can be written as a sum of its real and imaginary parts, and the guiding equation implies the that the velocity is zero for all real or imaginary wave functions, then, the criterion imposes  the non-measurability of velocity.}. Conceding this, then, \emph{any} measurement can be described by a POVM within such a de Broglie-Bohm universe.

\section{Questioning the strength of Absolute Uncertainty and the POVM theorem}

From the above two results the entire empirical content of pilot-wave theory is derived and complete empirical agreement with the standard quantum predictions is claimed. The uncertainty principle, as well as paradigmatic conclusions as the, `in principle', empirical inaccessibility to exact trajectories, to the absolute time ordering of space-like separated events, or even the non-detectability of the privileged spacetime foliation in relativistic extensions of pilot-wave theory, all rest on the validity and strength of absolute uncertainty or the POVM theorem \cite{durr1992quantum,Maudlin1996,PhysRevA.53.2062}. However, the validity and strength of these results depend, in turn, to the validity and justification of a key assumption on how information is encoded.  Here we will expose and analyze more deeply this assumption.

\subsection{The instantaneous configurations-only assumption}

Concerning absolute uncertainty, we saw above that from the fundamental conditional probability formula (\ref{absoluteuncertainty}) we can only claim absolute uncertainty to be valid if we further assume that: 
\begin{itemize}
	\item[\namedlabel{configurations-only}{IC}: ]   ``Information is always configurationally grounded''.
\end{itemize}
It is indeed the case that the formula (\ref{absoluteuncertainty}) is always true within the de Broglie-Bohm theory. That is, it is indeed the case that conditional just on $Y_t$, what can be said about the distribution of the particles of a target system is always exhausted by the Born rule. However, the assumption (\ref{configurations-only}), goes beyond first principles of pilot-wave theory. It asserts that: 
\begin{itemize}
\item[\ref{configurations-only}] All information (or knowledge) about the state of a system, must be contained in (or gathered from), the instantaneous \emph{positions} of the particles of the environment of the system (or agents external to it). 
\end{itemize}
Let us emphasize that, only if we assume this constraint on how the information can be encoded (\ref{configurations-only}), the simple formula (\ref{absoluteuncertainty}) turns into an in principle limitation on the possibility of acquiring knowledge about something postulated to be real and objective within the de Broglie-Bohm theory.

The assumption (\ref{configurations-only}) forbids, for example, something like `velocity patterns' as degrees of freedom responsible for encoding information---and, more generally, non-configurational degrees of freedom. It is easy to see that one can escape the \emph{knowledge limitation} by dispensing with this prohibition. Note that, in general, $\dot{Y}_t$ is a function of $X_t$, so knowing $\dot{Y}_t$ may give us additional information about $X_t$.

A concrete example might be more helpful. Lets consider two particles in one dimension, with $x$ and $y$ their respective positions, in the state 
\begin{equation}\label{examplestate}
	\Psi(x,y) = \frac{1}{\sqrt{2}} \left[\phi_a(x) \varphi_p(y) + \phi_{-a}(x) \varphi_{-p}(y) \right]
\end{equation} 
with $\phi_{\pm a}$ a wave function for the $x-$particle with highly localized position around $\pm a$ and $\varphi_{\pm p}$ a wave function for the $y-$particle with highly localized momentum around $\pm p$. According to the formula (\ref{absoluteuncertainty}), knowledge only about $Y$ would not provide more information about $X$ than what is already exhausted from knowing the quantum state. In particular, the state (\ref{examplestate}) implies a 50/50 chance for the position of the $x-$particle to be $\pm a$. Notwithstanding, we can break the limitation to knowledge if we knew $\dot{Y}$; as this would allow us to determine whether $a$ or $-a$ is the case, with absolute \emph{certainty}.

 One could argue that in pilot-wave theory, possessing first-order equations, where the wave function and the positions at time $t$ fully determine all velocities at the same time, all information about $\dot{Y}$ is already contained in $Y$. This, however, is not the case, as $X$ is needed to compute $\dot{Y}$, according to the non-locality of the theory when there is entanglement between the system and the environment. Thus, knowledge about $Y$, does not generically confer knowledge about $\dot{Y}$. Of course, without entanglement between a system and its environment, all that can be said about the configuration of the system by an external agent, is exhausted by the Born rule. Thus, in general, quantum equilibrium does not convey a limit on knowledge.
 
 Let us now tackle the case of the POVM theorem. It is only through the assumption that all the relevant information of a target system is encoded in the \emph{position} of the pointer in the measuring apparatus (i.e., (\ref{configurations-only})), that we can write the outcomes as $Z=F(Q_T)$ ---and not in terms of other properties like velocity--- so that quantum equilibrium becomes relevant, providing the quadratic map $\psi \rightarrow\rho_{\psi}^{Z}$, and leads to the POVM theorem. If, on the other hand, the result $Z$ is allowed to be encoded in terms of, say, the velocity of the pointer, the conditional probability formula (\ref{absoluteuncertainty}) does not convey maximal information (i.e., absolute uncertainty), so the map from the initial wave function to the final prediction is not necessarily quadratic, and the POVM is not derived.
 
 Here, one could object that, even if we allow for the result of a measurement to be recorded, at some time, in terms of both, the position and velocity of the pointer, it would be possible to record all that information only in terms of the instantaneous position of another pointer. If this were valid, then, the assumption that all the measurement records can be encoded in the positions would be justified. In more concrete terms, suppose that a measuring apparatus $M$, encodes its results of the interaction at a time $T_1$ with a system $S$, in terms of its degrees of freedom $Y_{T_1}$ and $\dot{Y}_{T_1}$. The question is if it could be possible, in general, to make $M$ interact with another apparatus $N$, such that, at a later time $T_2$, its configuration $Z_{T_2}$ contains the information of $Y_{T_1}$ and $\dot{Y}_{T_1}$. If it is possible, then, we could say that the actual measurement ended at $T_2$, and at $T_1$ it was not a completed measurement.
 
 However, as can be seen from the analysis of measurability in \cite[section 7.1]{durr2004quantum}, it is possible to show that, in general, $Y_{T_1}$ and $\dot{Y}_{T_1}$ cannot be recorded in terms of $Z_{T_2}$. This is because, it is a dynamical fact that, the properties fulfill the so-called measurability criterion (\ref{Meas-Crit}) are precisely the ones that can be encoded solely in terms of positions \footnote{If $\dot{Y}_{T_1}$ could be recorded only in terms of $Z_{T_2}$, then, to every distinct value of the former, there would correspond a distinct value of the latter, otherwise, those values would not be distinguishable by knowing $Z_{T_2}$.}. In particular, given that non-configurational degrees of freedom, like velocity, does not fulfill the criterion and the POVM theorem, such things like velocity patterns cannot be translated into configuration patterns.
 
 We conclude then that the validity and strength of absolute uncertainty and the POVM theorem, crucially rests on the validity of the assumption (\ref{configurations-only}): that all information must always be configurationally grounded. We shall now assess if there is any justification for the configurations-only assumption.
 
 \subsection{Is there a justification for (\ref{configurations-only})?}\label{ssect:CI}
 
 As we discussed in the previous chapter in section \S \ref{section:detectability}, on the general characteristics about detections, it is only through the \emph{dynamics} of a given theory that we can decide which degrees of freedom are responsible for encoding records. Recall the simple example of a universe whose particles are postulated to have $A-$properties and $B-$properties (as equally real and objective). If the dynamics in this simple universe is such that there is no inter-type interactions between the $A$ and $B-$properties, then, a statement placing the $A-$properties as preferred concerning the encoding of records would be simply an unjustified \emph{imposition}. And to accept as valid such an assumption could not lead, of course, to an in principle limitation of knowledge, regarding for example, the information about $B-$properties. Rather, with such dynamics, we should only conclude that a measuring apparatus that can give results just in terms of the $A-$properties, is the wrong choice to detect something about the $B-$properties\footnote{As for another contrasting example, consider Classical Mechanics. Can we say that all information is grounded in particle positions? In this case, it seems we indeed can, as it is allowed by the dynamics. That is, the dynamical equations makes it possible to form the proper correlations between a property like velocity and the positions. In this sense, the velocity of a system (e.g. a bike) can be recorded in terms of the position of another system (e.g. the pointer of a tachometer), so, speedometers can be constructed in this theory.}.

With the above in mind, can we assume within the pilot-wave theory that all the information that an external agent can have about a system is encoded in the configuration of its environment? The dynamics of the theory, i.e., the guide equation and the Schrödinger equation, have already answered this for us. We argued at the end of section \S \ref{ssect:CI} that velocity patterns can not be translated into configuration patterns. Thus, in line with our discussion on detections, to assume (\ref{configurations-only}) would amount to an unjustified imposition. What we can actually conclude from the POVM theorem and absolute uncertainty is that measuring apparatuses whose results are just in terms of positions, are bad apparatuses to measure non-configurational degrees of freedom. 


Another line of defense for assuming (\ref{configurations-only}) appeals to our experience in our own world, which might not be of the de Broglie–Bohm type. For instance, the authors of \cite{durr1992quantum} fully endorse Bell's statement:
\begin{quote}
	...in physics the only observations we must consider are position observations, if only the positions of instrument pointers. It is a great merit of the de Broglie-Bohm picture to force us to consider this fact. (\cite{Bell1982-BELOTI})
\end{quote}
Certainly, it might be natural and easiest to think of real measurement results by means of instantaneous `configurational representations', however, to be physically meaningful to us they also have to be `stable' through an interval of time \cite{BeckLazarovici2025}. A printed picture of a thermometer can suffer deterioration over time and still convey what what the temperature of the milk was. However, we should bear in mind that these are common conceptions of our daily life in \emph{our} world. We cannot take for granted that a particular theory engenders a world like ours without any empirical disagreement. That is precisely the question. To artificially restrict a theory so that its objective properties can only be detected or encoded in the ways that are most natural to us would be to beg the question.

In this sense, de Broglie–Bohm theory forces us into a choice: we must impose (\ref{configurations-only}) if we want to enforce, by hand, that pilot-wave predictions are fully concordant with the empirically verified standard quantum predictions in cases where the latter are unambiguous. This, of course, can be done, but it should not be confused with derivations or explanations from first principles of the theory. Otherwise, by dispensing with (\ref{configurations-only}), we can look for new empirical results that are not possible within the artificially restricted theory, nor within standard quantum mechanics.

Moreover, we are not really sure that, in actual practice, we are only capable of observing the instantaneous positions of things; it also seems that we feel accelerations, see velocities, or `sense' the passing of time. All of these matters remain under debate, with no sign of a consensus being reached anytime soon. So, when \cite{durr1992quantum} claims in footnote 9 that:
\begin{quotation}
	\noindent ...results of measurements must always be at least potentially grounded configurationally, in the sense that we can arrange that they be recorded in configurational terms without affecting the result. Otherwise we could hardly regard the process leading to the original result as a completed measurement.
\end{quotation}
It is difficult not to question it. Our perceptions might not supervene on instantaneous brain states. Our brains might store information in non-configurational terms.

Therefore, with all that has being said, we can further state another conclusion concerning the de Broglie-Bohm universe saying that: \emph{it could be a prediction of the pilot-wave theory that there are things that can be known but can not be communicable just through instantaneous configurations}. If this turns out to be the case, and if it turns out that our own world does not allow for this, then, it will be the theory with (\ref{configurations-only}) the one in trouble.

\section{Consequences}
\label{Section:ConsequencesBohm}

Absolute uncertainty is said to impose an in principle limit on what can be known. Similarly, the POVM theorem is argued to severely restrict the set of properties that can be measured. Both results are essential to ``hide'' the extra structure within the pilot-wave framework and deliver an empirical agreement between the theory and standard quantum mechanics. We argued, however, that this results depend upon the assumption that information is always configurationally grounded---an assumption that cannot be justified from first principles. We grant that, with the configurations-only assumption in place, the predictions of pilot-wave theory do coincide with those of standard quantum mechanics, but when such key assumption is lifted, then all bets are off. Below, we briefly explore the potential consequences of letting go of absolute uncertainty and the POVM theorem.

\subsection{Certainty: violation to Heisenberg's principle}
Without the assumption of configurationally grounded
detections, we can gather more information about the actual configuration $X_t$ of a subsystem than what is contained in the configuration $Y_t$ of its environment. The possibility of complete measurements whose records are not strictly configurational opens the possibility to obtain information about, for example, the velocities $\dot{Y}$ of the environment. This, through the guide equation (\ref{GuideEq}) and its explicit non-locality, give us additional
information about the positions $X_t$ than what is contained in $|\psi_t(x)|^2$. For instance, in a two-path experiment this makes it possible to know which path was actually taken by a given particle \footnote{See the state (9) of the concrete example considered in section 4.1.} That is, in
pilot-wave theories the particle trajectories are not so hidden after all. Thus, we can see that the assertion in \cite{durr1992quantum}, that 
\begin{quotation}
	...from the perspective of a Bohmian universe the uncertainty principle is sharp and clear. In particular, from such a perspective it makes no sense to try to devise \emph{thought} experiments by means of which the uncertainty principle can be evaded, since this principle
	is a mathematical consequence of Bohmian mechanics itself. One could, of course, imagine a universe governed by different laws, in which the uncertainty principle, and a great deal else, \emph{would} be violated, but there can be no universe governed by Bohmian mechanics---and in quantum equilibrium--- which fails to embody absolute uncertainty and the uncertainty principle which it entails.
\end{quotation}
is not correct. According to our analysis, \emph{in} such a universe, the uncertainty principle is not a mathematical consequence; particle configurations can be in quantum equilibrium and still, knowledge of the configuration of a subsystem may not be strictly mediated by its wave function, meaning that $\psi$ does not represent maximal information about its configuration.

\subsection{The possibility of sending faster-than-light signals}

It has also been commonly claimed that faster-than-light signaling is not possible within pilot-wave theory. For instance, in \cite{valentini1991b} it is claimed that this is true if, and only if, quantum equilibrium holds, as in the case of the uncertainty principle. However, this claim (at least the converse implication) necessarily depends on absolute uncertainty and, therefore, on the assumption that all measurement outcomes are configurationally grounded. To see this, it should simply be noted again that, in the alleged proofs showing that quantum equilibrium implies no-signaling, it is already assumed that  $\psi_t(x)$ conveys maximal information about the configurations $X$ (although sometimes this is not recognized explicitly).  Provided that it is not justified to impose the configurations-only assumption, the actual position of the particles can be measured to a better degree than what the Born rule dictates and nothing prevents us from assuming that probability distributions could generally depend on other degrees of freedom that are not configurationally grounded. And with this, to be able to devise, within Pilot-wave universe, a protocol to send superluminal signals.

Think, for example, in an EPR-type setting, where a pair of entangled particles in a state like (9) are sent, one to Alice and the other one to Bob. It is standard in the literature the claim that pilot-wave theory explains Bell's correlations, without letting us know if Alice or Bob measured first, (i.e., hiding the absolute time-ordering structure). The reasoning behind alludes to the inaccessibility to the actual particle trajectories.


\subsection{Detectability of the privileged foliation in relativistic extensions of Pilot-wave theory}

To construct a relativistic pilot-wave model, where a wave function or functional, satisfying an appropriate relativistic wave equation, ``guides'' the motion of either particles or fields, the most natural way to write the dynamical laws (and the only known way that could recover standard quantum predictions), is to add first a preferred time foliation $\mathcal{F}$ to the spacetime $M$. This foliation plays the role that absolute simultaneity plays in the non-relativistic theory, a fact that seems to imply a violation of Lorentz invariance. It has been argued, though, that the proposed relativistic pilot-wave models are such that a precise kind of ``conspiracy'' within them can be shown to hide from observation any violation to relativity, thus making the preferred foliation completely undetectable. This, it is said, implies that these models are fully Lorentz invariant at the empirical level. 

However, based on the previous sections, we believe that all arguments in favor of the unobservability of the preferred foliation are deficient, since they take for granted: absolute uncertainty, the POVM theorem, the configurations-only assumption, the invariance under Lorentz transformations of the possible measurement outcomes, or a combination of all these. For concreteness, below we develop our argument in the context a particle theory, but an analogous argument would also run with fields.


Consider, then, an N particle relativistic pilot-wave model, with wave function $\psi$ and particle world-lines $Q_{i}(\tau)$ for all $i\in(1,\dots,N)$.  Then, we can make sense of the relativistic generalization for the guidance equation considering the spatial location of the $N$ particles in a ``simultaneity'' slice  $\Sigma_{\tau}$ contained in a privileged foliation $\mathcal{F}$. For simplicity take $M$ to be the Minkowski spacetime, foliated into spacelike parallel hypersurfaces $\Sigma=\Sigma_{\tau}$, parameterized by a time function $\tau$ in a Lorentz frame\footnote{Thus, $Q_i(\tau_0)$ denotes the intersection of the $i-th$ world-line with $\Sigma_{\tau_0}$ and the guide equation for the $i-th$ particle reads as
	\begin{equation}\label{relbohm}
		\frac{d Q^{\mu_i}}{d \tau}=\frac{j^{\mu_i}}{\psi^*\psi}(Q_1(\tau_0),...,Q_N(\tau_0)), \ \ \ \ \ \ \  \textrm{with}\ \ \mu_i=1,2,3.
	\end{equation} 
	Where in the case of $N$ Dirac particles $j^{\mu_i}=\psi^* \mathbf{\alpha}_i \psi$, with the matrices $\mathbf{\alpha}_i$ acting on the spin index of particle $i$. See chapter $7$, section $7.6$ of \cite{tumulka2022foundations}}. Here, ``simultaneous'' means ``lying on the same leaf of $\mathcal{F}$''.

As mentioned above, the outcomes of any detection or measurement performed by the system must end up encoded as a function of its objective features, which in this case means $Z= Z(\psi,Q_{i}(\tau))$. Note that, given the presence of the preferred foliation $\mathcal{F}$, the angle between the tangent to the world-line of each particle, when it crosses the hypersurface $\Sigma_{\tau}$, and the normal to $\Sigma_{\tau}$ at the intersection, is an objective non-configurational feature of the system on which detections might depend on. 

Suppose, now, that the system undergoes an active Lorentz transformation $\Lambda$. The transformed system would then be described by a wave function $\psi^{\Lambda}= \Lambda \psi$ and world-lines $Q_{i}^{\Lambda}(\tau)= \Lambda Q_{i}(\tau)$. Given that the wave function is assumed to satisfy a Lorentz invariant, relativistic wave equation, the transformed wave function would be a solution to such an equation. It is not obvious, though, given the presence of $\mathcal{F}$, why would it influence the trajectories of the particles, but not the wave function. And, of course, if that would be the case, then the transformed wave function would not necessarily be a solution to the dynamics, and the preferred foliation could be observable. Although the assumption that the wave function satisfies a Lorentz invariant equation is standard, one should note that it corresponds to the conditional wave function defined for arbitrary subsystems and there is no guarantee that it would be a solution to a Lorentz invariant equation. Of course, one might respond to this argument saying that the conditional wave function only depends on configurational degrees of freedom, which, as we shall see in a moment, are Lorentz invariant. Therefore, the argument goes, the conditional wave function is Lorentz invariant. However, for reasons that will become apparent in due time, nothing prevents us from assuming the wave function to generally depend on additional non-configurational, Lorentz-variant, objective quantities of the system. Regarding the particle trajectories, given that those do depend on $\mathcal{F}$, for them it is, indeed, not the case that the transformed world-lines are a solution of the equations of motion. Still, the claim that the preferred foliation is undetectable is argued for as follows. 

It can be shown that if one assumes the density of crossings satisfies the standard quantum statistics on an initial leaf of the foliation, $\Sigma_{0}$, then this will hold for all leaves of the privileged foliation. This is not true, though, for all hypersurfaces. In fact, in \cite{PhysRevA.53.2062} it is shown that the quantum distribution cannot simultaneously be realized in all Lorentz frames. In spite of this, what can be shown is that, when \emph{measurements} are performed along any hypersuface, the presence of detectors produces that the density of crossings measured always corresponds to the standard quantum results.\footnote{Note that \cite{PhysRevA.53.2062} show this for a particular type of experiment (called Hardy's experiment), where parts of the wavefunction are recombined and measurements are performed along any hypersuface. However, this result is generalized by \cite{Lienert-Tumulka} for any conceivable experiment; they mainly claim that in the presence of ``ideal detectors'' on $\Sigma$, the joint distribution of the N intersection points is a curved Born distribution regardless of whether or not $\Sigma$ belongs to $\mathcal{F}$.} From this, together with absolute uncertainty, it is concluded that the foliation is undetectable. It must be noted, though, that, in order to conclude this, they assume that all recorded outcomes should be \emph{stable}, meaning that these outcomes should be Lorentz invariant \cite[14]{Lienert-Tumulka}. One, certainly, can simply postulate for this to be the case, but then one would not be showing the foliation to be undetectable, but would be, instead, only putting the feature in by hand.  

Thus, in order to conclude that the foliation is undetectable, it is not enough to show that the density of measured crossings always corresponds to the standard quantum results; one must also show that all measurement outcomes are Lorentz invariant. In more specific terms, one must demonstrate that measurements performed by the transformed system considered above, from the point of view of the transformed frame, are equal to the corresponding detections by the original system, from the point of view of the original frame, namely, that $Z(\psi,Q_{i}(\tau)) = Z^{\Lambda'}(\psi^{\Lambda'},Q_{i}^{\Lambda'}(\tau))$ for all $Z$. As shown in (DGZ, 1992), the analogous of this equation, for the case of non-relativistic pilot-wave theory, $Z(\psi,Q) = Z^{b'}(\psi^{b'},Q^{b'})$, seems to be true. But this occurs precisely because \emph{it is taken for granted without any justification that all measurement outcomes are configurationally grounded} (i.e., that any detection, measurement or perception performed by the system must end up encoded solely as a function of $\psi$ and $Q$, namely, $Z=Z(\psi,Q)$ for all results $Z$ and for a given $\psi$). It follows from this assumption that the original and transformed systems would behave in the exact same way under all circumstances, meaning that the non-relativistic theory is Galilean invariant. However, without the assumption that all measurement outcomes are configurationally grounded, the above equation cannot be true in non-relativistic pilot-wave theory and, furthermore, it could not be true for the relativistic theory under consideration. This is because $Z$, being a function of the objective features of the system, might very well be also a function of non-configurational degrees of freedom, such as the angle between the tangent to the world-line of the particles, when they cross $\Sigma_{\tau}$, and the normal to $\Sigma_{\tau}$ at such crossings. This objective feature is clearly not invariant under Lorentz transformations. Therefore, there is no guarantee that detections made by the transformed system---when viewed from the transformed frame---will, in general, coincide with those made by the original system in the original frame. And, of course, without this equivalence, one cannot claim that the foliation is undetectable.

\subsection{Arrival-time experiments}

Recently, there has been intense discussion within the pilot-wave community concerning simple time-of-flight experiments that admit a natural description in pilot-wave theory and that, according to some, could lead to significant deviations from the predictions of standard quantum mechanics \cite{das2019arrival,goldstein2024spin,das2023comment}. Such deviations include outcome statistics excluded from the POVM theorem, as well as the possibility of superluminal signaling. Here, I will clarify the relation of these results to our own findings.

In quantum mechanics, it is common to think of measurements involving the probability of finding particles in a given volume around a position at a fixed time. However, it is uncommon, to think of the probability of finding particles at a fixed position in an interval of time. The reason for this is that standard quantum mechanics does not has the dynamical resources to cleanly analyze such type of experiments. However, pilot-wave theories naturally can. The at-all-times defined positions together with the guide equation permits us to devise an experiment where a particle is initially trapped in a potential well, released at a given time to move, and compute the time it would take for it to arrive to a detector at a certain distance.  In \cite{das2019arrival} a pilot-wave formula for the first arrival time of spin-1/2 particles was devised. Their simulations produced statistical results that contrasted with the semiclassical predictions of standard quantum mechanics and could not be captured by any POVM\footnote{Note, also, that if the arrival-times experiments were carried and the results agree with the predictions in \cite{das2019arrival}, then, a protocol could be devised to send signals faster than light.}. However, it should be noted that the influence of the detectors was neglected for simplicity. And it has been suggested that if they were properly modeled, their influence would make the statistical results to be in accordance to the POVM theorem \cite{goldstein2024spin}. Although this suggestion has also been strongly debated \cite{das2023comment}. 

Our analysis showed that the instantaneous configurations-only assumption was not properly justified, so dropping it could lead to deviating predictions from the POVM theorem. However, note that such predictions are in terms of non-configurational degrees of freedom that, due to the pilot-wave dynamics, can not be encoded configurationally. That is, it appears as if the contrasting results could not be `written in a paper', as it is naturally expected in \cite{das2019arrival} and in common experimental scenarios. Thus, if we were to empirically verify such results, it seems that we could not record and communicate them in the usual way. Notwithstanding, it is far from clear that using `ink-on-paper' to encode the information of the \emph{different times} of arrival of particles in these experiments actually amounts to using instantaneous configurations-only. To conclude this, one should model the communication process using the pilot-wave dynamics to ensure that the information one can gather from ink-on-paper in these scenarios is encoded only in the configurations at an instant. Alternatively, one could try to derive a stronger POVM theorem by also adding an assumption about the \textit{stability of measurement outcomes} \cite{BeckLazarovici2025}. We saw in the last subsection an issue with this in the relativistic case; one would incur in question-begging by imposing this in order to prove the non-detectability of the privileged foliation. In the non-relativistic case, one could question the stability assumption by noting that in experimental scenarios like those of arrival times, where the ink-on-paper at the end of the experiment contains information about distinct temporal intervals, there is no reason to employ it. Non-configurational degrees of freedom might still be communicable through ink-on-paper via non-configurational degrees of freedom within the ink-on-paper communication process. Take, for instance, the case of Homer's \emph{Odyssey}: would you be able to understand the sufferings throughout the journey of Ulysses without reading each page at successive, ordered intervals of time?

Of course, I do not take this as a conclusive argument. Much more has to be analyzed in more concrete terms to opt for a full POVMian description of our world or otherwise. The only thing one can be sure to extract from all this is the significance of performing the experiments proposed by Siddhant Das and collaborators following \cite{das2019arrival}.

\section{Conclusions}\label{ConclusionsBohm}

The de Broglie-Bohm theory is indeed a successful resolution to the foundational problems of the standard formalism of quantum mechanics. Its ontological and dynamical clarity, contrary to the standard formalism, allow for a proper description of physical situations. However, apart from having clarity regarding \emph{what} exist and \emph{how} it behaves, it is also expected from a physical theory to be in agreement with actual experiments. Given that situations where the standard formalism can make predictions have been empirically verified, a quantum theory as the de Broglie-Bohm theory, should be able to explain this success. In this work, we reviewed the two main results to argue for such an empirical agreement. The proponents of these claims take it to be a \emph{complete} agreement. That is, there is no place for any deviations when both, the pilot--wave theory and the standard formalism, can make predictions. However, we stressed that those results, absolute uncertainty and the POVM theorem, strongly relied on an assumption concerning `knowledge'. If taken for granted, such results can be interpreted as proofs of limitations to knowledge and restrictions to what can be measured and how to predict the outcomes respectively. Of course, if the assumption on which they depend, already restricts from the beginning what and how can something physical be known, those seemingly profound results are already expected. 

It has been argued that the assumption we called instantaneous configurations-only is not justified. Absolute uncertainty and the POVM theorem thus beg the question on the empirical agreement with standard quantum mechanics. It has been explained in detail why restricting our knowledge of a system to the configuration of its environment is unfounded from first principles, and why defending such a restriction by appealing to empirical evidence is circular. Moreover, such a restriction may not even hold in our world.

Thus, some of the possibilities that arise in a de Broglie–Bohm universe where such an assumption is not imposed by hand were explored. In this context, we found, for instance, superluminal signaling, violations to the Heisenberg uncertainty principle, and the detection of a preferred frame as potential consequences. These results, in sum, call into question the empirical agreement between pilot-wave theory and the standard formalism. However, we must let the empirical evidence be the one that determines whether the de Broglie–Bohm theory is the most viable description of our world or not.

This may point toward further developments in physics. In recent years, proposals have been made to perform more precise arrival-time experiments. Siddhant Das and collaborators have developed models for experimental setups in which one expects to find an empirical distinction between the predictions of pilot-wave theory without the configurations-only assumption, standard quantum mechanics, and pilot-wave theory with the configurations-only assumption. Interestingly, even in these setups---far simpler than extreme situations such as black-hole mergers or other quantum-gravity scenarios---there is still no clear consensus on what the standard quantum formalism actually predicts.

\subsection{A final note about `underdetermination'}

Finally, the general analysis outlined in the first chapter regarding what it means, within a given theory, for a physical quantity or property to be detected has been applied in these two chapters to the cases of Newtonian mechanics and de Broglie–Bohm theory. We saw that when it is claimed that some postulated entity is real but nonetheless non-detectable according to the relevant theory, the arguments do not rest on its fundamental principles but instead on arbitrary and unjustified restrictions concerning which degrees of freedom are allowed to register detection outcomes. In Newtonian theory, only relational quantities are permitted to encode such results; in de Broglie–Bohm theory, only particle positions are allowed.  However, in both theories there are further \textit{real} properties and entities which, without artificially hiding them, the dynamics in each case would allow to have the same characteristics regarding detectability as relational quantities or particle positions. In this sense, it is an artificial restriction---an unjustified imposition by hand employed to `save the phenomena' of the theory.

It could be argued that such restrictions should be part of our theories, since our observations of the world should guide theory construction. This argument, however, contains a gap. While empirical data should indeed guide us, we should not be guided to postulate, at the level of first principles, real entities or properties that must then be artificially hidden. Take, for instance, the construction of modern Galilean spacetime, where absolute space is no longer a real entity that is merely artificially hidden, as it was in Newton's original theory.

Then, I suggest that the strength of arguments appealing to the underdetermination of physical theories on the basis of supposed undetectability should be reconsidered. According to the analysis of detection developed here, one can no longer simply assume that two competing theoretical descriptions of the world might, in principle, yield empirically indistinguishable predictions. This suggests that, whenever the existence of some real entity, structure, or property is posited, there is no principled way to rule out its detectability---at least within the universe associated with that theory.

Therefore, if a theory posits the existence of certain entities or properties but no corresponding detections have been made in our actual world, we are left with two natural options: either to question the validity and empirical adequacy of the theory, or to regard the situation as involving novel predictions that await new experimental results consistent with the theory's claims.
\part*{\hypertarget{PartII}{Part II}: \\ On Explaining the Approach to Thermodynamic Equilibrium }
\addcontentsline{toc}{part}{\protect\numberline{}Part II: On Explaining the Approach to Thermodynamic Equilibrium}

\chapter{The universal approach to thermodynamic equilibrium }\label{Chapter:Equilibrium}

\section{Introduction}\label{Section:IntroEquil}

Thermodynamics is a phenomenological science that codifies our observations and understanding of macroscopic systems and their interactions in terms of a few macroscopic variables or thermodynamic properties. The ultimate nature of such macroscopic systems, as part of the physical universe,  as well as their observed behavior, must ultimately be a consequence of the fundamental nature of the world's elementary constituents and the laws that govern them.

Today, we know that these constituents are quantum in nature. However, the standard formalism of quantum mechanics faces fundamental issues that impede a clear and unambiguous description of the underlying entities and their behavior independently of the act of observation. In contrast, physicists have historically aimed for a \textit{physics} of natural processes that does not depend on ``highly qualified measurer[s]---with a PhD'' (as John Bell famously put it \cite{bell2001quantum}), a view in which, as typically presupposed in thermodynamics, coffee cups left alone unambiguously get colder, and distant stars objectively burn their fuel. To achieve this aim, one should begin by considering fundamental physical theories with a clearly specified ontology and well-defined dynamics. In a slogan, these are theories that can tell us \textit{what there is and how it behaves} \cite{maudlin2018ontological}.

On the other hand, the universally observed time-asymmetric evolution towards a (macro-)state of equilibrium, or `maximum entropy', shown by macroscopic systems, has resisted a general, coherent and objective, explanation from our most fundamental physics. Boltzmann was the first to look for such an explanation from several fronts, when classical mechanics was our most fundamental description of matter in motion. Broadly speaking, he produced two main approaches, namely the H-theorem and the ergodic hypothesis. Although both are insightful and remain highly influential in the study of equilibration, they rest on assumptions that are not grounded in fundamental principles and are applicable only to highly idealized systems.

Moreover, the advent of quantum phenomena makes it indispensable to tackle such reductive project in quantum terms. If the world is quantum, our explanations of its properties should also be. However, almost every modern approach to the subject of `thermalization' treats the properties of the wavefunction and its behavior considering only the unitary evolution part of the standard quantum formalism. Notwithstanding, the macroscopic objects we observe to equilibrate are never in superpositions of macro-states. The fundamental problems of the quantum formalism have not been taken into account. To address this issue, David Albert made a `semi-classical' suggestion employing a quantum collapse theory. In particular, he suggested that the state of a system in classical phase space undergoes stochastic perturbations due to the modified dynamics of quantum collapse theories. These perturbations are then responsible for the trajectories in phase space that purportedly explain the approach to equilibrium. His motivation relied on the fundamentally time-asymmetric dynamics of such theories, which appear to offer a potential advantage for explaining equilibration.

The present work goes a step further. Here, the purported advantage of collapse theories will be concretely assessed. As will be seen, some of the issues of the previous approaches, like being restricted to highly idealized systems or the use of unjustified assumptions, can be partly alleviated.  However, although our proposed explanation of the approach to equilibrium has collapse models as the underlying fundamental dynamics of matter, we shall see that equilibration process is, after all, an approximately classical behavior that quantum macroscopic systems follow. 

This chapter is organized as follows. In \S \ref{RemarksEquil} the general project of explaining the approach to thermodynamic equilibrium will be concretely stated and motivated, and in \S \ref{RealAppSection} a quick response will be given against the objection to take an objective and realist attitude towards accounting, from a fundamental theory, for the universal approach to equilibrium. Then, in \S \ref{ClassicalSection} I will give a brief account of the classical explanations, mainly motivated by Boltzmann, including its objections and limitations. Afterwards, in \S \ref{Section:QuantumCase} I will start motivating the need for a quantum treatment of the subject. Here, it will be argued that a clean description can only be achieved if we first address the fundamental problems of the standard formalism of quantum mechanics. These problems will be briefly reviewed in \S \ref{QMproblemSection}. Then, before describing Albert's `semi-classical' suggestion in \S \ref{Albertsection}, we will review von Neumann's quantum statistical framework in \S \ref{vnFsection}, exposing also some of its issues and limitations as an explanation of the approach to equilibrium. Finally, in \S \ref{Section:OurEquilProp} I will present our own approach in terms of objective collapse theories, where I will try to address the mentioned issues afflicting the previous approaches and I shall be as clear as possible about the persisting issues in our approach. Finally, the conclusions will be stated in \S \ref{Conclusion:Equilibrium} and some further interesting research topics concerning the proposed approach will be mentioned.

 \section{Remarks on the Second Law of thermodynamics and the approach to equilibrium}\label{RemarksEquil}
 
The best-known proposals---mostly developed by Boltzmann---to account for thermodynamic behaviour from fundamental physics attempt to derive the desired phenomena using classical mechanics supplemented with unjustified (i.e., non-fundamental) assumptions that are valid only for a very limited class of systems. This restriction confines their applicability to highly simplified or unrealistic models \cite{Frigg2008Guide}. As a result, classical proposals suffer from a problem of generality, leaving parts of the foundations and philosophy of physics community dissatisfied.
 
 First of all, let's clarify a few things. In this work we are interested in the physical explanation of a specific universally observed macroscopic behavior, which is: the evolution of macroscopic systems, \textit{left on their own}, towards a particular (idealized) macroscopic state---understood as the list of values assigned to a finite set of macroscopic properties (e.g. temperature, volume, pressure, entropy, etc)---or `macro-state', specifically the one known as the \textit{equilibrium} macro-state. This behavior through time is usually imputed to the second law of thermodynamics. In whose terms, the endeavour amounts to the project of identifying an appropriate statistical definition of non-equilibrium entropy and to show that it increases forward in time. However, as is thoroughly stressed in \cite{myrvold2020explaining}, the thermodynamic formulations of the second law and the various definitions of thermodynamic entropy employ a crucial distinction between work and heat, which depend ambiguously on \textit{our} capabilities and \textit{knowledge}. If someone does know how to exploit the energy in a certain system, he would call it work, otherwise he would label it as heat. And, as far as the second law of thermodynamics concerns, equilibrium states \textit{maximize} entropy, but, this says nothing specifically about \textit{how} the time evolution drives the systems towards their equilibrium macro-state.  Moreover, it is possible to maintain the validity of the second law, in its thermodynamic formulation, and assume, say, that isolated systems never change their macro-state in time. They can persist in non-equilibrium states, or even some of them could cycle between different states, as long as they have the same entropy \footnote{See \cite{myrvold2020explaining} and \cite{myrvold2021beyond} chapter 6 for better and extensive exposition of this point}.  Due to this, I agree with \cite{myrvold2020explaining} that this explanatory project can be clearer if we distinguish the second law from an \textit{equilibration principle}\footnote{This principle is  also known as the `minus first law' of thermodynamics (since Brown and Uffink's work \cite{BROWN2001525}.} \cite{BROWN2001525}): 
 \begin{center}
 	An isolated system in an arbitrary initial state within a finite fixed volume will spontaneously attain a unique state of equilibrium.
 \end{center}
 In this terms, one can seek an account of such general time-asymmetric behavior from a fundamental theory, before (or even without) defining a non-equilibrium statistical entropy\footnote{I think this is an advantage worth the conceptual distinction so it will be employed to construct our own proposal. In spite of this, the argumentative story to be expelled in this work can also be interpreted in the `increasing non-equilibrium entropy' language.}. 
 
Another, less frequently acknowledged but crucially important point in accounting for the time-asymmetrical process of equilibration in isolated systems is that we now know the world to be quantum. Consequently, the collective behaviour of the atoms composing a cup of coffee, or a massive star, should ultimately be explainable in terms of quantum dynamical laws. However, the majority of approaches to the subject analyze the behavior of the wavefunction $\psi$ (or the associated density matrix $\rho$) considering only the unitary evolution and forget about the collapse postulate in the standard quantum formalism \cite{deutsch2018eigenstate,linden2009quantum,rigol2008thermalization}. These works find numerous interesting results which are supposed to be related to the thermalisation of quantum macroscopic systems. But, what does it physically mean for $\psi$ or $\rho$ to behave in such and such way? This of course, depends on the meaning of the wavefunction, so the question can only be answered with a clearly posed quantum ontology. Furthermore, it is well known that considering just the unitary evolution we simply can not explain the absence of macroscopic superpositions or `Schrödinger cats' \cite{Maudlin1995-MAUTMP}. And surely, the macroscopic objects we observe to equilibrate are assumed not to be in a superposed macroscopic state if we do not want to deal with Many-Worlds. In other words, if we want to give any clear physical meaning to those mathematical results about thermalisation due to the dynamics of $\psi$ (or $\rho$), we need to solve the fundamental problems of the standard quantum formalism first. That is, we have to give a clear quantum ontology and avoid the `measurement problem' with unambiguous dynamics.
 
 The general motivation in this work is to outline a proposal covering the two general issues expelled above. That is, the issue of generality and the fundamental issue with the standard quantum descriptions. For this, one shall begin by considering clear quantum dynamical laws (i.e. solving the measurement problem) that are applicable to any system. In particular, objective quantum collapse models will be considered, which can be associated with a `flash ontology' or `mass ontology' \cite{bell2001there,maudlin2019philosophy,norsen2017foundations,tumulka2022foundations,bricmont2016making,RevModPhys.85.471}. These models possess a fundamental time-asymmetrical law, which induced David Albert to suggest their possible advantage in the project of accounting for the irreversible behavior of thermodynamic systems \cite{Albert,Albert2021}. The present work develops a concrete argument along these lines in order to fill some of the gaps left by previous approaches. In the quantum case, John von Neumann made the first and  one of the cleanest approaches to the subject, where the physical motivations, assumptions, and limitations, can be clearly identified \cite{von_Neumann_2010,goldstein2010normal}. There, the thermodynamic systems, their properties, behavior, and related concepts, can be understood quantum-statistically. 
 
 In essence, von Neumann constructed a (quantum) framework analogous to Boltzmann's approach; where macroscopic systems can be analyzed with macroscopic variables defined in terms of the fundamental (quantum) theory, and proposed a dynamical property (`quantum ergodicity') for those systems, which ensured that non-equilibrium states evolve, with very high probability, into the equilibrium state (described by the microcanonical density matrix $\rho_{mc}$).
 
 As will be seen, some objections to this approach can be avoided with our proposal, and a much clearer understanding of what can actually be proved could be achieved. This will be regarding for example, the regime where the distinct (quantum or classical) dynamical properties show themselves and merge together to describe a collective behavior e.g. an equilibration process. In brief, we will argue that our analysis must be understood in the following way: Firstly, that for every initial quantum state represented by $\psi_{t_0}$ (or a density operator $\rho_{t_0}$), associated to macroscopic isolated systems, the dynamical laws provided by collapse models ensure that with overwhelmingly high probability, the state will be objectively driven, in a future finite small time, into a well defined \textit{non-equilibrium classical} macrostate. Secondly, that this ``highly'' localized macrostate of the system, according to the dynamical rules of collapse models, follows an approximately classical trajectory, so that Boltzmannian-type arguments for equilibration can be employed from here.  This would mean that the probabilities concerning with the equilibration process (i.e., at the macroscopic regime) are determined only in small part by the fundamental probabilistic quantum laws governing the collapse of the wavefunction---such as those posited in dynamical reduction models like GRW or CSL \cite{ghirardi1986unified,bell2001there}---which drive the system into a future well-defined initial macrostate. To a much greater extent, the probabilities associated with thermal equilibration are determined by the emergence of an approximate classical dynamics for quantum macroscopic bodies within objective collapse models.
 
 In other words, this work aims to provide an argumentative basis for the idea that the familiar and well-recognized macroscopic tendency toward future thermalisation---such as coffee cups cooling down, their aroma spreading through a room, and milk mixing into coffee---can ultimately be understood as a manifestation of underlying fundamental quantum stochastic processes, albeit in a more subtle way than that originally proposed by Albert.
 
 \section{Understanding equilibration as a real and objective macroscopic behavior }\label{RealAppSection}
 
 Some criticism to an objective and realist approach to explain the equilibration process focuses on the \textit{anthropocentric} content of the thermodynamic properties (used to define macro-states and equilibrium). These of course, are defined by us human scientists, so (in principle) they could be different for other kinds of intelligent beings for example, and also, they depend on the precision of our apparatuses. The idea is that even if we aim for an objective account of the observed equilibration, our work remains inevitably marked by subjectivity---since the very definitions and uses of macroscopic properties depend on it. These properties are, in a sense, as ``subjective'' as a microscope is: the device must have an eyepiece through which we can see, or it must be connected to a screen that produces light in a way we can interpret. Yet, whenever we adopt a realist stance in the natural sciences, we implicitly assume that the experimental apparatuses used to gather information about the world yield approximately the same results under the same conditions, \textit{independently} of the observer.
 
 That is, a thermometer should always reveal the same value if we prepare the measured system in the same way independently of the agent who uses it. Given that a thermometer is a physical system, this is simply a specific instance of the natural assumption that all systems behave identically under the same conditions and laws of nature\footnote{This also applies for nondeterministic laws. According to a realist stance, physical systems should always follow the laws of nature, although probabilistic.}. Concerning the precision-dependence, from a realist stance, even using the most rudimentary apparatuses, it is already assumed that the state of a cup of coffee, just after brewing, is objectively different than the state revealed by a thermometer after half and hour left in a cold room. One assumes that processes like this could be repeated objectively by any agent and its happening does not depend on someone defining the concept of temperature, or on observing the experiment with more refined thermometers. Of course, if our thermometers happen to be so bad as to assign the same state to our coffee when it feels cold and when it burns our mouth, then, we would not use them for scientific purposes until we construct better equipment.
 
 In other words, physical phenomena is not subjective just because we use (theoretical and experimental) tools  to describe it. When we do natural science, we commonly assume it is the other way around; we think that nature and its processes are objective because we can construct theoretical and experimental tools to analyze them---and to meaningfully compare our results with someone else's results\footnote{Some theoretical proposals, notably perspectival approaches to quantum mechanics, challenge the very notion of comparing distinct physical descriptions. However, in light of the additional difficulties they introduce, these approaches are not considered in this work (see for instance \cite{MucinoRQM}).}. That is, we define some particular macroscopic properties, and construct apparatuses to measure them, because we want to distinguish between distinct objective states so that we can study the laws that govern those changes of state.   
 
 Having clarified the type of explanation we seek, let us briefly review the classical approaches of this type. Broadly speaking, the most well-known are the time-irreversible H-theorem approach and the time-reversible ergodic approach.

\chapter{The Classical Approach to Equilibrium}\label{ClassicalSection}

	In this section, we will briefly present the classical approaches to give a fundamental description of the time-directed behavior towards equilibrium shown by macroscopic systems. Both of these projects have their roots in the work of Boltzmann, who was an atomist and believed that natural phenomena, from thermodynamics to life, should be explainable in terms of (large aggregates of) the fundamental constituents of the world.

\section{Boltzmann's irreversible approach: The H-Theorem}

In 1872, Boltzmann presented the first approach, known as the $H$-theorem, where he tried to argue that every initial state of a particular type of gas would inevitably evolve to a equilibrium state \cite{FurtherStudiesLBoltz,sep-statphys-Boltzmann}. Boltzmann's argument applied only to monoatomic diluted gases modeled as a system of hard spheres contained in a box of elastic walls. After one of the well-known assumptions, the \textit{stosszahlansatz}, for the distribution of the gas molecules, he arrives at what has come to be known as Boltzmann's equation. This is an integro-differential equation determining the time evolution for the distribution $f(q,v,t)$ of a system of $N$ particles, with $q$ and $v$ their positions and velocities, from any initial state. His further analysis of the solutions $f$ to this equation, showed that a function $H$, defined as
\begin{equation}
	H(f(q,v,t)) := \int f(q,v,t)  \log f(q,v,t) dq dv,
\end{equation}
approaches to a minimum value, given that $f(q,v,t)$ tends to the Maxwellian distribution, i.e., 
\begin{equation}
	f(q,v,t) \underset{t}{\longrightarrow}  f_{eq}\propto e^{-\frac{1}{2} \frac{mv^2}{kT}},
\end{equation} also known as the \textit{equilibrium distribution}, where $k$ is Boltzmann's constant and $T$ the temperature of the gas. Due to this suggestive tendency of the system to an equilibrium distribution, Bolztmann thought of $H$ as representing the evolution of non-equilibrium entropy to its maximum value and defined $S(t)=-H(t)$ as the statistical entropy. In conclusion, it appeared as if Boltzmann had shown, at least for a particular type of systems, that the approach to equilibrium followed from the fundamental classical mechanics.

Of course, this explanation caused a lot of turmoil back then, how can the irreversible behaviour of macroscopic systems be possibly reduced to the time-reversible Newtonian dynamics of particles? Lochsdmidt and Zermelo made influential arguments claiming that this could not be right. They answered Boltzmann with their `Reversibility' and `Recurrence' objections which, in sum, showed that: i) a mechanical system of $N$ particles whose velocities were inverted, would travel along the same phase-space trajectory but in the opposite direction and that due to the work of Poincaré, ii) a mechanical system left to itself for long enough time, would trace a trajectory that comes back, as close as we wish, to the initial conditions. In other words, they consistently argued that the entropy of mechanical systems can and will decrease.

These criticisms against Boltzmann brought the question about what can be wrong with his proof. In fact, his analysis was correct, but he made an assumption not contained in Newtonian mechanics. Thus, he did not showed that reversible classical mechanics explained the irreversible behaviour in thermodynamics. Instead, he showed that classical mechanics added with a time-asymmetric statistical-dynamics assumption, the \textit{stosszahlansatz}\footnote{Sometimes also known as the \textit{molecular chaos} assumption}, recovered the desired behavior. As, Huw Price puts it \textit{no asymmetry in no asymmetry out} \cite{price1996time}. This assumption in Boltzmann's approach, as suspected, carries the asymmetry \textit{in} the construction\footnote{\label{footSZA}It can be stated as follows: For a pair of particles labeled $1$ and $2$, their respective velocities $\vec{v}^1$ and  $\vec{v}_1^2$, are independent of each other (uncorrelated) \textit{before} the collision. That is, \cite{liboff2006kinetic} \cite{brown2009boltzmann} \cite{sep-statphys-Boltzmann},
	\begin{equation}
		f_2(\vec{v}^1,\vec{v}_1^2) \propto f(\vec{v}^1)f(\vec{v}_1^2)
	\end{equation}
	where $f_2$ is the distribution for the pair of particles. Of course, as assumed by Maxwell and Boltzmann, \textit{after} the collision, the states of the particles, determined by their velocities, are indeed correlated \cite{davies1977physics}.}.

Although we can think for justifications for the \textit{stosszahlansatz}, like claiming that it is reasonable for all practical purposes, this assumption is not a fundamental principle nor it can be derivable just from the fundamental classical mechanics. As it is evident\footnote{See footnote \ref{footSZA}}, it implies an asymmetric treatment of time, it distinguish the evolution to the future from the evolution to the past, in contrast to classical mechanics.

\section{Boltzmann's reversible approach: The ergodic hypothesis}

The 1877 approach has almost nothing to do with the one from 1872. Instead of constructing a kinetic theory of gases with a plausible assumption (molecular chaos), statistical and combinatorial arguments are made. It is said that J. C. Maxwell first noted that the second law of thermodynamics can not be always true. Maxwell's demon shows intuitively that there is always in principle, a small \textit{probability} that the second law fails\footnote{Maxwell's thought experiment starts with a demon placed in the division of a system of two gas subsystems. The demon knows the velocities and positions of all the particles in the two gases and decides to let specific particles from one side to cross to the other to make the difference in temperature increase in time, violating the thermodynamic tendency to equilibrium. Then, it sounds plausible to think that if instead of a demon, there is a slit that opens randomly whenever a particle approaches, it becomes a matter of probability that the second law of thermodynamics will fail. }. It is also said that because of this argument, Boltzmann thought that the Second law must be weakened to a probabilistic law that is followed with ``high probability''. 

Thus, after the criticism from Lochsmidt and Zermelo, Boltzmann looked for a cleaner and more general approach, entirely in terms of the then fundamental mechanics. In this new approach, he proposed a statistical framework where a universal dynamical behavior could be defined, characterizing the highly likeliness of mechanical macroscopic systems to approach to equilibrium from almost every initial state. I will present here a brief overview of this type of approach in its modern form (For much more detailed presentations see in particular \cite{Frigg2008Guide,frigg2011explaining}). 

The idea is this. Let's think of macroscopic systems made of $N$ particles, confined to a finite volume $V$, framed in the classical $6N-dimensional$ phase space $\Gamma_{tot}$, and consider the \textit{microstates} of such systems as the points $X=(q,p)$ in $\Gamma_{tot}$. Now, the thermodynamic properties or \textit{macrovariables} can be defined in a mechanical-statistical way, as a finite family $i \in I$ of coarse-grained functions $M_i:\Gamma_{total} \rightarrow \mathbb{R}$. Thus, the \textit{macrostate} $\nu_t$ of the system is represented by the values of the family of macro-properties at time $t$,
\begin{equation}
	\nu_t=(M_1(t), M_2(t),..., M_I(t) ).
\end{equation} Given the coarse graining of these functions, many different microstates correspond to the same value for a given macrostate. This represents the intuitive idea that, exchanging the position of any two particles of a (e.g. gaseous) system of $\sim 10^{23}$ particles, should correspond to the same assigned macroscopic state. Then, any macrostate $\nu$ has an associated subset $\Gamma_{\nu}$ of microstates, so we can use them to partition the phase space $\Gamma_{tot}$ into subsets (also called macro-regions)\footnote{Restricting the phase space $\Gamma_{tot}$ to an energy shell $\Gamma:(E, E+\delta E)$, where the energy $E$ is one of the macrovariables, $M_1=E$.} 
\begin{equation}
	\Gamma= \cup_{\nu=2}^{I} \Gamma_\nu. 
\end{equation}
Now, Boltzmann suggested a particular dynamical property for such macroscopic systems which ensured that, starting from \textit{almost any} initial microstate, the subsequent microstates of the system are chosen ``with equal weight''. Following this intuition, all the microstates are ``equally probable'', if chosen at \textit{random}. Thus, macro-states associated to much more microstates, that is bigger macro-regions, would be  ``the most probable'' regions. And, if we associate the equilibrium to an overwhelmingly big macro-region $\Gamma_{\nu'}$,  evolving to the future would lead, most probably, to the equilibrium state. 

There is a manner to make mathematically more precise this set of ideas with the famous dynamical property known as \textit{ergodicity}. A confined system in its (restricted to an energy shell) phase-space $\Gamma$, with the associated Lebesgue measure $\mu_L$,  is said to be ergodic, if and only if, for every measurable subsets $A \subset \Gamma$ and for every $X \in \Gamma$ (except for a zero-measure subset $\mu_L(B)=0$), the time average $T_A(X)$ of a solution $\phi_t(X)$ satisfies\footnote{	Where the time average is defined by $$T_A(X)=lim_{t \to \infty} \frac{1}{t}\int_0^t \mathbbm{1}_A(\phi_{\tau}(X))d\tau.$$ With $\mathbbm{1}_A$ the characteristic function of $A$.} \footnote{Furthermore, a solution $s_X(t)=\phi_t(X)$ is defined to be ergodic with respect to $A$, if and only if the proportion of time it stays within $A$ is equal to the size of $A$ with respect to the Lebesgue measure, i.e. $\mu_L(A)$.} \cite{frigg2011explaining} \cite{Frigg2008Guide}:
\begin{equation}\label{ergoBC}
	T_A(X)=\mu_L(A).
\end{equation}
Therefore, for the mechanical systems which can be proved to be ergodic we can say, that a distribution representing the initial possible microstates of a system, will tend to appear in the long time average as a uniform probability distribution also associated to equilibrium: e.g. the microcanonical distribution with respect to the Lebesgue measure on phase space
\begin{equation}
	\rho_{mc}(x)= \frac{\mathbbm{1}_{\Gamma}(x)}{\mu_L(\Gamma)}.
\end{equation}
Moreover, when there is an equilibirum macro-region $\Gamma_{eq}$ occupying almost the entire energy shell $\Gamma$, it is overwhelmingly probable that for microstates $X_t$ out of $\Gamma_{eq}$, the dynamical evolution would lead them into $\Gamma_{eq}$, and they will stay there for very long times\footnote{Normally, $\Gamma$ is normalized with respect to the uniform (Lebesgue) measure, $\mu_L(\Gamma)=1$, so that we can talk of a probability measure.}. Thus, the project translates into finding arguments to support the idea that typical macroscopic systems are ergodic.

Of course, there have been objections to the ergodic approach. In simple words, it has been argued that many systems which we expect to equilibrate do not posses this dynamical property or that it is highly limited to very special systems so that it could not count as a general explanation of the approach to equilibrium\footnote{For instance, some systems which are expected to equilibrate does not have an overwhelmingly big equilibrium macrostate.} \cite{goldstein2001boltzmann,goldstein2012typicality}. And, although some answers to these objections have been given \cite{frigg2011explaining}, again, our world is quantum. The reductive project we are looking for should, at least, start in quantum terms and then proceed to make room for this type of classic approaches, drawing a clear story of how the thermodynamic behavior can arise from the quantum nature of matter. Another possibility could be to give an explanation from a quantum theory without the need of merging it at some point with the classical approach. This last option was first explored by von Neumann with his `quantum ergodic theorem'\cite{von_Neumann_2010}. 

Another historically famous objection started noting that now, we no longer have a time-asymmetric dynamics (as in the $H$-theorem due to `molecular chaos' assumptions). Then, this explanation works as well for the future, as for the past, but this appears to clash with our records. We don't see systems equilibrating to the past (or milk and coffee spontaneously un-mixing to the future).  


\section{The Past Hypothesis}\label{HP}

As we saw in the last section, Boltzmann's ergodic argument framed in a probability language appears to explain a tendency in the future to equilibrium; but, the underlying classical theory does not distinguish between temporal directions. Thus, the ergodic argument applies to the future as well to the past.

To see this, take increasing $t$ as a future-directed evolution and consider a macroscopic system whose microstate $X_t$ is out of equilibrium. Then, it is highly probable for $X_t$ to evolve in a time-interval $\Delta t$ towards a bigger macro-region until it reaches the equilibrium. But there is no restriction within classical mechanics for $\Delta t$ to be positive. The dynamical laws apply as well for $\Delta t < 0$.  If our macroscopic system is a frozen ice cream at $t$, we should say that at $t+\Delta t$ it will likely be unfrozen. Of course, this is a good prediction when  $\Delta t > 0$ (i.e, to the future), but it is a `disastrous retrodiction' for  $\Delta t < 0$ (i.e., to the past)\cite{Albert}. 

 In other words, although Boltzmann's reversible argument could correctly tell us that marcoscopic systems equilibrate (with high probability) towards the future, as we can commonly observe, it also tell us that all of our memories and records of the past are wrong (also with high probability).  For instance, if I remember my ice cream to be more frozen in the past, Boltzmann's reversible argument considers that as a very unlikely state. 

 In order to prevent this `disastrous retrodictions', one has to have an explanation about why something like the ergodic approach applies only towards the future and not towards the past. To add this temporal-asymmetry to the dynamics as a mere restriction, motivated by our daily observations, would be to impose, not to explain. Then again, one may ask how temporal asymmetry can, in principle, arise within a theory whose fundamental dynamics are time-reversible.
 
 Fortunately, there is a well-established proposal, widely recognized within the community working on the foundations of thermodynamics since the work of Albert \cite{Albert}. The idea---apparently also stemming from one of Boltzmann's suggestions in response to critics---is simply to introduce time asymmetry as an initial condition on the macroscopic state of the universe. These are boundary conditions, similar to solving hamiltonian systems. Particularly, initial conditions are chosen for the entire universe, assuming the laws of thermodynamics apply to the entire universe as a closed system (as was formulated by Clausius). However, instead of fixing initial conditions for the universal microstate $X_0^U$, like its position and velocity, a condition for the \textit{initial macrostate} $\Gamma_{\nu_0}$ of the universe is implemented. Given that what we are trying to explain is the irreversible macroscopic behaviour toward equilibrium, the initial macro-condition of the universe is supposed to be exaggeratedly far from equilibrium, or to have `very low entropy'. This is what is now known as \textit{the past hypothesis} \footnote{The Past Hypothesis has also been under scrutiny pointing to: its use of non-fundamental concepts while also having a status like a fundamental law and being too ad-hoc or fine tuned, making it `very unlikely' that the universe actually started in such state. More criticism can be found in \cite{Earman2006}}\cite{sep-time-thermo,Albert}. 

The last ingredient for explaining equilibration is to tell us how to choose the actual initial microstate of the universe. For this, the Lebesgue measure over $\Gamma$ is used, and it is assumed that the initial $X_0^U$ is chosen `randomly' to be inside of  $\Gamma_{\nu_0}$. This is called, in \cite{Albert},  the  \textit{statistical postulate} (See also \cite{allori2020some}).

\subsection{Further remarks and summary of the objections}

In this way, Boltzmann's reversible argument, equipped with the past hypothesis and the statistical postulate conforms the most widely accepted explanation of the tendency of entropy to increase toward the future or, more precisely, of the approach to equilibrium. However, as has been emphasized in the preceding sections, there remain serious objections and difficulties in regarding this explanation as fully satisfactory. In summary, let us highlight the following points, along with some further remarks:

\begin{enumerate}
	\item The macroscopic systems that have been shown to posses the appropiate dynamical property (say, ergodicity or a similar characteristic) are very few and too idealized to consider them as actually representing the real macroscopic systems we see equilibrating every day.
	
	\item The past hypothesis is a non-dynamical macroscopic assumption with an ad-hoc flavor and an implication as strong as if it were a fundamental \textit{law of nature}. Currently, there is no consensus of whether it is untenable \cite{Earman2006}. In any case, one could also consider the possibility to \textit{derive}, instead of impose, the initial macro-condition of the universe from the dynamics of a more fundamental theory.
	
	\item The statistical postulate is also imposed by hand with no justification beyond our desire to account for familiar observations. Together with the Past Hypothesis, it is taken to be fundamentally required at the `beginning' of the universe in order to explain a \textit{universal} behaviour. However, both rely on the \textit{non-fundamental} notion of a macrostate, which in turn depends on the definitions that we, as coarse-grained macroscopic creatures, assign to certain macrovariables useful to us.
	
	\item Classical theories of motion are no longer regarded as describing the fundamental nature of matter, which is now understood to be quantum. Thus, a satisfactory explanation of such a universal behaviour as the approach to equilibrium should be sought in terms of our fundamental quantum theories and be capable of being stated \textit{objectively}---that is, independently of our subjective perspectives.
\end{enumerate} 

The brief review of the classical approach to explaining equilibration ends with this and now, it is the turn of the quantum approaches. In the next chapter, I will present first, what I take to be the clearest quantum `Boltzmannian' approach pioneered by von Neumann in \cite{von_Neumann_2010} and further clarified in \cite{goldstein2010normal}, and second, the more modern `semi-classical' suggestion in \cite{Albert} \cite{Albert2021} to consider objective quantum collapse theories.

\chapter{The Quantum Approach to Equilibrium}\label{Section:QuantumCase}

An explanatory account of the equilibration process that takes seriously both the quantum nature of the world and the absence of macroscopic quantum superpositions (effectively captured by the collapse postulate in standard quantum mechanics) while also maintaining that coffee cups and similar macroscopic objects thermalize objectively has, unfortunately, been missing. Such an account can only be achieved within quantum mechanics if we clarify what the wavefunction is supposed to represent, what exists (the ontology), and how it behaves (the dynamics), independently of `qualified measurers'. 

This chapter is organized as follows. First, in section \S \ref{QMproblemSection}, I will further motivate the overlap between the foundations of quantum mechanics and the foundations of thermodynamics by briefly describing the measurement problem of quantum mechanics, together with some proposed solutions, and by discussing the necessity of addressing this fundamental problem in the standard formalism of quantum mechanics before undertaking the task of explaining the thermodynamic `arrow of time'. Then, in \S \ref{vnFsection}, I will review von Neumann's project, closely following \cite{goldstein2010normal}, where some objections to this project will be discussed, as well as the modern and stronger reconstruction offered by the same authors. Before developing our own proposal in \S \ref{Section:OurEquilProp}, its motivation will be outlined in \S \ref{Albertsection}, presenting David Albert's `semi-classical' suggestion to consider the potential advantages of objective collapse theories---which feature time-asymmetric dynamics---in explaining the time-asymmetric process of thermodynamic equilibration. The conclusions are stated in \S \ref{Conclusion:Equilibrium}.

\section{The missing fundamental quantum parts}\label{QMproblemSection}

As was already discussed in \S \ref{RemarksEquil}, there have been many studies of the ``thermal'' behavior of the wavefunction in recent decades; however, almost none of them say anything about the relation of $\psi$ to the real world \cite{deutsch2018eigenstate,linden2009quantum,rigol2008thermalization,goldstein2010normal}. If, for instance, someone claims that the wavefunction represents only our knowledge and nothing more, then these ``quantum thermalization'' results concern the thermalization of our knowledge---whatever that might mean---rather than the objective mixing of soups or the consumption of fuel by stars. If, on the other hand, we assume within the standard quantum formalism that the wavefunction represents something real about fundamental entities (or structures, etc.), we can only understand how $\psi$ relates to the observed world if we also include the Born rule and the collapse postulate. Again, if we do not want to invoke many worlds, we can safely affirm that we only observe definite outcomes with associated definite values after any kind of experiment, including those in which we observe equilibration. Yet we would not be able to explain this outcome definiteness using only the unitary Schrödinger evolution \cite{Maudlin1995-MAUTMP}. Thus, quantum thermalization studies that consider only the unitary part of quantum evolution remain incomplete, at least with respect to their relation to the real world we observe. This is why it is indispensable to first address the fundamental problems of standard quantum mechanics before seeking a general, coherent, and objective explanation of equilibration that fully acknowledges the quantum nature of the world.

As is well known today (at least within the foundations of physics community), a solution to the fundamental problems of quantum mechanics takes the form of a quantum theory with a clear ontology and well-specified dynamics. Both of these are missing in the standard quantum formalism, where the latter issue pertains to the infamous \textit{measurement problem}. Concerning the former, a theoretical model may postulate the same dynamical rules while adopting different ontologies, and thus yield distinct physical theories\footnote{For example, Newton postulated an ontology of particles with absolute position and velocity (i.e., absolute space), whereas neo-Newtonian mechanics postulates a Galilean spacetime with relative velocities. Both theories differ because they make different claims about the world \cite{maudlin2019philosophy}.}, simply because they posit different kinds of real entities\footnote{And of course, one can evaluate which ontology is most appropriate.}.

The three most prominent approaches to solving these fundamental problems are the following types of quantum theories: (A) those that assume the wavefunction evolving according to Schrödinger's equation provides a correct but incomplete description, known as hidden-variables theories (e.g.\ Bohmian mechanics); (B) those that, by contrast, assume the wavefunction provides a complete description but that Schrödinger dynamics is incorrect and must be modified, such as objective collapse theories (e.g.\ GRW, CSL, etc.); and finally, (C) those proposals that deny that we actually observe definite outcomes, such as many-worlds theories \cite{sep-qm-manyworlds}. In this work, we will evaluate the advantages (and difficulties) that the second type of quantum theory may offer in constructing an explanation of equilibration.

Given the properties of collapse theories, David Albert \cite{Albert,Albert2021} has suggested that they could be advantageous in the project of explaining equilibration. In particular, these models postulate a fundamental time-asymmetrical evolution. They have concrete rules that specify probability distributions towards the future, instead of the past. These probabilities are approximately equal to the Born rule probabilities which convey the collapse of the wavefunction after a `measurement' in the standard formalism. In contrast, collapse models implement concretely such process by modifying the unitary evolution with a non-linear stochastic equation for the wavefunction. What Albert proposed was that the stochastic disturbances produced by the collapse mechanism in the GRW model would rapidly take `anti-themodynamic microstates' \footnote{\label{antithermo}These `anti-thermodynamic' microstates are thought as the phase-space points $X$ in $\Gamma_{tot}$ which move farther away from the  equilibrium macro-region $\Gamma_{eq}$. Conversely, `thermodynamic microstates' are the $X$'s that evolve in time into bigger macro-regions $\Gamma_{\nu}$.}, into `thermodynamic microstates'. However, as advertised, this is just a semi-classical suggestion. There is no classical phase space of points $(q, p)$ in the quantum world due to the uncertainty relations and there is no proof that such particular behavior happens. 

The aim of this essay is to fill the gaps in such direction. To construct a proposal that starts from a fundamental (quantum) theory, where clear dynamical laws (addressing the measurement problem) for a concrete ontology, allow us to describe how general macroscopic systems, which behave approximately as classical, thermalize independently of us observing them.

\section{Towards a completely quantum explanation of equilibration}\label{TowardsCom}

Before stating the main proposal of this work, it is interesting to present another more ambitious historical approach in the quantum case. Where, if it succeeded completely, no classical notions like phase-space or trajectories in it are needed. 

First, if the aim is to provide a general and concrete reductive account of thermodynamic equilibration within a proper statistical framework, then a translation from thermodynamics into the probabilistic language of the fundamental theory in question should be provided. For this, we have to define macrostates, thermodynamic properties (i.e. macro-variables), microstates, and the arena where their thermal behaviour is supposed to unfold. 

As we saw in chapter \ref{ClassicalSection}, when classical mechanics was thought to be our most fundamental theory, Boltzmann proposed a classical statistical-mechanical framework in which microstates were represented by points $X$, and their thermal behavior unfolded in their proper arena, namely the total phase space $\Gamma_{tot}$. The macrovariables and all other relevant notions where also properly defined there. But when quantum mechanics was already settled, von Neumann took the challenge of constructing an analogous explanation to that of Boltzmann but completely in quantum terms.  He succeeded in erecting a concrete quantum statistical framework where all the relevant thermo-statistical notions can be clearly defined \cite{von_Neumann_2010}. However, we think that the gaps in his explanation does not permit one to really say why quantum macroscopic systems tend to thermodynamic equilibrium. In particular, we should not carry on the fundamental problems of the standard formalism into this project. As we have said, a Schrödinger cat will simply undermine any argument for the equilibration of the cat, if the underlying theory cannot properly explain the emergence of definite macroscopic states for the cat. 

What von Neumann did was define a particular quantum dynamical property that apparently ensures the equilibration of wavefunctions of macroscopic systems. This property was motivated by von Neumann through an analogy with classical ergodicity. In essence, he defines ``quantum-ergodicity'' as a property for macroscopic systems such that in the long-run, the quantum state (a pure density matrix) ascribed to the system in question, gives us the same approximate probabilities for the outcomes with respect to the macrovariables (previously defined), as the probabilities given by its   `equilibrium' distribution (the quantum microcanonical density matrix).

Goldstein et. al \cite{goldstein2010normal} have reconstructed von Neumann's framework improving his proofs and clarifying its statements. They call this particular dynamical property `normal typicality', instead of quantum-ergodicity. To prove that quantum macroscopic isolated systems behave `normal' or `ergodic' in \cite{von_Neumann_2010} and \cite{goldstein2010normal}, one has to further assume that there are no resonances, nor degeneracies in the hamiltonian of the systems. Of course, this restricts a lot the type of systems that would come out `normal'. And again, such proofs treat only the unitary part of the quantum evolution. Goldstein et. al in \cite{goldstein2020gibbs} are aware that an approach like that of von Neumann still needs to address the fundamental problems of standard quantum mechanics. They also give a first discussion of how pilot-wave or objective collapse theories could be used to supplement von Neumann's framework to obtain the complete story of how the approach to equilibrium occurs. Interestingly, also von Neumann explicitly leaves open the relation that collapses could play towards the approach to equilibrium:

\begin{quote}
	However, quantum mechanics does know an irreversible elementary process: the measurement. It is irreversible [...], but whether it is relevant to the irreversibility of reality we leave open. In this work, we do not deal with measurement. (\cite{von_Neumann_2010}, p. 10)
\end{quote}


\section{von Neumann's Quantum Statistical Framework}\label{vnFsection}

A short time after the chaotic construction of Quantum Mechanics, a `Proof of the ergodic theorem and the H-theorem in Quantum Mechanics' was published by von Neumann in 1929 (English translation by Roderich Tumulka in \cite{von_Neumann_2010}). The object of his paper was a discussion and clarification of:
\begin{quote}
	In particular, first, how the peculiar, seemingly irreversible behavior of entropy arises, and second, why the statistical properties of the (fictitious) micro-canonical ensemble can be attributed to the incompletely known (real) system. \cite{von_Neumann_2010}
\end{quote} 
To this end, as advertised in his abstract, 
\begin{quote}
	The main notions of statistical mechanics are re-interpreted in a quantum-mechanical way, the ergodic theorem and the H-theorem are formulated and proven (without ``assumptions of disorder''). \cite{von_Neumann_2010}
\end{quote}

In other words, what von Neumann actually had in mind was just a reconstruction of Boltzmann's ergodic approach, redefining the main notions used in that argument, concerning the system's state behavior across the partition of its phase-space, but entirely in new quantum terms \footnote{Although von Neumann takes the H-theorem as a motivation, he rejects the validity of such approach based on the assumption of `molecular chaos' ( or `disorder' in von Neumann's words). His rejection has to do with the unknown nature of such assumptions and their incompatibility with the fundamental theory at work.}. For this, von Neumann starts outlining the difficulty to traduce the phase-space tool to the quantum language given the Heisenberg uncertainty relations. Therefore, it is necessary to implement a new `physical-states' space in which the `peculiar' thermodynamic behaviour of the system state unfolds. Noting that in standard quantum mechanics we can think of the wave function $\psi$, describing the state of a quantum system, as a vector in the Hilbert space $\mathcal{H}$, he then takes this $\mathcal{H}$, to play the role analogous to classical phase space, as the new arena where some desired quantum statistical property unfolds. I will call his construction von Neumann's framework (vNf). 

In vNf, one decomposes the Hilbert space and defines an energy shell spanned by the eigenfunctions of the quantum energy (i.e. the Hamiltionian of the system $H$). Like in Classical Statistical Mechanics, a partition (decomposition in the quantum case) is provided by defining the thermo-statistical functions first. In this case, we should first define the quantum macro-observables associated to the thermo-statistical properties of the system, like the temperature, average number of particles, pressure, and `macroscopic energy', etcetera. 

In particular, one considers the microscopic energy of a quantum macroscopic system of $N$ particles, isolated and confined to a finite volume $V$, represented by the hamiltonian $H$ whose eigenfunctions  $\phi_n$ span the total Hilbert space $\mathcal{H}_{tot}$. The wave function  $\psi_t=\psi(q_1,...,q_N, t)$ follows the Schrödinger equation
\begin{equation}
	i\hbar \pdv{\psi}{t}= H\psi_t,
\end{equation}
and $H$ has a discrete spectrum of energy levels $E_{\alpha}$ due to the confinement to $V$. Thus, the eigenfunctions $\phi_{\alpha}$ are such that,
\begin{equation}
	H\phi_{\alpha}=E_{\alpha}\phi_{\alpha}.
\end{equation}
Then, one considers a \textit{coarse graining} function $f_1$ of the energy levels $E_{\alpha}$, with $f_1$ the step function \footnote{Considering only Hamiltonians bounded from below, we can order the energy levels as $E_0\leq E_1\leq E_2 \leq ...$}
\begin{equation}
	f_1(E)= \frac{E_a+E_{a+1}}{2},\ \ \textrm{for}\ \ E \in \mathcal{I}_a=[E_a, E_{a+1}).
\end{equation}
$\mathcal{I}_a$ are the disjoint `micro-canonical' intervals \footnote{Note that the new index $a$ in  $\mathcal{I}_a$ is introduced to label each one of the different intervals.}, or in other words the (macroscopic) energy surfaces, which are \textit{large in the microscopic scale and small in the macroscopic interval} in the sense that: any interval contains many eigenvalues $E_{\alpha}$ but any of those distinct eigenvalues in the same interval cannot be distinguished macroscopically. With this, von Neumann constructs the   ``macroscopic energy'' represented by the macro-observable operator $M_1$, which is obtained from coarse-graining $H$ by means of $f_1$ and the micro-canonical intervals $\mathcal{I}_a$, i.e.,
\begin{equation}
	M_1=\sum_{\alpha} f_1(E_{\alpha})\ket{\phi_{\alpha}}\bra{\phi_{\alpha}}.
\end{equation}
Then, in vNf it is considered that the other macroscopic relevant properties, apart from the macroscopic-energy, can be represented by operators $M_j$ constructed in the same coarse-graining way as $M_1$. Importantly, von Neumann assumes that such operators can be constructed in a way that the group of $k$ macro-observables ${M_1, M_2,..., M_j,..., M_k}$, commute with each other\footnote{This is a strong supposition that could be wrong and needs to be defended. von Neumann gave a small argument for the case of two observables and gave plausible suggestions for more general cases. See \cite{goldstein2010normal} and \cite{goldstein2020gibbs} section 9.1 for more details.}.

With all this, a macro-state is characterized by the list
\begin{equation}
	\nu=(m_1, m_2, ..., m_k),
\end{equation}
of the $m_j$ eigenvalues of $M_j$. Each list $\nu$ corresponds to a subspace $\mathcal{H}_\nu \subseteq \mathcal{H}_{tot}$, spanned by the simultaneous eigenvectors of the set $\{M_j\}$ with eigenvalues $m_j$. We will now call the microcanonical \textit{energy shell} the subspace $\mathcal{H}_{\mathcal{I}_a} \subseteq \mathcal{H}_{tot}$ spanned by the $\phi_{\alpha}$ with $E_{\alpha} \in \mathcal{I}_a$. Then, vNf with its assumptions so far is such that the micro-canonical energy shell $\mathcal{H}_{\mathcal{I}_a}:=\mathcal{H}$ gets decomposed into the mutually orthogonal subspaces $\mathcal{H}_{\nu}$,
\begin{equation}
	\mathcal{H}=\bigoplus\limits_{\nu}\mathcal{H}_{\nu}.
\end{equation}
Each `macro-space' $\mathcal{H}_{\nu}$ corresponds to a different macro-state $\nu$. This is analogous to the partition (of the energy shell) of phase space $\Gamma$ into disjoint macro regions $\Gamma_{\nu}$ in the classical Boltzmannian approach. A particular decomposition of  $\mathcal{H}$ arises from a family of macro-states, $\mathcal{D}:=\{\mathcal{H}_{\nu}\}$. The dimensions of each subspace will be denoted by $d_{\nu}=\textrm{dim}\mathcal{H}_{\nu}$ and $D=\textrm{dim}\mathcal{H}$.

Thus, as is now usual, the \textit{microcanonical density matrix}  $\rho_{mc}$ is given by 
\begin{equation}
	\rho_{mc}=\frac{1}{D}\sum_{\alpha:\phi_{\alpha}\in \mathcal{H}} \ket{\phi_{\alpha}} \bra{\phi_{\alpha}},
\end{equation}
and the microcanonical average of an observable $O$ on $\mathcal{H}$ will be
\begin{equation}
	\textrm{Tr}(\rho_{mc}O)=\frac{\textrm{Tr}O}{D}.
\end{equation}

\subsection{Quantum Ergodicity or Normality}

The next great and simple insight in vNf begins by considering that, in the classical ergodic treatment, the behavior giving rise to the approach to equilibrium can be summarized by Eq. (\ref{ergoBC}). Here, $T_A(X)$ can be understood as the time average of a delta measure $\delta_{X_t}$ concentrated at $X(t)$, representing the pure state of a classical system. Since $\mu_L(A)$ is the uniform distribution, it can be associated with the microcanonical measure $\mu_{mc}$ over a subset of phase space. That is, Eq. (\ref{ergoBC}) can be rewritten as \cite{goldstein2010normal}
\begin{equation}\label{ergoBC2}
	\overline{\delta_{X_t}}=\mu_{mc}.
\end{equation}
Now, the first naive quantum analogue that one could think of Eq. (\ref{ergoBC2}) would be to write $\ket{\psi_t}\bra{\psi_t}$ instead of the pure state $\delta_{X_t}$ and the quantum micro-canonical distribution $\rho_{mc}$ instead of the uniform measure $\mu_{mc}$, i.e.,
\begin{equation}\label{naiveErgo}
	\overline{\ket{\psi_t}\bra{\psi_t}}=\rho_{mc}.
\end{equation}
However, this strict equality misses the entire macroscopic perspective we wish to include, namely the above scheme of quantum thermodynamic macro-states and macro-properties. Moreover, Eq. (\ref{naiveErgo}) holds only for very special wave functions, leaving aside many physically relevant cases. Therefore, von Neumann looked for something similar to Eq. (\ref{naiveErgo}) but that would hold approximately once the macroscopic perspective is incorporated. To this end, one introduces the notion of \textit{macroscopic equivalence}, as it is called in \cite{goldstein2010normal}, and denotes it by $\sim^{\mathcal{D}}$ to make explicit its dependence on a particular decomposition $\mathcal{D}$, arising from the set of macrostates associated with the macro-spaces $\mathcal{H}_{\nu}$. What we are intuitively looking for is that two microstates associated with a given quantum system be macroscopically indistinguishable if and only if the probabilities associated with the outcomes of measurements of the thermo-statistical properties, used to define the macrostates, are approximately equal for both microstates. In this sense, we are going to say that two density matrices are macroscopically equivalent, $\rho\  \sim^{\mathcal{D}} \rho'$, if and only if 
\begin{equation}\label{MacroEquivTrace}
	\textrm{Tr}(\rho P_{\nu})\approx\textrm{Tr}(\rho' P_{\nu}) \ \ \ \ \forall \nu.
\end{equation}

In brief, what von Neumann achieved was to show that\footnote{See \cite{goldstein2010normal} for an extensive exposition of the meaning of this relation and the measures involved when talking about the majority of decompositions $ \mathcal{D}$. Note in particular that for $\rho=\overline{\ket{\psi_t}\bra{\psi_t}}$, one side of expression (\ref{MacroEquivTrace}) becomes: $\textrm{Tr}(\rho P_{\nu})= \overline{\norm{P_{\nu}\psi_t}^2}$, which has physical meaning if we assume the Born rule in standard quantum mechanics.} 
\begin{equation}\label{MacroEquiv}
	\overline{\ket{\psi_t}\bra{\psi_t}}\ \  \sim^{\mathcal{D}} \  \rho_{mc}, \ \ \ \forall \psi_{t_0} \text{ and the majority of } \mathcal{D}. 
\end{equation}
In \cite{goldstein2010normal} it is argued that this proof in vNf is, compared to the classical Boltzmannian ergodic argument, weaker in the sense that (\ref{MacroEquiv}) is only an approximate agreement, and stronger in the sense that it typically follows for \textit{every} pure state (in contrast to the zero-measure exceptions in the classical argument). Furthermore, in \cite{Goldstein2011Approach} it is argued that ``not just the time average but even $\ket{\psi_t}\bra{\psi_t}$ itself is macroscopically equivalent to $\rho_{mc}$ for most times $t$ in the long run, i.e., 
\begin{equation}\label{normal}
	\textrm{Tr}(\rho_t P_{\nu}) =\norm{P_{\nu}\psi_t}^2 \approx \frac{d_{\nu}}{D}=	\textrm{Tr}(\rho_{mc} P_{\nu}),
\end{equation}
for all $\nu$ for most $t$.''\footnote{Systems defined by  $H,\ \mathcal{D}$ and $\psi_0$ that satisfy (\ref{normal}) are the ones the authors of  \cite{Goldstein2011Approach} call \textit{normal}.} Now, as in the classical case, these results at first seems very suggestive because in (\ref{MacroEquiv}) we have a relation between the (quantum) state evolving over long times and (the density matrix that represents) the equilibrium in our (quantum statistical mechanics) framework.

However, as has already been stressed in this work, for (\ref{MacroEquiv}) to have any physical meaning we must assume something about how $\psi$ relates to real entities in the world and specify how probabilities are to be extracted from it. Recall also the objective sense in which equilibration is to be understood according to this work. Thus, the probabilities involved will not be useful if they emerge from a $\psi$ taken epistemically, representing merely our incomplete knowledge of systems. If, on the other hand, we associate quantum probabilities with something real about systems, we must remember that in standard quantum mechanics they are given by the Born rule, which already assumes that a collapse of the wavefunction has taken place. And if we do not want to ``put ourselves in the equation'' as those primarily responsible for the collapse, we need a resolution of the measurement problem. Thus, a result like (\ref{MacroEquiv}) or (\ref{normal}) should be justifiable and understandable from the concrete dynamical laws of a quantum theory without the measurement problem\footnote{For example, collapse theories in particular could be useful in justifying a relation like (\ref{normal}) in a more straightforward way, without the need for further assumptions such as von Neumann's no-resonances condition or non-degeneracy of the Hamiltonian. We initially attempted to evaluate this possibility using the general operators formalism that emerges naturally from collapse models \cite{bassi2007hilbert}. However, we found a potential flaw in such an approach independently of the underlying quantum theory in use (see the following paragraph).}.

Although this approach might seem promising, this work will follow a different path. This is because it is not entirely clear that quantum ergodicity or `normality' can be understood as the dynamical property giving rise to thermodynamic equilibration. The reason is that such results arise mainly from the unitary evolution of the quantum state, whereas thermodynamic equilibration seems more like a macroscopic process unfolding after the collapse of the wave function has occurred and macroscopic properties are well-defined at all times. Thus, I propose that a description of how systems equilibrate, although fundamentally quantum, should also include the emergence of approximate classical trajectories. This is the next goal.   

\section{The GRW theory and irreversibility in thermodynamics}\label{Albertsection}

In 1986,  GianCarlo Ghirardi, Alberto Rimini and Tulio Weber proposed a precise theoretical model of spontaneous collapses of the wavefunction (now known as the GRW model) with the aim of unifying the description of the quantum behaviour of microscopic systems and the classic behavior of macroscopic systems \cite{ghirardi1986unified}. In their proposal, the preferential rol of the observer gets banished from the formalism. GRW is clear, precisely because when and how collapses occur does not depend on the notion of measurement or on whether measurements are performed. GRW collapses happen randomly in time with a (postulated) fixed probability per unit time for every ``particle'' in the universe. Their objective was to explain mathematically, rather than with vague words, the absence of macroscopic superpositions.

The GRW model can be summarized by the following postulates \cite{norsen2017foundations}:
\begin{enumerate}
	\item The wavefunction is informatively complete; that is, it fully describes the quantum state.
	\item Every particle in a system of $n$ particles undergoes, at random times distributed according to a Poisson process with mean frequency $\lambda_0$, a spontaneous localization. 
	\item The effect of the collapse. The localization process of the i-th particle is modeled mathematically by multiplying the wavefunction of a system of $n$ particles by a Gaussian distribution whose form depends on $\alpha$, a new constant of nature. That is,
	\begin{equation}
		\psi(\vec{x_1},..., \vec{x_n})\xrightarrow[\text{loc}]{t}  N( \frac{\alpha}{\pi} )^{\frac{3}{4}} e^{-(\frac{\alpha}{2})(\vec{x_i}-\vec{x_0})^2}\psi(\vec{x_1}, ..., \vec{x_n}) \equiv N \phi_0(\vec{x_1}, ..., \vec{x_n}),
	\end{equation}
	where $N$ guarantees normalization.
	\item Relation to the Born Rule probabilities. The probability density for the occurrence of a localization around $x_0$ is (in approximate agreement with the Born Rule)
	\begin{equation}
		P(x_0)= \int_{-\infty}^{\infty} |\phi_0(x)|^2 dx. 
	\end{equation}
	\item Lastly, between consecutive collapses, the system evolves according to the Schrödinger equation.
\end{enumerate}
In this model, $\lambda_0$ and $\alpha$ are new constants of nature whose values are chosen such that, for microscopic systems, we recover the same predictions as the standard quantum formalism over long times (such as interference phenomena), while for macroscopic systems localization occurs over very short times. At the same time, the effect of the collapse should not localize systems to such a small volume as to produce ionization problems. The values initially chosen by the GRW authors were $\lambda_0 \approx 10^{-16},\mathrm{s}^{-1}$ and $1/\sqrt{\alpha} \approx 10^{-5},\mathrm{cm}$\footnote{In this way, a single particle will suffer a collapse once every $100{,}000{,}000$ years, and the width of the collapse Gaussian is of the order of $200$ times the size of the hydrogen atom \cite{bell2001there}.}. Thus, GRW provides a theoretical framework that recovers the empirical success of standard quantum mechanics while alleviating the ambiguity in its dynamical rules by means of a mathematically clear model. 

Of course, as has been emphasized in this chapter and in many other expositions of the topic (see \cite{maudlin2018ontological}), to have a full quantum theory we need to specify what sorts of \textit{real} things are supposed to behave as our dynamical model dictates. That is, we need to specify the ontology. The most famous proposals for GRW are the ontology of ``flashes'' and the mass-density ontology.

The first was proposed by John Bell himself in \cite{bell2001there}, where he takes the spatial points at which the centers of the collapse Gaussians occur as the constituents of the ontology (or \textit{local beables}, in his words). These spatial points are now known as `flashes' in the foundations of physics literature. Thus, as Bell wrote, ``a piece of matter is a galaxy of events''.  

The second proposal goes as follows. Take the wave function of a system of $N$ particles, which can be considered for simplicity as a scalar in configuration space, i.e. $\psi(\vec{x_1},...,\vec{x_N}) \equiv \bra{\vec{x_1},...,\vec{x_N}}\ket{\psi}$. Then, it is assumed that
\begin{equation}
	m(\vec{x},t) \equiv \sum_{n=1}^{N} m_n \int d^3 x_1 ... d^3 x_N \delta^{(3)}(\vec{x_n}-\vec{x})|\psi(\vec{x_1},...,\vec{x_N})|^2
\end{equation}
describes the \textit{mass density} distribution of the system in three-dimensional space as a function of time. Thus, instead of a discrete ontology of flashes, the local beables are now represented by the continuous distribution of mass in space and time.

\subsection{Albert's Suggestion: Analysis and Evaluation}

In \cite{Albert}, Albert suggested that GRW's irreversible dynamics could explain the irreversible approach to equilibrium in thermodynamics. His suggestion rests on the belief that the stochastic perturbations a system undergoes due to quantum collapses would invariably drive any anti-thermodynamic microstate\footnote{See footnote \ref{antithermo} in \S \ref{QMproblemSection}.} into a thermodynamic one. He presents a ``semi-classical'' plausibility argument in which, instead of the wave function $\psi$, the points $X$ in a classical phase space $\Gamma$ are taken to represent the microstates of systems.

He assumes that macroscopic systems, even if fundamentally quantum, can be well described thermodynamically using the classical partition of $\Gamma$ into macroregions $\Gamma_{\nu}$ associated with macrostates $\nu$. As is usual in such frameworks, it is assumed that there is a privileged macroregion $\Gamma_{eq}$, much larger than all the others, so that it can be associated with equilibrium\footnote{The size is measured with respect to the uniform Lebesgue measure.}. Then, microstates $X$ outside $\Gamma_{eq}$ should almost surely evolve into $\Gamma_{eq}$. There are, however, microstates that do not do this, called anti-thermodynamic. Why have we not observed them? Albert responds that this is explicable by the properties of quantum collapse theories like GRW. Assuming (among other things) that anti-thermodynamic states are distributed in phase space in such a way that any one of them is surrounded by many thermodynamic states, Albert claims that the perturbations produced by the collapse mechanism will quickly drive an anti-thermodynamic microstate into a thermodynamic one. Since such perturbations are caused by the time-asymmetric dynamics of the GRW model, the time-asymmetrical behavior of our macroscopic world would thus be explained, at least in part, by the collapse mechanism. 

However, for Albert's suggestion to be physically relevant, it must be the case that in thermodynamic scenarios---where macroscopic systems are well described by the classical trajectories of their microstates $X(t)$ in phase space $\Gamma$, partitioned by the coarse-graining into macrostates $\nu$---quantum collapses produce stochastic perturbations significant enough to influence the approach to equilibrium. In other words, the effects of collapses should be sufficiently strong to influence system dynamics at the macroscopic level, where classical trajectories provide an approximately good description. This can be assessed more concretely. For instance in \cite{bassi2010long} three temporal regimes are identified and described as ``more or less well separated'', depending on the collapse-model parameter $\lambda$ and the mass $m$ of, for example, a free particle (or an isolated system described by its center of mass). Collapse effects influence the dynamics only in the first and the last regimes. In the first, a wavefunction initially spread in space becomes well localized due to the collapse mechanism. In the second, called the classical regime by the authors of \cite{bassi2010long}, it is claimed that after collapse a macroscopic particle with large mass has a well-localized wavefunction behaving ``for all practical purposes, like a point moving deterministically in space according to Newton's laws.'' Of course, this seems a necessary property if collapse models are to explain the classical motions we observe in macroscopic objects. In the third regime, quantum fluctuations produce departures from classical dynamics.

What is important to note with this is that, once we are in the classical regime, we can not expect the collapses to keep influencing the dynamics. Many macroscopic systems which are observed to equilibrate follow approximately classical trajectories. If one of these systems starts in an `anti-thermodynamic' microstate, it seems that we cannot ask collapse models to explain, at the same time, both its approximately classical dynamics and a perturbation significant enough to convert it into a thermodynamic state. This reasoning seems to provide the missing explanation behind the negative results of the numerical tests of Albert's suggestion in \cite{te2021master}.

\section{A `not so novel' quantum approach }\label{Section:OurEquilProp}

In the previous sections, two different approaches to describing the evolution toward equilibrium have been reviewed, assuming that the fundamental dynamics is given by a quantum theory. Von Neumann's explanatory scheme is motivated by the classical ``Boltzmannian'' argument, which constructs a statistical framework where thermodynamic notions and properties can be defined in the language of the fundamental theory. Here, the general idea is that the distributions assigned to the states of macroscopic systems show a particular evolution, due to the (fundamental) dynamical rules, toward an ``equilibrium'' distribution.

In the second approach, the underlying quantum (collapse) dynamics is used to explain how macroscopic systems that follow approximately classical trajectories are perturbed toward trajectories that do equilibrate. Their potential drawbacks have also been discussed, particularly concerning the temporal regimes in which we expect certain behavior. In von Neumann's approach, uncollapsed wave functions are supposed to correctly describe equilibrating macroscopic systems with well-defined position and momentum. In Albert's approach, equilibrating systems with already well-defined position and momentum are supposed to still be influenced by wave function collapses. Given this, we assert that the story should be different, although more similar to Albert's proposal.

\subsection{ From quantum (collapse) dynamics to classical equilibration}

Here, a more coherent and complete argumentative story is proposed as follows. Big or small, matter must be quantum so we can start considering that an isolated macroscopic system $\mathcal{M}$, the one we expect to equilibrate, of say $N$ particles\footnote{For illustration, the system $\mathcal{M}$ can be thought as composed by a measuring apparatus $A$ and a subsystem $S$. $A$ is such that its spatial degrees of freedom, for example, permits us to distinguish between distinct (previously defined) macrostates and $S$ can be thought as the rest of the system. For example, $A$ could be a macroscopic closed box equipped with mercury thermometers, initially containing a gas $S$ at temperature $T_i$ in one side of the box. In the end, the way in which we think the entire system is composed in subsystems is not substantial to the explanation.} be describable by a quantum collapse theory, for example. We will represent the quantum state of the entire system with the normalized vector $\ket{\psi}$ in $\mathcal{H}$, where $\mathcal{H}$ is the Hilbert space associated to the entire system $\mathcal{M}$. Lets also assume that the system is confined to a finite region $\Lambda \subset \mathbb{R}^{3}$ and, for concreteness, that the dynamical behavior is ruled by the GRW model\footnote{The story that we will present should also apply for general objective collapse models.}. In density operators notation, the GRW collapse can be represented epistemically by 
\begin{equation}\label{GRWcollapse}
	\rho_t= \ket{\psi_t}\bra{\psi_t} \xrightarrow{collapse} \int d^3 x L_n(\textbf{\textrm{x}})\ket{\psi_t}\bra{\psi_t}L_n(\textbf{\textrm{x}}) 
\end{equation} 
with $\ket{\psi_t}$ the wavefuntion just before the occurrence of the collapse and $L_n$ is the collapse operator associated to the $n-$th particle. The center of the collapse is unknown around $\textbf{\textrm{x}}$. Now, we will use the characterization and results for the operator formalism of collapse theories in \cite{bassi2007hilbert} and \cite{bassi2010long}. 

In collapse models the position of a macroscopic object can be represented ideally by its center of mass localized in the collapse region, 
\begin{equation}
	\textbf{\textrm{q}}_t=\bra{\psi_t}\textbf{\textrm{q}}\ket{\psi_t}\ \ \ \in \mathbb{R}^3,
\end{equation}
where $\textbf{\textrm{q}}$ is the center of mass position operator. Then it can be showed \cite{bassi2007hilbert} that the probability $\mathbb{P}_{\textbf{\textrm{q}}_t}(\Delta)$ that a collapsed wavefunction lies within an interval $\Delta$ of the real axis goes, in approximate accordance to the Born rule as
\begin{equation}
	\mathbb{P}_{\textbf{\textrm{q}}_t}(\Delta)\approxeq ||P_{\Delta}\psi_t||^2,
\end{equation}
with $P_{\Delta}(x)$ the characteristic function of $\Delta$. Idealizing the situation, we can think for example that $\textbf{\textrm{q}}_t$ describes a pointer that moves in one spatial dimension and measures the average kinetic energy of our macroscopic system. Such that when the collapsed wave function is approximately contained in $\Delta$, the system has a temperature $T_{\Delta}$ corresponding to the macrostate $\nu_{\Delta}$. Now, in general $\nu_{\Delta}$ won't be the equilibrium macrostate and it is not even the case that it would be selected with high probability, because in this regime the measure of `most probable' is given by the quantum rules. Thus, the collapse produces a particularly out of equilibrium `initial' macrocondition which in classical presentations is commonly just assumed to be the case\footnote{Actually, another objection raised against classical equilibration arguments notes a tension between, on the one hand, claiming that the equilibrium macro-region is overwhelmingly large according to the uniform measure being used, and, on the other hand, observing that we commonly find systems out of equilibrium. Of course, one can resolve this by postulating a very special initial macrocondition for the universe (See \S \ref{HP}). Still, there is a further objection: even if this explains why the universe starts far from equilibrium and approaches it over time, it does not necessarily explain why subsystems of the universe---such as boxes filled with gas---also do so.}.

However, as concretely argued in \cite{bassi2010long} and as was briefly said in the last section, the long time-behavior of the quantum state represented by $\psi$ can be divided in three regimes (for an idealized macroscopic quantum system): 
\begin{enumerate}
	\item \textit{Collapse regime:} A $\psi$ having an initial large spread, localizes in space, in agreement with the Born rule.
	\item  \textit{Classical regime:} The localized $\psi$ moves in space like a classical free particle. ``The fluctuations [...] can be safely ignored''.
	\item  \textit{Diffusive regime:} Eventually the random fluctuations dominate and $\psi$ starts to diffuse appreciably.
\end{enumerate} 
Then, when we arrive at the initial macrostate $\nu_{\Delta}$, the system enters to the classical regime where it will now behave approximately according to Newtonian dynamics; with a well-defined trajectory across a phase space $\Gamma$, that can be partitioned into disjoint macro-regions $\Gamma_{\nu}$. The system's microstate can now be thought (for all practical purposes) as a point $X\in \Gamma$ which will start moving approximately classically from its associated initial macro-region $\Gamma_{\nu_{\Delta}}$. Thus, we can now use the uniform measure in $\Gamma$ to compare the sizes of the distinct $\Gamma_{\nu}$'s. 

Moreover, as in the classical arguments, we now have enough theoretical resources to describe a dynamical property (e.g. ergodicity) for macroscopic systems in this regime, such that their phase flow $\phi(X(t))$ selects subsequent $X(t')$'s uniformly, or `randomly' if you wish. And, for systems for which there is an overwhelmingly big (relative to the uniform measure) macro-region $\Gamma_{eq}$; we can say that the dynamics, emerging in the classical regime of quantum collapse theories, will drive an initial microstate $X_0\in \Gamma_{\nu_{\Delta}}$, with very high probability, into to its equilibrium macro-region. Of course, one can still raise some of the objections to the classical arguments to this new approach, concerning for example, the idealization to systems for which an overwhelmingly big macro-region actually exists.

However, there are some advantages. We do not need Albert's type of argument for `anti-thermodynamic' microstates. Such microstates evolving away from equilibrium can exist. But, the uniform measure appearing in the classical regime permits us to say that the majority of $X$'s will be of the `thermodynamic' ones\footnote{Since the uniform measure is associated to all of $\Gamma$, including every subset of it, we can also say that in every neighborhood of any anti-thermodynamic microstate, we will have a majority of thermodynamic microstates. Thus, anti-thermodynamic states are dispersed across $\Gamma$, or stable as Albert calls them\cite{Albert2021}.}. Therefore, macroscopic systems evolving anti-thermodynamically are probable, but very unlikely. And the variety of arguments and technical results studying classical systems approaching to equilibrium, or maximizing entropy (See for a general overview for example \cite{goldstein2001boltzmann,frigg2011explaining,Frigg2008Guide}), become relevant in this proposed scheme, which started from a fundamentally quantum theory of matter. Also, this proposal is more general than von Neumann's argument because we did not need to assume particular properties for the hamiltonian like no-degeneracy. In our scheme, quantum systems just need to be sufficiently big for the collapses to produce a classical regime. Notwithstanding this, we still have some of the problems classical equilibration arguments have. For instance, it has been said that there are not sufficient proofs for general systems ensuring that they behave accordingly, say for example ergodically \cite{goldstein2001boltzmann}.


\section{Conclusion}\label{Conclusion:Equilibrium}

As was previously advertised, there is a lack of argumentative stories taking a fundamental theory to explain the common irreversible equilibration processes seen in our macroscopic world. The classical Boltzmannian approaches can not obviously be considered as relying on a fundamental theory anymore. As we saw in section \ref{ClassicalSection}, they are accused of applying to very idealized systems, or of making additional suppositions (`\textit{stosszahlansatz}')\footnote{It might also be conjectured that the \textit{stosszahlansatz} is valid at all times and for all practical purposes because dynamical quantum collapses could destroy the velocity correlations before collisions, but after collisions  correlations can be formed. Of course, a much more serious analysis is needed.} not contained in the underlying theory. In the quantum case, almost every proposal forgets about the foundational problems of quantum mechanics, and constructs an argument for the wavefunction, totally disconnected from the real world. This is simply because they inherit such disconnection from the standard formalism of quantum mechanics\footnote{If we just throw off the Born rule and the collapse postulate from the quantum formalism we have a total disconnection from the real world. If we include them we have an ambiguous connection due to the measurement problem. Cats would die and coffee cups would get cold because we decided to measure them.}. To evade some of these problems we have argued it is better to start with a quantum theory with concrete ontology and clear dynamics, and we have shown one way to do it with collapse theories.

A further problem has been noted in modern quantum proposals concerning the temporal regimes where the thermodynamic behavior of systems can be distinguished. Von Neumann's type of argument is a result for the wave function, and seems insufficient to describe macroscopic systems commonly seen equilibrating following classical paths. And Albert's proposal includes quantum effects in a regime where we only expected classical behavior \footnote{Our conclusions may thus explain the conclusion in \cite{te2021master} where it is found, through numerical simulations, that the collapses of the wavefuntion does not affect the classical trajectories.}. Our proposal also evades these problems but takes into account Albert's suggestion for the possible advantage of employing quantum collapse theories. In brief, we presented a description of macroscopic systems equilibrating that begins by considering the quantum nature of matter in such a way that classical behavior can be shown to emerge from here and the approach to equilibrium occurs at this classical level.

In the proposal outlined in this work, as in the classical phase-space case, there is still an `anthopocentric' element due to the macroscopic variables used to construct the partition of $\Gamma$ into $\Gamma_\nu$'s. But this seems unproblematic as long as these notions only enter at this level and not in the underlying theory. Quantum theories addressing the fundamental problems of the standard quantum formalism are, of course, free of such \textit{macroscopic} concepts. However, we still need to employ macroscopic notions---such as macro-variables, macro-regions, and equilibrium---to describe the process of equilibration. Crucially, the process occurs objectively, not \textit{because} of these notions. This is, I believe, analogous to any measurement scenario. For instance, when we observe the trajectory of a stellar object through a telescope's eyepiece, we assume that the object's behavior occurs independently of the anthropocentric tool used to describe it. Even if quantum properties are more relevant in a given scenario, the theories that resolve the measurement problem provide theoretical and objective explanations for when and how measuring systems affect the results. The outcomes are not subjective.




\part*{\hypertarget{PartIII}{Part III}: \\ Relativistic Collapse Theories and a self-consistent model of semiclassical gravity }
\addcontentsline{toc}{part}{\protect\numberline{}Part III: Relativistic Collapse Theories and a self-consistent model of semiclassical gravity}\label{PartIII}

\chapter{Towards Relativistic Quantum Theories }\label{Ch: RelatQuantumTheo}

\section{Introduction}

As is well-known, the standard non-relativistic quantum mechanics formalism has some fundamental issues that can be summarized as the fact that it is not a proper physical theory. In more concrete terms, we will maintain that a physical theory should posses a clear ontology (what exist in the world) and concrete dynamical rules (how it behaves)\footnote{See \cite{MaudlinOntological} for an illustrative exposition of this point and \cite{maudlin2019philosophy} for quantum theories that fulfill this requirement.}. Let us take first the \textit{dynamical problem}, which appears from the fact that rules of standard quantum formalism are ambiguous, they depend on external observers---an artificial distinction between what is being measured and what is not, and on the meaning of the macroscopic notion of `measurement'. This has also been historically known as the measurement problem and famously exemplified in Schrödinger's cat thought experiment \cite{bell1990against,norsen2017foundations,maudlin1995three}. Second, it is far from clear which real entities these rules are supposed to refer to; there is no concrete commitment to what supposedly exist in nature (e.g. particles, fields, a quantum state, etc) \cite{MaudlinOntological}. This \textit{ontological problem} has led many physicists to suggest that quantum mechanics should be understood merely as a recipe for making predictions. However, even if one adopts this perspective, the question remains: why is the quantum recipe empirically successful? Naturally, such a question calls for an explanation from a more fundamental theory.
And what happens to all this discussion when we consider the lessons of relativity? Can we have clear and self-consistent relativistic ontologies and dynamics? Although we, ultimately want to undertake these questions, it is useful to first understand the non-relativistic versions and resolutions to the ontological and dynamical problems, which can be done satisfactorily and independently of the relativistic versions of the quantum recipe (like quantum field theory); to consider relativity from the beginning would only make it harder to solve both problems \cite{bell1995quantum,bell2001beables}.

Nowadays, we already have well-known resolutions to these fundamental problems of the non-relativistic standard formalism. These resolutions come in the form of `realists' quantum theories, equipped with an ontological description and proposing that `either the quantum state, as given by the Schrödinger equation, is not everything, or it is not right', as Bell said \cite{bell1995there}. That the quantum state `is not everything' is proposed in pilot-wave theories, which add variables and dynamical rules for them, and that `it is not right', is proposed by collapse theories that modify the Schrödinger equation in a non-linear stochastic manner \cite{bell1982impossible,ghirardi1986unified}. 

Although the non-relativistic versions of these theories have been recognized as resolutions to the fundamental problems of standard quantum mechanics, it is commonly claimed that, in any case, if you consider relativity, both present further problems. For example, it has been argued that attempts to extend collapse theories to a relativistic setting face several challenges: difficulties in constructing Lorentz-invariant models, tensions between indeterminism, the block-universe picture and foliation-dependent properties, and the risk of enabling faster-than-light signaling. And this is without even taking gravity into account, which introduces additional complications, especially in semiclassical approaches where spacetime is treated as classical and matter as quantum.  

In this chapter, we will focus primarily on the relativistic extensions of quantum collapse theories. Our first aim is to demonstrate that these theories present no such conflicts and leave no room for mystery.
A second, independent aim is to introduce the general framework of semiclassical gravity, which is motivated independently of any specific quantum theory. We will review the common objections it has faced, including claims of empirical inadequacy, internal inconsistency, and a more recent concern involving faster-than-light signaling. These alleged issues have been raised against the general framework of semiclassical gravity. However, we will see that the strength of such accusations actually depends on the underlying quantum theory. Finally, we will propose a novel, self-consistent, empirically viable, semiclassical gravity framework, in which the expectation value of a quantum field, evolving via a relativistic objective collapse dynamics couples to a wholly classical Einstein tensor. We  will present the general framework, a concrete example, and briefly explore possible empirical consequences of our model.

\section{Quantum Mechanics and Relativity}

On one side, it has been said countless times that relativity conflicts, in principle, with the formulation of a `realist' quantum theory \cite{aharonov1981can}. For instance, both routes mentioned earlier to solve the problems of the standard formalism have been criticized when extended to a relativistic framework. 

Non-relativistic pilot-wave theories consider the position, as a function of time, of the particles composing the universe, as an additional variable, and a `guiding' equation that correlates the velocity of a particle, at a certain time, with the position of all the other particles, at the same time. Thus, when a relativistic setting is considered, the most natural way to make sense of a guiding equation is to postulate a privileged foliation on the structure of spacetime, so the notion of simultaneous positions is with respect to this foliation. This feature has been criticized as going against the `spirit of relativity'\footnote{See \cite{struyve2024lorentz} for a survey a the proposals within pilot-wave theory to construct a relativistic extension.}.  With respect to the relativistic extension of collapse theories, we have already mentioned the sort of objections raised, from the inconsistency between the fundamental indeterminism they postulate and the block universe picture that relativity suggest, to superluminal signaling. 

The intuition behind the antipathy to both theories in a relativistic setting might have something to do we the old prejudices regarding what counts as a relativistic theory and our common notions of `locality'. It is true that both, pilot-wave and collapse theories, are fundamentally non-local, in the sense of Bell's theorem \footnote{Bell regarded a model as local if it satisfied his principle of `local  causality', meaning that the probability the model assigns to the value of a beable $b_x$, at a spacetime event $x$, satisfies:
	\begin{equation}
		P(b_x|\lambda_{\sigma})=	P(b_x|\lambda_{\sigma}, b_{y}),
	\end{equation}
	where $\lambda_{\sigma}$ is a complete specification of the state at $\sigma$, a space-like slice across the entire past light cone of $x$, and $b_y$ is the value of any beable at a spacetime event $y$, space-like separated from $x$ and outside of the causal future of $\sigma$. See \cite{ciepielewski2023superdeterministic} for more details.
}. which together with the empirical data, taught us that any theory that accounts for the quantum correlations, must be non-local \footnote{There are some alternatives to this conclusion, but the price to pay is much higher than accepting the non-locality in quantum theories. See also \cite{ciepielewski2023superdeterministic} for a discussion on this.}. However, this purported conflict is just, as we said, a prejudice. Einstein's theory says nothing about Bell's principle and there exist working relativistic extensions of both theories which are self-consistent and prevent the possibility of faster-than-light signaling. See \cite{tumulka2022foundations} for a survey of these theories.

Now, on the other not so popular side, it has been suggested that the relativistic extensions of these `realist' quantum theories could provide some suggestions, or point to some missing pieces, towards the construction of a quantum theory of gravity, given the higher degree of clarity in the dynamical rules and postulates of these approaches. See for example \cite{okon2014benefits} or \cite{pinto2019bohmian}. In this work we will endorse this view, and we will try to give further arguments to follow this approach within collapse theories.

Quantum collapse theories present a unified universal dynamics and their predictions agree with those of the standard quantum formalism without the use of external agents, or a distinction between the micro and macroscopic world\footnote{An interesting feature of collapse theories is that they are testable and falsifiable. At the mesoscopic scale their predictions differ from those of the standard formalism.}. As we have said, we sustain that a physical theory should not only present clear dynamical rules to make empirical predictions, but it should also posses concrete representations of the real entities that are supposed to exist in the world. The latter, known as the ontology, can come in distinct parts: local entities (like point particles or fields) and non-local entities (like a quantum state). Whereas the dynamical laws are supposed to describe or govern how these entities behave and relate to each other, like Newton's second law.

As we saw in section \S\ref{Albertsection} of the previous chapter, the most famous proposals for the ontology of quantum collapse theories are known as the `\textit{flashes}', associated with the center $x$ of the collapse of the wavefunction, and the \textit{mass density}, which also depends on the wavefunction. Both are local entities living in our physical space, in contrast to the quantum state, represented by the wavefunction, a non-local element of the ontology that lives in a higher dimensional configuration space.

The distinction between local and non-local `beables' (in Bell's terminology for the elements of the ontology) will be crucial to alleviating apparent conflicts between a quantum theory and relativity \cite{Bell_Aspect_2004localbeables}. This distinction is related to the notion of \textit{separability}---it is roughly said that an entity is separable if we can give a complete physical description of it just by giving its physical properties at every point in the space it occupies, if we can not give this description, the entity will be non-separable. It turns out that in classical physics all the entities we encountered were separable. The mass of a bike can be given adding the mass of the parts that compose the bike. Thus, when General Relativity arrived and given that matter is considered classical in Einstein's equations, it gave the strong impression that for a theory of matter to be relativistic, separability must be an ingredient, engendering one of the purported tensions with Quantum Mechanics. If we consider the quantum state as a beable of our quantum theory, it can only be a non-separable entity, or more precisely a non-local beable. The singlet state of two electrons can not be reduced just to the states of the individual electrons.

One of the first questions that must be addressed is: how can we construct a relativistic dynamical description of a non-local beable like the quantum state? This was already addressed since the Schwinger-Tomonaga formalism considering only a unitary evolution for the quantum state, but this is not enough to explain the appearance of single outcomes in measurement situations if the quantum state is considered to be complete\cite{maudlin1995three}. As we will explain in section \ref{SectionRelCollapse}, there is also a natural way to do this for the quantum state if we include spontaneous collapses, as was clearly and generally outlined by Myrvold in \cite{myrvold2003relativistic} and applied in a concrete relativistic collapse model in \cite{bedingham2011relativistic}.

Concerning the local beables of these theories, successful relativistic proposals have been published a bit more recently, for the mass density in \cite{bedingham2014matter} and for flashes in \cite{tumulka2021relativistic}.  Together with these, there has been other important works in the last decades treating various aspects of the relativistic extension of quantum collapse theories \cite{aharonov1984usual,
	fleming1986lorentz,
	fleming1988lorentz,
	myrvold2002peaceful,
	myrvold2003relativistic,
	tumulka2006relativistic,
	tumulka2022foundations}.

However, despite all these fruitful advances, there still persist, mostly in the theoretical physics community, the widespread idea that quantum collapse theories are problematic or even impossible, in principle, to merge with relativity. The main ideas of this sort are based on the non-local behavior of the collapse affecting a global quantum state. That is, the non-local beable of the theory gets affected instantaneously, which seems to require an absolute notion of simultaneity (potentially opening the possibility of faster-than-light-signaling \cite{sorkin1993impossible}), otherwise, to `relativize' the effect of the collapse--- making it dependent to the foliation---would bring conceptual problems, the objectors would say. Regarding this last sort of objections, it has been said, for example, that a photon could have a non-determined polarized state with respect to some foliation and a determined polarized state with respect to another \cite{maudlin2011quantum}. Similar influential arguments identified observers with foliations and argued that, when for an observer, a distant (spacelike separated) collapse is already in the past, for another observer next to the first, the same collapse has not already occurred \cite{Rietdijk,Putnam1967-PUTTAP}. These ideas are related to the more general supposed conflict between the block-universe picture of completely determined events that emerges from relativity and the undetermined possible future states that a quantum collapse theory can only give us \cite{delSanto2021relativity}. In section \ref{SectionObjections} we will carefully review these objections. 

\section{Objections to a Relativistic Quantum Collapse Theory}\label{SectionObjections}

Given that the objections to relativistic collapse theories were raised before the successful proposals were even constructed or published, we can review in this section the discomfort and motivation leading to these allegations before presenting the concrete relativistic framework of collapse theories that alleviates the issues and clarifies the misunderstandings. For this, it will only be necessary to recall some general features of non-relativistic collapse theories: First, as in standard non-relativistic quantum mechanics, to every time $t$ (a slice of absolute simultaneity in Galilean spacetime) corresponds an associated quantum state. Second, the quantum state exhaust everything than can physically be said about the system. Finally, given the quantum state at time $t$, the most that can (in principle) be said about the system's future is a list of the allowed states and their respective objective probabilities.

\subsection{The block-universe picture against indeterminism}

On the one hand, and as we will explicitly see in an EPR setting in the following subsection, in collapse theories  we can think of the world as fundamentally indeterministic, where the future states are treated differently than past and present states. One might think that future states are only as 'real' as the objective probabilities assigned to them by the laws of nature, whereas the present state certainly exists, and the past is already settled.\footnote{But bear in mind that this is a metaphysical position independent of collapse theories. In this regard, non-relativistic collapse theories, or even collapse theories with a preferred foliation of spacetime, could appear to fit well with the intuitive notions of `becoming', the `openness of the future' or the `passage of time'.} 

On the other hand, when we take into account the lessons of relativity, it seems natural to adopt a four-dimensional `block universe' view, in which there is no preferred foliation of spacetime and all events are fully determined. This has led many people to believe that indeterministic theories of matter, like those of quantum collapse, conflict fundamentally with relativity. See \cite{delSanto2021relativity} for another exposition of this general objection. Some authors have exposed particular faces of this objection to show the alleged problem between indeterminism and relativity.
In \cite{Rietdijk}, it is argued that special relativity implies that any event in the future of one observer is already in the past (and therefore determined) for another. From this, it is concluded that 'there is determinism, also in micro-physics.' However, this conclusion rests on the assumption that inertial reference frames can be identified with observers. A similar line of reasoning was defended in \cite{Putnam1967-PUTTAP}.

However, it is worth noting that even in the non-relativistic case of quantum collapse theories, one can conceive of a block universe in which the entire state history of the world is determined. In this picture, at every instant of absolute time $t$, the state at a later time $t_+$ is already fixed. Nevertheless, the dynamical laws of nature are such that, at any given instant $t$, they provide at most objective probabilities for the possible future states. These probabilities can be interpreted objectively because they do not depend on agents having any information, nor in an incomplete specification of the state. In other words, the rules for computing the probable future states are not epistemic, they are as objective as the laws of nature can be up to the time $t$. If for example, at $t_+$ a collapse occurs which changes an initially entangled state in some basis onto an eigenstate of the same basis, then, what at $t$ appeared as undetermined, with respect to $t_+$ appears as determined. There is anything mysterious about this. The state of undeterminacy depends on the absolute time, or `slice of simultaneity'.  

An analogous thing is going to happen in the relativistic case, with the difference that slices of simultaneity become foliation dependent. With this in mind, we can start to feel how the mysteries will dissolve when we arrive to the relativistic context.

\subsection{The objection of undefined properties: The EPR example}

A common example to show explicitly the different faces of this apparent issue, between quantum indeterministic collapses and the alleged determinism in relativity, is the EPR experiment. The setup of the experiment is this: Suppose that two observes, Alice and Bob, have arranged to measure the $z$-spin component of two spacelike separated $\frac{1}{2}$-spin particles with a previously prepared entangled quantum state  $$\ket{\psi}=\frac{1}{\sqrt{2}}(\ket{z+}_A\otimes\ket{z-}_B+\ket{z-}_A\otimes\ket{z+}_B)$$ (ignoring spatial degrees of freedom), respectively. For simplicity and without loss of generality we will assume that the spatial displacement of the particles is only in the $x$ direction. Suppose that Alice measures the left particle and finds it with $z$-spin  $+1/2$\footnote{Strictly speaking, what we mean when we say Alice finds her particle with $z$-spin  $+1/2$ is that the outcome displayed by her detector is associated with that value of the spin.}. 

\begin{figure}[ht]
	\centering
	\includegraphics[width=5cm]{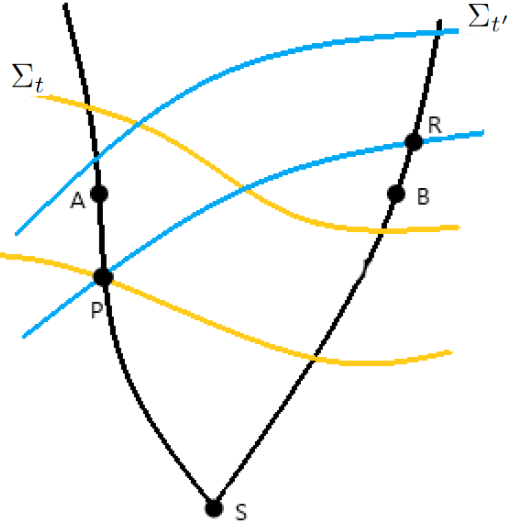}
	\caption{{\small Spacetime diagram of the EPR setting with two foliations: in yellow th one with hypersurfaces $\Sigma_{t}$ and in blue the one with hypersurfaces $\Sigma_{t'}$. $A$ and $B$ represent the events associated with Alice and Bob's measurements in the $z-$spin basis respectively. $P$ is an event in Alice's worldline, before she makes her measurement, and $R$ is an event in Bob's worldline, after he makes his measurement.}}\label{2foliations}
	
\end{figure}

Let us analyze this experiment to display some of the apparent problems that collapse theories in a relativistic setting could have. For this, consider the two different foliations,  $\Sigma_t$ and $\Sigma_{t'}$, depicted in figure \ref{2foliations} with their respective slices of (relative) simultaneity. 

 Consider, on one hand, Alice's description of her particle around the spacetime point $P$, using the (yellow) foliation of slices of (relative) simultaneity labeled by $t$. Then, for the yellow slice crossing $P$, no measurement (hers or Bob's) has occurred, so she must assign the same initial quantum state represented by $\ket{\psi}$. The value of Alice's $z$-spin measurement remains undetermined (with a $50$-$50\%$ chance for each outcome). On the other hand, we can also think of the description of Alice's particle at the same point $P$ using the (blue) foliation of slices of (relative) simultaneity labeled by $t'$. Then, all the spacetime events along the blue slice crossing $P$ are simultaneous. In particular, the event $R$ is in the crossing of this slice with Bob's worldline after his measurement at $B$. Thus, the associated quantum state should have already collapsed to obtain a single outcome, for example $\ket{z+}_A\otimes\ket{z-}_B$, associated to a determined value $+1/2$ for the $z-$spin of Alice's measurement. Therefore, we arrive at something that could seem as a terrible contradiction between state descriptions: how could the spin of a particle be well defined and undefined at the same spacetime point? How can a measurement outcome be determined and undetermined at $P$ \cite{maudlin2011quantum}? 

In \cite{myrvold2002peaceful}, one of the first clear and general attempts to dissolve the apparent contradiction between relativistic quantum collapse theories and relativity was presented. In brief, it is argued that we should not forget that, contrary to classical intuition, quantum mechanics features non-local properties that cannot be described separably---such as spin. In a relativistic context, spin turns out to be a property of an extended beable: the quantum state. This state not only depends on its associated hypersurface but is also, as we have noted, non-separable. We will explain how this framework makes the job with a concrete relativistic collapse theory in section \ref{SectionRelCollapse}. For now, let us keep in mind that for a theory of this sort to be satisfactory, apart from providing a clear relativistic description of its non-local beables and its dynamics, it should also provide a relativistic description of its local beables and their relation to the non-local ones in a way that permits Bell-type correlations without faster-than-light signaling.

\subsection{Relativistic-invariance, Narratability and Becoming}

Another aspect of the above objection is formulated in terms of other classical, and somewhat ambiguous, notions as ``narratability'', ``coming into being'', or ``temporal becoming'', which are allegedly in conflict with relativity. For example, in \cite{AlbertSR2000}, it is argued that for a theory to be \textit{metaphysically compatible} with special relativity, it must `depict the world as unfolding in a four-dimensional Minkowskian space-time'. The author complains that, if we were to maintain this quantum collapse theory, `we have to let go of the idea of the world's having anything along the lines of a narratable story at all'. 

In particular, the author claims in \cite{AlbertSR2000} that it must be the case that if we know the history in a foliation (namely, the quantum states in every hypersurface of it) we can always `tell' in a covariant manner the history in another foliation. But it is argued that this can not be achieved in the relativistic extension of quantum collapses. To see this, consider again the setup of the EPR experiment above but without measurements. In this case is easy to see that the state in all the hypersurfaces of $\Sigma$ and $\Sigma^{'}$ is a singlet state. However, if a magnetic field is turned on `simultaneously' on $\Sigma$, with respect to this foliation the state will always be singlet state, but with respect to $\Sigma^{'}$, the state is not always a singlet (see for explicit calculations \cite{myrvold2002peaceful}). Thus, when a measurement of spin produces a collapse: according to $\Sigma$ the outcomes will certainly be opposite, but according to  $\Sigma^{'}$, there is a non-zero probability that the outcomes of Alice and Bob will match. This suggests that, even if we know all the history in one foliation, in general we can not obtain in a covariant manner the history described from another foliation. However, as we will see later, we can always achieve this if we also know the hamiltonian.

A similar objection is raised in \cite{GisinEsfeld}. There, the authors analyze a relativistic extension of a particular quantum collapse model, the GRW dynamics, with a `flash' ontology, or $rGRWf$ in short. After analyzing an EPR-setting and the flashes associated to the outcomes of Alice and Bob, they conclude that  `neither $rGRWf$, nor any other theory, can account for the occurrence of the Alice-flash and the occurrence of the Bob-flash in a Lorentz-invariant manner''. We will not get into much more detail of these arguments here, but what is needed for such conflicts to arise with quantum mechanics is: First, as explicitly stated in \cite{AlbertSR2000}, that `everything there is to say about the world can straightforwardly be read off of a catalogue of the local physical properties at every one of the continuous infinity of positions in a space-time'. And second, that `whatever law-like relations there may be between the values of those local properties are invariant under Lorentz-transformations'. 

In brief, the first petition is that separability is fulfilled. But separability seems more a property of classical theories of matter rather than of relativistic ones. As we have already mention in the introduction, it may be just a coincidence that the first relativistic theories that we conceived treated matter classically. It could actually be that nothing in principle forbids a relativistic description of non-separable entities, like the quantum state. In section $3$ we will show how this has already been done satisfactorily. In any case, the root of the supposed problems between the quantum non-separability added with collapse and determinism, can be exemplified with cases of quantities well-defined and undefined for some observer at the same spacetime event as we saw above with the EPR-setting.

\subsection{The faster-than-light signaling tension}

Finally, the last objection unrelated to gravity against relativistic collapse theories that we will mention was put forward in \cite{sorkin1993impossible}. There, the author devices a sort of no-go theorem for idealized measurements in quantum field theory. He considers a thought experiment analyzed under certain `natural' suppositions he expects about the relativistic generalization of quantum measurements. He finds a dependency of the expectation values in a bounded region, space-like separated, to the non-selective measurements carried out in another bounded region.

The set-up proposed is this, we have two space-like separated bounded regions $A$ and $B$, and a third region $C$ being a space-like hyperplane between $A$ and $B$. The idea is to arrange three experimenters Alice, Bob and Charlie to interact in a certain way with a quantum system in the regions $A$, $B$ and $C$, respectively. The quantum system is composed by two particles initially unentangled, one located in $A$ and the other in $B$. First, Alice carries out a procedure (a non-selective measurement) in her particle at $A$. Second, Charlie performs a projective measurement to the entire system at $C$. And third, Bob makes a measurement of a certain observable to his particle at $C$. It is shown that Bob's results (the expectation values of the observable) will depend on whether Alice decides to carry out her procedure or not. So, they have found a way to send signals faster-than-light.

We will see in \S \ref{SectionRelCollapse} that there are two crucial characteristics that relativistic collapse theories have---preventing them from superluminal signaling---that Sorkin's measurements do not have. In brief,  relativistic collapse theories will have: i) local beables, in the sense of Bell, which can not change due to space-like separated influences; and ii) an evolution for the non-local quantum state that, from one hypersurface to another, only depends on the collapses between them.

\chapter{Semiclassical gravity and its potential problems}\label{Chapter:SemiClass}

Semi   classical gravity is a formalism in which the gravitational field is treated as a classical relativistic space-time whereas matter is described by a quantum field on that space-time. The best known implementations would have the matter following the standard QFT dynamics, while space-time follows Einstein's semiclassical equations,
\begin{equation}\label{SemiClass}
	G_{ab}=8\pi G \bra{\psi}\hat{T}_{ab}\ket{\psi},
\end{equation} 
where the (renormalized) expectation value of the energy-momentum tensor of quantum matter fields in a specific choice of quantum state $\psi$, act as a source for gravity described by the classical Einstein tensor. However, it is a common belief in the community that semiclassical gravity has been shown to be untenable---both as an effective model and even more so as a fundamental theory---with a variety of influential arguments over the years pointing to its internal inconsistencies, potential for faster-than-light signaling, or empirical inadequacy. Note, to begin, that as a result of the Bianchi identities, the divergence of the left-hand-side of the equation (\ref{SemiClass}) is always zero. While the divergence of the expectation value on the right-hand-side is not generically guaranteed to vanish---particularly during quantum collapses. 

The fact that matter is acknowledged to posses a quantum nature implies the necessity to construct a theory of gravity that, unlike general relativity, acknowledges such a quantum character. This does not necessarily implies the metric itself to be quantized, contrary to widely popular sayings. However, a very likely implication of a theory in which the metric is quantized is that classical space-time should appear as an effective phenomenon that, at some level of approximation, emerges out of the fundamental, quantum gravity degrees of freedom. Although we do not yet have a complete theory of quantum gravity, most approaches to it do not pose spatio-temporal degrees of freedom at the fundamental level\footnote{They may pose degrees of freedom related to spatio-temporal notions but not in the standard way. For instance, purely spatial related degrees of freedom as in Loop Quantum Gravity, or some that merely represent causal relations in the Causal Sets approach.}. And, in order for them to be empirically viable, it is required from them to explain how, at some approximation, an essentially classical space-time emerges. That is, even if at the fundamental level quantum gravity theories have nothing resembling a classical space-time, they should be able to describe the semiclassical scenarios, producing an emergent space-time, in which they could be experimentally probed. Because of this, even in the absence of a quantum gravity theory, it can be said that, at some level, it must be well-approximated by a semiclassical scenario, in which matter fields are treated quantum mechanically, but space-time is treated classically.

Let us briefly review some of the main arguments against semiclassical gravity.

\section{Inconsistency and Empirical Inadequacy}

For instance, Eppley and Hannah \cite{eppley1977necessity}, composed a famous thought experiment in which the position of a quantum particle, assumed to have small uncertainty in momentum, is measured with a classical gravitational wave with small wavelength \textit{and} small momentum. They claimed to show that `the assumption that a classical gravitational field interacts with a quantum system' is untenable whether we assume that the interaction counts as a measurement that collapses the wave function or not, whatever the nature of the coupling is. In the former assumption, we would have two possible outcomes: an inconsistency due to a violation of either momentum conservation, or empirical inadequacy due to a violation of the uncertainty principle. More precisely, the first two possibilities are: i) the measurement localizes the quantum particle producing, according to the uncertainty principle, a state of very high momentum-- in which case momentum is not conserved; or conversely, ii) the measurement localizes the quantum particle to within the gravitational wave without transferring a large momentum--in which case Heinsenberg's uncertainty principle is violated.

However, the argument crucially depends on the interpretation of quantum mechanics you take, with the authors leaving it unspecified. This is why in \cite{huggett2001quantize} it is concluded that the argument in \cite{eppley1977necessity} falls short of a no-go theorem. Worse yet, it has the following issues: a)  it is not clear that the gravitational degrees of freedom are sufficient to conduct the measurement, that is, it might be the case that we can only access to the information codified in the gravitational degrees of freedom trough matter, whose quantumness would bring the uncertainties back in. b) As stressed in \cite{bahrami2014schrodinger}, the uncertainty relations are `a corollary of linear quantum theory' rather than a fundamental principle, not empirically probed in the proposed situation. And c) in  \cite{mattingly2005quantum} it is shown that the proposed experiment cannot be carried-out, even in principle. 

Another famous assessment of the viability and empirical adequacy of semiclassical gravity is carried-out by Page and Geilker in \cite{PageGeilker}, in which an actual experiment is reported. The experimental set-up employed uses a quantum event (e.g., the spin along $z$ of a spin-$\frac{1}{2}$ particle prepared spin-up along $x$) to determine wether a sphere of mass $M$ is displaced towards the right or left\footnote{The experimental set-up in \cite{PageGeilker} is actually more complex but the details are not relevant for the present discussion. }. Then, the gravitational field of the sphere is measured through a Cavendish-like torsion balance and a small test mass. The result, as expected, shows that when the quantum event presents a displacement towards one side, the Cavendish test mass experiences a gravitational field at the corresponding location.

Some advocates of the quantum gravity program take this enthusiastically as an indication of the non-viability of semiclassical gravity. Again, they basically arrive at two possible conclusions, or semiclassical gravity is not viable due to internal inconsistencies, or it is not empirically adequate. Their reasoning goes as follows: If the quantum state of the system does undergo some sort of collapse, then the divergence of $\bra{\Psi}\hat{T}_{ab}\ket{\Psi}$ during such process will not (generically) vanish as mandated by the Bianchi identities on the left-hand side of the semiclassical equation. On the other hand, if there is no state vector reduction, then the spin measurement will leave the spin-$\frac{1}{2}$ particle in a spin-up and spin-down superposition along $z$. Thus, the mass $M$ will also get into a superposition of being displaced towards the right and the left, so the gravitational effect of the sphere will retain its initial symmetry and the test mass will not move either way. This, however, is contradicted by the experimental result.  

As strong as the argument may seem, we do not take it as a conclusive indication that semiclassical gravity is useless (See \cite{carlip2008quantum} for another defense to the semiclassical program). Needless to say, \cite{PageGeilker} do indeed bring to the table two potential problems of the semiclassical framework. On the one hand, the equation might fail to be consistent due to the divergence of the expectation value of the energy-momentum tensor (particularly during collapses). On the other hand, the program might be empirically inadequate when encountered with macroscopic superpositions where the quantum state presents large dispersions in $\hat{T}_{ab}$.

Although there are possible responses available to both issues, we acknowledge that the first one constitutes a serious hurdle for taking semiclassical gravity as a \textit{fundamental} description. That being said, we do not see it as an impediment for looking it as a viable \textit{effective} description, with restricted but rather wide applicability. That is, we can expect the semiclassical description of space-time to breakdown in some situations, like those associated with what in the standard approach is known as the collapse postulate\footnote{A similar example of an effective theory is provided by the standard-reading of the Navier-Stokes equations as applied to fluid dynamics. Such equations provide a robust description of fluids under a wide set of circumstances. Yet, they break down under certain conditions like turbulence or the break of a wave in the ocean.}. In section \ref{Section:SCU} we will explore the issue with objective collapse theories. 

The conclusion drawn in \cite{PageGeilker} is that semiclassical gravity does not apply in such circumstances and that, to account for such situations, one need to resort to full quantum gravity. The problem we would be facing is of course that we have at this time no really satisfactory and workable theory of quantum gravity despite the enormous efforts along several apparently promising paths. For readers who believe otherwise and might readily point to their preferred theory, it is worth considering how such an approach would describe, for instance, the quantum state corresponding to the spacetime associated with matter in a quantum superposition of different locations. 

In view of the analysis of \cite{maudlin2020status} on the argument of \cite{PageGeilker}, there is an indication that, generically, all paths to address the conceptual problems that afflict quantum theory \cite{maudlin1995three}, while at the same time relying on semi-classical gravity, will entail suspension of the covariant conservation of the expectation of the quantum stress energy tensor at moments of collapse.

But then, at moments of collapse, the unitary time-evolution rule of quantum mechanics is suspended anyway so maybe one should not think of this suspension of covariant conservation as making things much worse than they anyway inevitably are when state-vector collapses happen.

Regarding the second potential issue presented in \cite{PageGeilker}, that is, the empirical inadequacy, we will also see that its strength depends on the underlying quantum theory or `interpretation'. This severely changes the answer to the following question: To what extent it is in fact possible to construct quantum states with large enough energy dispersions to make semiclassical gravity empirically inadequate? For example, it might be argued by some advocates of orthodox quantum interpretations that at the practical level, because of decoherence, the superposition of $M$ being at two distinct places considered in \cite{PageGeilker} never actually realizes. The problem, though, is that the vagueness present in orthodox interpretations like the standard quantum formalism, does not really allow for a general, rigorous assessment of the issue. In other words, the ambiguities associated with the collapse postulate in standard quantum mechanics get in the way of deciding if it is actually possible \textit{in practice} to prepare a macroscopic system in a quantum superposition of empirically relevant distinct positions. In section \ref{Section:aRelatQCTheory} we will explore relativistic collapse theories, which actually permit us to provide a concrete answer to this issue, consistent with the empirical result obtained in \cite{PageGeilker}, undermining the conclusion that, to account for such experiments, it is necessary to resort to full quantum gravity.

In contrast, what \cite{PageGeilker} concludes from their argument is that, ``\textit{in order to retain (\ref{SemiClass})[i.e. the semi-classical Einstein equations] as the simplest semiclassical theory of gravity, we must assume that the universal matter wave function $\Psi$ never collapses, as in the Everett formulation of quantum mechanics.''}.

At this point, we should note that, as discussed in \cite{maudlin2020status}, if one intents to adopt an Everettian type of approach, one must deal in a clear manner with the question of what does one take to be the space-time and the corresponding energy momentum expectation value associated with each of the Many World branches collectively represented by the quantum state. If we take the state of the gravitating matter to be the complete collective state then all branches of the wave function will be associated with the same space-time and in particular the proposal will fail to address the experimental results contemplated in \cite{PageGeilker}. On the other hand, if each branch is associated with its own spacetime---taking the gravitating quantum state to be the one corresponding to that branch---we would face the same problem of internal inconsistency as in the case where collapses occur.

\section{Signaling through gravitational influence}\label{Section:SGsignaling}

Now, recall that we said above that Eppley and Hanna's argument purportedly showed the untenability of a semiclassical framework whether we assumed the interaction of the classical gravitational field with quantum matter to produce a collapse of the wave function or not. We have discussed the first situation, which according to \cite{eppley1977necessity} resulted in i) inconsistency or ii) empirical inadequacy. If, on the other hand, we assume that the interaction does not induce a collapse, \cite{eppley1977necessity} claim instead that  iii) the measurement could be used for superluminal communication. 

To arrive to this conclusion they compose a different thought experiment in which the scattering of the gravitational wave gives information about the shape of the wave function. Using this in an EPR setting would allow for superluminal signaling. Here, we could object again that this crucially depends on the underlying quantum theory, that the gravitational degrees of freedom might not be sufficient to conduct a completed measurement, that this experiment might not be realizable in principle, and that matter together with collapse rules should be brought back in to avoid the issue. 

However,  there is a more elaborated and modern argument, with a similar flavor, that one could use against semi-classical gravity, implying faster-than-light signaling. This has to do with the characteristics of the Schrödinger-Newton equation, which, as shown in \cite{bahrami2014schrodinger}, together with the standard collapse postulate, leads to faster-than-light signaling. Moreover, in \cite{bahrami2014schrodinger} it is also shown that the weak-field non-relativistic limit of the semiclassical equations (\ref{SemiClass}) is precisely the Schrödinger-Newton equation 
\begin{equation}\label{SNeq}
	\mathrm{i} \hbar \partial_t \psi(t, \mathbf{r})=\left(-\frac{\hbar^2}{2 m} \nabla^2-G m^2 \int \mathrm{d}^3 \mathbf{r}^{\prime} \frac{\left|\psi\left(t, \mathbf{r}^{\prime}\right)\right|^2}{\left|\mathbf{r}-\mathbf{r}^{\prime}\right|}\right) \psi(t, \mathbf{r}).
\end{equation}
Thus, semiclassical gravity could be accused to lead to superluminal communication.

Concretely speaking, in \cite{bahrami2014schrodinger} the weak-field limit of equations (\ref{SemiClass}) is considered, that is, where classical gravity can be described by the space-time metric,
\begin{equation}
	g_{ab}=\eta_{ab}+h_{ab}.
\end{equation}
Then, the expansion on $h_{ab}$ leads to the gravitational wave equations also for the semiclassical case, and in the Newtonian limit (where $v\ll c$) the term $\braket{\psi|\hat{T}_{00}|\psi}$ is large compared to all other terms so (\ref{SemiClass}) reduces to the Poisson equation\footnote{This is analogous to standard general relativity in the Newtonian limit where the curvature of spacetime is reduced to a Newtonian potencial produced by the energy density term of the stress-energy-momentum tensor.}
\begin{equation}\label{Poisson}
	-\frac{c^2}{2}\nabla^2 h_{00}=\frac{4\pi G}{c^2}\braket{\psi|\hat{T}_{00}|\psi}.
\end{equation}
This Newtonian potential, together with the term  $\hat{T}_{00}$, are the dominant contributions to the interaction hamiltonian\footnote{Here, the assumption that only matter is quantized appears again, with the perturbation $h_{\mu\nu}$ mantaining its classicality. In contrast to what should happen in a theory of quantum gravity.}
\cite{maggiore2007gravitational}
\begin{equation}
	\hat{H}_{int}=-\frac{1}{2}\int d^3 r\  h_{\mu\nu}\hat{T}^{\mu\nu}\ \stackrel{N}{=}\ -G\int d^3 r\  d^3 r'\ \frac{\braket{\psi|\hat{\rho}(\textrm{\textbf{r}}')|\psi} }{|\textrm{\textbf{r}}-\textrm{\textbf{r}}'|} \hat{\rho}(\textrm{\textbf{r}}),
\end{equation}
where ``$\stackrel{N}{=}$'' means the Newtonian limit, the expression $\hat{T}_{00}=c^2\hat{\rho}$ has been used, and the equation (\ref{Poisson}) has been integrated. Considering $\hat{\rho}$ to be a mass density operator defined as $m|\psi|^2$, for only one species of particle, we can arrive to the Schrödinger-Newton equation (\ref{SNeq}).  

Now, faster than light communication can be achieved in the following way. Note that the second term on the right hand side of eq. (\ref{SNeq}) implies a \textit{self-attraction} of the wave packets. A single wave packet tends to self-focus and different wave packets tend to attract each other. Consider an EPR-type scenario where a pair of entangled particles in the singlet state are sent to Alice and Bob, both of whom have a Stern-Gerlach apparatus to test the spin of their particles. Alice will measure first and she can choose to orient her apparatus in the $z$ or $x$ direction, while Bob has his apparatus fixed in the $z$ direction. Due to the singlet state, Bob's particle will end up prepared in different states, i.e., $\ket{x\pm}$ or $\ket{z\pm}$, depending on Alice's choice, and outcome. 

If Alice chooses the $z$ orientation, Bob's particle will be deflected either to the $+z$, or to the $-z$ direction, ending in a displacement $\pm d$ from the center of a detection screen in Bob's lab. However, if Alice chooses the $x$ orientation, Bob's particle enters the apparatus in a superposition of $\ket{z+}$ and $\ket{z-}$. The two wave packets deflect in the respective directions, but now, the second term of the Schrödinger-Newton equation (\ref{SNeq}) produces an attraction between these two wave packets. This, in turn, implies that the detections in the screen will be displaced by a smaller amount $\pm d'$. Thus, Bob is able to receive messages from Alice faster-than-light \cite{bahrami2014schrodinger}.

Here, we can argue that the above possible argument against semiclassical gravity, again, crucially depends on the underlying quantum theory. In particular, on the specific dynamical rules producing the collapse of the wavefunction and on what entities are postulated to be the source of gravity. We will see in the next section that proper relativistic collapse theories, not only involve covariant evolution rules for the quantum state, they also have a relativistic prescription of the local beables, the part of the ontology assumed to be the responsible of influencing the spacetime geometry in the semiclassical equations (\ref{SemiClass}). A concrete recipe in \cite{bedingham2011relativistic}, defines a `mass density function', or in this case, an energy-momentum density tensor
\begin{equation}\label{E-Mdensitytensor}
	\mathcal{T}_{ab}(x)=\Bigl\langle \psi_{PLC(x)} \Big|  \hat{T}_{ab} \Big| \psi_{PLC(x)} \Bigl\rangle.
\end{equation}
Where $\psi_{PLC(x)}$ means the quantum state on the past-light-cone of $x$. Note that the right hand side of the semiclassical equations (\ref{SemiClass}) need a specification of the hypersurface $\Sigma$ to which the state $\ket{\psi}$ is associated. If $\Sigma$ is space-like, changes on $\hat{T}_{ab}$ will have a non-local influence, as needed for the superluminal signaling argument. However, if we use instead (\ref{E-Mdensitytensor}), the local beables in Bob's lab won't notice any difference until it is captured in its past light cone, thus avoiding faster than light communication.

\chapter{Relativistic Collapse Theories}

\section{Objective Quantum Collapse Theories}\label{SectionRelCollapse}

Here, we will consider quantum collapse theories assuming that the quantum state is a complete description of the system, that is, not as an epistemic tool used by an agent to encode relevant information of the system. These theories consists of two main ingredients: The first is a dynamical non-linear stochastic rule for the quantum state. The motivation for this non linear character is to avoid superpositions for macroscopic systems by localizing them objectively in the position basis,  but the effect should be negligible for microscopic systems, where we know the linear Schrodinger equation makes satisfactory predictions. Stochasticity is necessary in these theories to approximate the Born-rule statistics and avoid superluminal signaling \cite{gisin1989stochastic,gisin1990weinberg}.
The second ingredient is a mathematical rule that represents the real entities existing in our four dimensional world (i.e. the `local beables' in Bell's terminology, like ``collapse flashes'' or the mass density distribution). The definition of the local entities can be related to the quantum state (a non-local beable which in general pertains to a higher dimensional space) in such a way to accommodate the Bell-type correlations.

The first fully consistent collapse theories were constructed in a non-relativistic context \cite{ghirardi1986unified,ghirardi1990markov}, then based on those, relativistic extensions were developed, which will be the main object of this section.

\subsection{Non-relativistic collapse theories}

In the non-relativistic case, the dynamical rule and the quantum state depend on the universal time parameter $t$ of Galilean spacetime, in such a way that if we know the quantum state at some initial time $t_0$, lets say $\ket{\psi_{t_0}}$, the dynamical rules gives for each future time, $t$, a random variable $\ket{\psi_t}$ for the state of the system, that is, the solution is a stochastic process, in contrast to the deterministic Schrodinger equation which gives a unique state for each $t$.

For example, as was previously presented in subsection \S \ref{Albertsection}, in the GRW theory, if at an initial time $t_0$ the wavefunction $\Psi_{t_0}(\textbf{\textrm{x}}_1, \textbf{\textrm{x}}_2,..., \textbf{\textrm{x}}_N)$ represents the quantum state of an $N$ particle system with $\textbf{\textrm{x}}_i$ the position of the $i-th$ particle, the stochastic dynamical rule gives us objective probabilities for the state to become at a future time $t$ one represented by 
\begin{equation} \Psi_{t}(\textbf{\textrm{x}}_1, \textbf{\textrm{x}}_2,..., \textbf{\textrm{x}}_N) =\frac{L_n(\textbf{\textrm{x}}) \Psi_{t_0}(\textbf{\textrm{x}}_1, \textbf{\textrm{x}}_2,..., \textbf{\textrm{x}}_N)}{||L_n(\textbf{\textrm{x}})\Psi_{t_0}(\textbf{\textrm{x}}_1, \textbf{\textrm{x}}_2,..., \textbf{\textrm{x}}_N)||}.
\end{equation}
Here, $L_n(\textbf{\textrm{x}})$ is the collapse operator associated to the $n$-th particle with the center of the collapse represented by the random variable  $\textbf{\textrm{x}}$ and has the gaussian form
\begin{equation}
	L_n(\textbf{\textrm{x}})=\frac{1}{(\pi r^2_C)^{3/4}}e^{-(\textbf{\textrm{q}}_n-\textbf{\textrm{x}})^2/2r^2_C},
\end{equation} 
with $r_C$ a new constant of nature which fixes the width of the localization process, $\textbf{\textrm{q}}_n$ the position operator of the $n$-th particle, with probability density for the collapse to occur at $\textbf{\textrm{x}}$ given by
\begin{equation}
	p_n(\textbf{\textrm{x}}):=|| L_n(\textbf{\textrm{x}})  \Psi_{t}(\textbf{\textrm{x}}_1, \textbf{\textrm{x}}_2,..., \textbf{\textrm{x}}_N)
	||^2.
\end{equation}
The collapses are distributed in time like a Poissonian process with a frequency given by another new constant of nature $\lambda_{GRW}$. Between two collapses the state evolves according to the Schrödinger equation (see \cite{bassi2003dynamical,Bassi2013Models} for a more detailed exposition of non relativistic GRW). Regarding the local beables, we will assume that matter is continuously distributed in spacetime and described by the mass density function
\begin{equation}\label{massdensity-NR}
	m(\mathbf{x},t):=\sum_i^N m_i\int d^3\mathbf{x}_1...d^3\mathbf{x}_N\delta^3(\mathbf{x}_i-\mathbf{x})||\Psi_{t}(\textbf{\textrm{x}}_1, \textbf{\textrm{x}}_2,..., \textbf{\textrm{x}}_N) ||^2.
\end{equation}
where $m_i$ is a constant usually
called the mass of particle $i$. Let us now consider relativity.

\subsection{Towards a relativistic ontic theory?}\label{Section:relatontictheory}

It is not a trivial issue to characterize what a relativistic theory exactly means and what kind of constraints it implies. There are many (possibly non-equivalent) ways to define what makes a theory relativistic---some arguably more so than others---and a substantial body of literature on the subject\footnote{See for example \cite{struyve2024lorentz} for a recent discussion of this in the context, including the proposals for relativistic extensions of realist quantum theories, \cite{tumulka2022foundations} chapter $7$, section $7.5$ for another summary of relativistic properties an ontic theory could have.}. In this work, we will not engage deeply with that discussion. Instead, we will adopt a general criterion: that the only intrinsic object related to spacetime appearing in the laws of physics is the spacetime metric $g_{ab}$. And particularly, we will assume that a relativistic theory of (quantum) matter with events `unfolding' should be constructed as such to satisfy the following\footnote{See \cite{myrvold2003relativistic} for the first general and pleasantly clear exposition of this type of approach.}: The whole physical stuff along any hypersurface has a complete ontic state such that along any foliation there is an objective account of the world unfolding in successive states on that foliation. There is no privileged foliation that gives the unique correct state history\footnote{Thus, different foliations are associated to different state histories simply because different foliations join up spacetime events in different ways to form their hypersurfaces of simultaneity, on which global states are defined.} Therefore, a complete state history given in any foliation, uniquely determines\footnote{This is provided that we also have the complete rules for evolving the state over hypersurfaces. In quantum theories this includes for example, the hamiltonian of the system and relevant operators associated to the evolution of the quantum state.} the state history along any other foliation.

\begin{figure}[h]
	\centering
	\includegraphics[width=12cm]{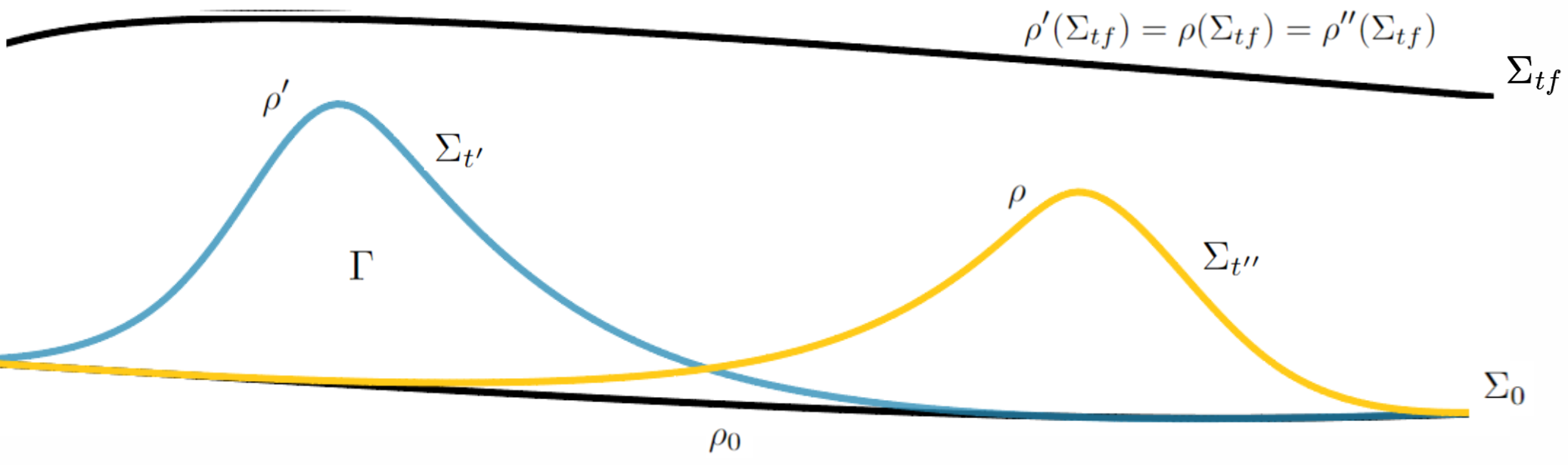}
	\caption{ {\small Different foliations sharing the same hypersurface $\Sigma_{tf}$. Independently of which foliation is used to arrive to the state in $\Sigma_{tf}$, e.g. the one that contains $\Sigma_{t'}$ with the associated state $\rho'$, or $\Sigma_{t''}$ with $\rho''$, the theory should assign the same $\rho(\Sigma_{tf})$ to $\Sigma_{tf}$. And any differences between the states $\rho'$ and $\rho_0$ are due to what happens in $\Gamma$.}}
	\label{onticfoli}
\end{figure}

More concretely, we will assume that our relativistic quantum collapse theories assign objective (ontic) states to hypersurfaces and that the dynamical laws let us assign (possible) states to subsequent hypersurfaces of the same foliation. For the assignment to be consistent, i) when some hypersurfaces, e.g. $\Sigma_0$ and $\Sigma_{t'}$ share the same spacetime events apart from the ones in a bounded region $\Gamma$ (see Figure \ref{onticfoli}), to arrive from the pure state $\rho_0$ over $\Sigma_0$ to $\rho'$ (of the set of possible pure states specified by the theory) over $\Sigma_{t'}$, the only relevant information is given by operators localized in $\Gamma$. This includes the hamiltonian density, the collapses, and any other `physical operation' that occurs in $\Gamma$, representable by pure operations local to $\Gamma$, which in turn, are represented by Kraus operators belonging to the associated algebra $R(\Gamma)$. See \cite{Clifton_2001} and \cite{myrvold2003relativistic} for more details on this. And ii) whenever any of these operators, let's say $\hat{O}_A$ and $\hat{O}_B$, belong to $R(\Gamma_A)$ and $R(\Gamma_B)$ with $\Gamma_A$ and $\Gamma_B$ space-like separated regions, they should satisfy the following commutation relations\footnote{This is also know as the microcausality condition, an assumption needed to prove the no-signaling theorem.}:
\begin{equation}
	[\hat{O}_A,\hat{O}_B]=0.
\end{equation}
Otherwise, we could arrive at different state assignments at $\Sigma_{tf}$, evolving the same pure state $\rho_0$ on $\Sigma_0$ considering in one case, first the relevant operators in $R(\Gamma_A)$ and then the ones in $R(\Gamma_B)$, and the inverted process in another case.

All this implies that when distinct foliations share the same hypersurface $\Sigma_{tf}$, the theory assigns the same state $\rho(\Sigma_{tf})$ on that hypersurface independently of the foliation. On the contrary, if our theory were to allow for more than one distinct state to be assigned to the same hypersurface, such states could not both be objective or pure. This would imply an epistemic rule for state assignment that depends on the foliation (or on the observer), which would be in direct conflict with our original requirement of a relativistic ontic theory.

With all this said, let us now get a real sense in the next section of how the non-local quantum state and the local beables in objective collapse theories would behave on relativistic spacetimes and what are the consequences for our understanding in simple examples.

\section{A Relativistic Quantum Collapse Theory}\label{Section:aRelatQCTheory}

When we consider general relativistic spacetimes, we seek for relativistic formulations of collapse theories where of course there is no absolute notion of simultaneity, or a preferred foliation of spacetime, to describe the evolution of global states. Thus, we will dispense the use of the universal time that non-relativistic collapse theories use. Then, as we have already said, we will consider a relativistic collapse theory that assigns an ontic quantum state to every hypersurface and its evolution to depend on the foliation. To this end, a consistent dynamical rule for this new formalism must be employed. For situations which do not include collapse, a consistent dynamical rule evolution was developed by Tomonaga and Schwinger almost 80 years ago. The idea is the following: suppose that (working in the interaction picture) we want to know the state on a hypersuperface $\Sigma$ that differs by a infinitesimal spacetime volume $\Delta V$ from a hypersuperface $\Sigma_0$ that has no points in the future of $\Sigma$. If we know the state, $\ket{\psi_{\Sigma_0}}$, on $\Sigma_0$ we can approximate $\ket{\psi_{\Sigma}}$ by
\begin{equation}
	\ket{\psi_{\Sigma}}\simeq [1+\mathcal{H}_{int}(x')\Delta V]\ket{\psi_{\Sigma_0}}
\end{equation}
where $x'\in\Delta V$ and $\mathcal{H}_{int}$ is the Hamiltonian density. Integrating this equation we can find the state in any hypersurface, $\Sigma$, lying in the future of $\Sigma_0$ by
\begin{equation}
	\ket{\psi_{\Sigma}}=Te^{-i\int_{\Sigma_0}^{\Sigma}\mathcal{H}(x)d^4x}\ket{\psi_{\Sigma_0}},
\end{equation}
with $T$ the time ordering operator. Or, equivalently written, we will assume that the Tomonaga-Schwinger is satisfied
\begin{equation}\label{TomonagaSchw}
	i \frac{\delta \ket{\Psi_\Sigma}}{\delta \Sigma(x)} = \mathcal{H}_{\text{int}} (x) \ket{\Psi_\Sigma}.
\end{equation}
 To make sure that advancing of the hypersurface across
two different points ($x$ and $y$) spacelike separated is independent of the order in which this is done it is sufficent that
\begin{equation}
	[\mathcal{H}(x),\mathcal{H}(y)]=0.
\end{equation}
In particular, this assures that if two foliations share the same hypersurface, the state associated with that hypersurface is the same in both foliations.
With this formalism the state assignment is made in a covariant way, that is, independently of coordinate frames, observers, or a privileged foliation. This also implies that we can get to a solution of the theory in some foliation from solutions in any other foliation.

Now, the main problem when trying to adapt the collapse evolution into this formalism is its global character given in the non-relativistic description, in contrast to the unitary evolution  determined by the local operators $\mathcal{H}(x)$. But this issue precisely suggest the solution: As was generally discussed in the last subsection \S \ref{Section:relatontictheory}, we should describe the collapse evolution also by local operators satisfying suitable microcausality conditions. An immediate consequence is that for each foliation the collapse ``occurs'' along the spacelike hypersurface that passes through the collapse region. We emphasize that this is a straightforward consequence of associating quantum states with each hypersurface and requiring that the collapse evolution be governed by local operators. It is not a contrived or ad hoc prescription intended to reconcile quantum collapse with relativity, as some authors have suggested \cite{AlbertSR2000,maudlin2011quantum}.

Thus, concretely speaking, with respect to the collapses we will assume a behavior similar to that in GRW: the quantum state undergoes discrete collapses associated
to randomly selected spacetime events. In particular, when an hypersurface $\Sigma$ passes through a collapse at $x$, for a moment, the state stops satisfying the Tomonaga-Schwinger equation and evolves, instead,  according to
\begin{equation}\label{CollRule}
	\ket{\Psi_\Sigma} \rightarrow \ket{\Psi_{\Sigma^+}} = \hat{L}_x (Z_x) \ket{\Psi_\Sigma},
\end{equation}
where $\hat{L}_x$ is the collapse operator at $x$ and $Z_x$ is a random variable, the `outcome', selected according to the probability distribution 
\begin{equation}\label{Prob}
	P(Z_x|\ket{\Psi_\Sigma}) = \frac{\bra{\Psi_\Sigma} |\hat{L}_x (Z_x) |^2 \ket{\Psi_\Sigma} }{\braket{\Psi_\Sigma}}.
\end{equation}
It is assumed that there is a constant probability of collapses per unit 4-volume, defined in a covariant manner (i.e., not depending on any specific choice of foliation) and that the collapse operators satisfy
\begin{equation}
	\label{Comp}
	\int dZ |\hat{L} (Z) |^2 =1.
\end{equation}
Moreover, to have a covariant, foliation-independent, and complete framework of relativistic collapses, one has to propose (in accordance with subsection \S \ref{Section:relatontictheory}) collapse operators satisfying
\begin{equation}
	\label{MC}
	[\hat{L_x} (Z_x), \hat{L_y} (Z_y)]=0, \quad \text{and} \quad [\hat{L_x} (Z_x), \mathcal{H}_{\text{int}} (y)]=0
\end{equation}
for spacelike separated $x$ and $y$, given hypersurfaces $\Sigma_i$ and $\Sigma_f$, with no point in $\Sigma_i$ to the future of $\Sigma_f$. In this manner, the dynamics assigns a foliation-independent state to $\Sigma_f$, and a foliation-independent probability of the complete set of $Z$s between $\Sigma_i$ and $\Sigma_i$, given the state on $\Sigma_i$. A first concrete proposal for the collapse operators satisfying Eqs. (\ref{MC}) is 
\begin{equation}\label{CollapseOpe}
	\hat{L_x} (Z_x) = \frac{1}{(2 \pi \sigma^2)^{1/4}} \exp{- \frac{ \left( \hat{A}(x)-Z_x \right)^2}{4\sigma^2}},
\end{equation}
with $\hat{A}(x)$ an hermitian operator and $\sigma$ a new fundamental constant\footnote{These operators automatically satisfy Eq. \ref{Comp} too.}. These operators can be viewed as quasi projections into approximate eigenstates of $\hat{A}(x)$, so that the cumulative effect of many collapses would lead the system into eigenstates of $\hat{A}(x)$.

A complication in this framework arises when attempting to compute the expected change in energy as the hypersurface $\Sigma$ passes through a collapse event: the result diverges due to spatial discontinuities in the field configuration caused by the point-like nature of the collapse. To address this issue, one can introduce an smearing procedure in an appropriate manner to maintain the covariance of the model as is done in \cite{bedingham2011relativistic}.

With respect to the nature of $\hat{A}(x)$, we saw that at the non-relativistic level the mass density function (\ref{massdensity-NR}) plays a central role, which can be written in terms of a mass density operator $\hat{M}(\textrm{x})$ as
\begin{equation}
	m(\textrm{x},t)= \langle\Psi_{t}| \hat{M}(\textrm{x})  |\Psi_{t}\rangle \equiv \sum_i^N m_i\int d^3\mathbf{x}_1...d^3\mathbf{x}_N\delta^3(\mathbf{x}_i-\mathbf{x})||\Psi_{t}(\textbf{\textrm{x}}_1, \textbf{\textrm{x}}_2,..., \textbf{\textrm{x}}_N) ||^2.
	\end{equation}
In the relativistic context, this suggests that the collapse operator should be associated with the energy-momentum tensor. If so, the dynamics could avoid large uncertainties in the expectation value of $\hat{T}_{ab}$, which, as discussed in \S\ref{Chapter:SemiClass}, could lead to conflicts with observations.

\subsection{The local beable}\label{ssection:RelCollocalbeable}

Finally, as with the non-relativistic models, in order to complete the description of this framework it is necessary to be explicit about its ontology. In this regard, following \cite{bedingham2011relativistic}, we will read the theory as postulating the existence in spacetime of an \textit{energy-momentum density} $\mathcal{T}_{ab}(x)$: as we already suggested in \S \ref{Section:SGsignaling} it is defined through the explicitly Lorentz-invariant prescription 
\begin{equation}\label{E-Mdensitytensor2}
\mathcal{T}_{ab}(x)\equiv \langle \psi |  \hat{T}_{ab} | \psi \rangle_{\partial J^-(x)}, 
\end{equation}
with $\partial J^-(x)$ the past null cone of $x$. It might be objected that, since the past null cone is not a Cauchy surface, the quantum state is not, in general, defined there, and the recipe is therefore ill-defined. However, note that instead of considering the state on $\partial J^-(x)$, one could use any hypersurface passing through $x$ and construct a non-physical state on it by considering only the collapses that occur within the causal past of $x$.

It is important to note that, since the dynamics under consideration includes objective collapses, it is not always the case that $\nabla^a T_{ab}(x) = 0$. In particular, this equation holds everywhere except at the collapse events and along their future null cones. In the next chapter, we explain how these considerations can be used to construct a self-consistent semiclassical framework.

To get a sense of how this formalism operates, consider the same EPR setup and the two foliations depicted in Figure \ref{2foliations}. Here, both foliations are equally valid in relativistic collapse theories and there is a quantum state associated to each of the hypersurfaces $\Sigma_{t}$ and $\Sigma_{t'}$, which in general would be different. However, the following curious situation now emerges naturally from this picture. 

On the hypersurface that intersects Alice's worldline at $P$, the quantum state assigns an undetermined value to the $z-$spin measurements of both Alice and Bob---e.g., a  $50-50\%$ probability for each outcome. In contrast, on the hypersurface that intersects Bob's worldline at $R$ (after his measurement at $B$) and also intersects Alice's worldline at $P$, the associated quantum state reflects a determined value for the $z-$spin measurements. That is, undeterminacy becomes also foliation dependent. Notwithstanding the above, situations of this type can not be used to find contradicting values associated to the same events (e.g. when Alice and Bob compare their results in the future) if our theories come equipped with a clear ontology. This is because we can not think of the different foliations as representing the `present' for different observers. We will illustrate this more concretely in the following example.

Aside from this, there also persists the widespread belief that collapse is incompatible with relativity, since collapse theories are thought to allow the possibility of superluminal signaling. To tackle this, we will make use of another simple example first.

\subsection{An example with superposition of masses}

Now we will show an example where it becomes much more evident the importance of understanding the difference between the local and non-local beables and between the \textit{epistemic states} used to make predictions and the \textit{ontic states} assumed to exist in the theory, independently of agents. We will see how all these elements conjugate to avoid superluminal signaling but mantaining the quantum non-local correlations and in accordance with experimental results. Again, we will assume that the local beable of the theory is the mass distribution given according to equation \ref{E-Mdensitytensor2}.

\begin{figure}[ht]
	\centering
	\includegraphics[width=10.3cm]{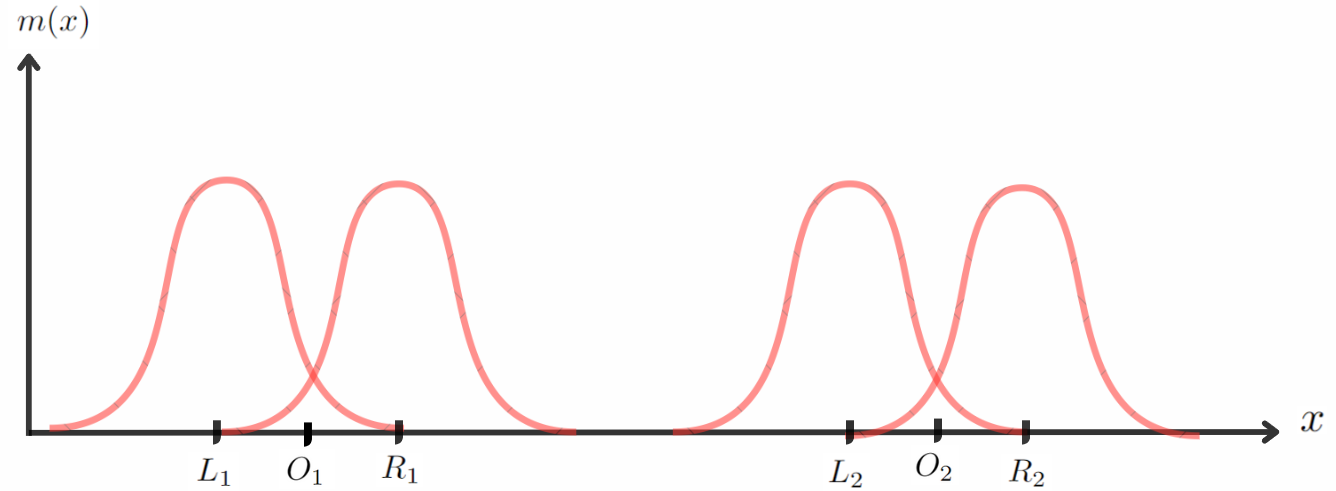}
	\caption{{\small Matter density of two pointers in the superposition (\ref{MassSuper}) showing just the $x$ spatial dimension of $\Sigma_0$.}}
	\label{MDsuperFig}
\end{figure}

Consider a system of two space-like separated idealized pointers of mass $m$\footnote{We are framing this example in terms of `pointers' in position states because as we will argue, we can generalize this entangled situation to any other relevant observable, e.g. spin, energy, etc, given the Hilbert operator formalism outlined in \cite{Bassi_2007}, which for all practical purposes, always lets us couple these operators to the position of `pointers' in collapse theories.}. The system is in an initially entangled ontic state of both  pointers being to the left or both to the right of their respective origins $O_1$ and $O_2$, over a hypersurface $\Sigma_0$, i.e.
\begin{equation}\label{MassSuper}
	\ket{\Psi_{\Sigma_0}}=\alpha\ket{L_1}\ket{L_2}+\beta\ket{R_1}\ket{R_2},
\end{equation}
where $\ket{L_i}$ is the state associated to the $i-th$ pointer being at a distance $L_i$ from its origin (analogously for $\ket{R_i}$).
Thus, it is easy to see that the local ontology over $\Sigma_0$ according to this theory, is conformed by a matter density distribution as the one depicted in figure \ref{MDsuperFig}, where every bump of $m(x)$ concentrates half the mass of the respective pointer.
Now suppose the positions of the pointers are measured by placing detectors in the future of their worldlines. The effect of the dynamical collapse will suppress the superposition, concentrating the mass to a single localized state ($\ket{L_1}\ket{L_2}$ or $\ket{R_1}\ket{R_2}$) after detections. And what the theory tells us about \textit{what} actually exist in the world, depends on \textit{where} in the world we are asking. 

\begin{figure}[ht]
	\centering
	\includegraphics[width=10cm]{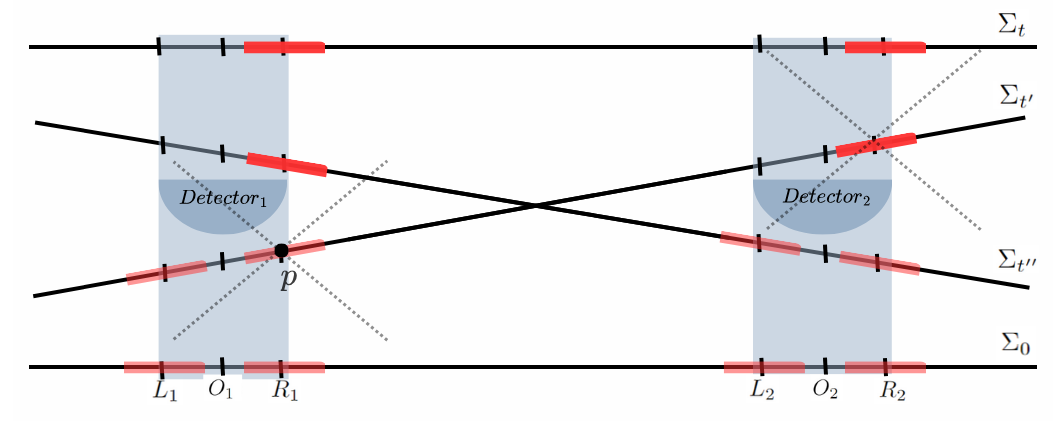}
	\caption{{\small Spacetime diagram showing the hypersurfaces $\Sigma_0$, $\Sigma_t$, $\Sigma_{t'}$ and $\Sigma_{t''}$. The mass density is shown in red with transparency for $m/2$ and full color for $m$ after detection, e.g. to the right, $R_1$ and $R_2$}.}
	\label{hypersurfMD}
\end{figure}

Take, for instance, the case where both pointers are simultaneously detected to the right of their origins with respect to $\Sigma_t$. Then, over $\Sigma_t$, we will have two bumps of mass $m$, each one associated to the pointers being to the right of their respective origins, as shown in figure \ref{hypersurfMD}, according to the past light cone (PLC) recipe of equation (\ref{E-Mdensitytensor2}).

Consider also the other two following hypersurfaces of different foliations depicted in Fig. \ref{hypersurfMD} (with the `$1^{st}$ pointer' being the one placed to the left-hand side of Fig. \ref{hypersurfMD} and the `$2^{nd}$ pointer' to the right-hand side). $\Sigma_{t'}$ is such that it intersects with the worldline of the $2^{nd}$ pointer after the detection, but with the $1^{st}$ pointer before the detection . And $\Sigma_{t''}$ has the inverted intersections, that is with the $2^{nd}$ pointer before detection and with the $1^{st}$ pointer after detection.

Now let us display formally the ontology of our relativistic collapse theory in each case, which includes the local and non-local beables (matter density and quantum states accordingly).

For simplicity and without loss of generality we will take $\alpha=\beta$ and assume a 2-D spacetime where we will put cartesian coordinates in the spacelike region where the experiment takes place in such a way that the state of the $i-th$ pointer can be localized at a distance $L_i$ or $R_i$ from the origin. 
The mass density functions over the $\Sigma$-hypersurfaces according to (\ref{E-Mdensitytensor2}) are given by\footnote{It is more realistic to associate gaussians rather than Dirac deltas, but for our arguments this will be irrelevant.}
\begin{align}
	m(\Sigma_0,x)&=\frac{m}{2}[\delta(x-L_1)+\delta(x-R_1)+\delta(x-L_2)+\delta(x-R_2)]\label{MDonSigma0}\\
	m(\Sigma_t,x)&=m[\delta(x-R_1)+\delta(x-R_2)]\\
	m(\Sigma_{t'},x)&=\frac{m}{2}[\delta(x-L_1)+\delta(x-R_1)]+m\delta(x-R_2)\label{MDonSigmat'}\\
	m(\Sigma_{t''},x)&=m\delta(x-R_1)+\frac{m}{2}[\delta(x-L_2)+\delta(x-R_2).]
\end{align}
As we have already seen, the mass density function over $\Sigma_0$ is going to give us four bumps, each associated with half the mass, $m/2$, centered at the left and right positions of both pointers. Over $\Sigma_{t''}$ we will get one $m$ bump to the right of the $1^{st}$ pointer and the two $m/2$ bumps for the $2^{nd}$ pointer. And finally, the inverted configuration for $\Sigma_{t'}$.

On the other hand, the non-local ontic states at each hypersurface are: 
\begin{align}
	\ket{\Psi_{\Sigma_0}}&=\alpha \ket{L_1}\ket{L_2}+ \beta \ket{R_1}\ket{R_2}\label{psihyp0}\\
	\ket{\Psi_{\Sigma_{t}}}& =\ket{R_1}\ket{R_2}\label{psihypt}\\
	\ket{\Psi_{\Sigma_{t'}}}&=\ket{R_1}\ket{R_2}\label{psihypt'}\\ \ket{\Psi_{\Sigma_{t''}}}&=\ket{R_1}\ket{R_2}.\label{psihypt''}
\end{align}
Now, we have to distinguish these from the epistemic states, which are the ones used by observers to make predictions and include the available and relevant information that the observers could have obtained, but exclude everything else. Non-selective measurements are employed for this and the epistemic states are then represented by reduced density matrices. For example, for an observer near the $1^{st}$ pointer the reduced density matrices are: 
\begin{align}
	\rho_{1}(t')= \alpha^2 \ket{L_1}\bra{L_1} +  \beta^2\ket{R_1}\bra{R_1}\label{rho1'} \ \ \ \ &\textrm{(before detection),}\\
	\rho_{1}(t'')= \ket{R_1}\bra{R_1}\ \ \ \ &\textrm{(after detection),}
\end{align}
and for an observer near the $2^{nd}$ pointer they are: 
\begin{align}
	\rho_{2}(t'')= \alpha^2 \ket{L_2}\bra{L_2} +  \beta^2\ket{R_2}\bra{R_2} \ \ \ \ &\textrm{(before detection),}\\
	\rho_{2}(t')= \ket{R_2}\bra{R_2}\label{rho2'}\ \ \ \ &\textrm{(after detection).}
\end{align}
Again, epistemic states are relative to observers and do not represent the ``real'' states, because different observers associate different states with the same hypersurface depending on their knowledge. For exampĺe, the reduced states $\rho_{1}(t')$ and $\rho_{2}(t')$ in equations (\ref{rho1'} and (\ref{rho2'}), are associated to different observers on the same hypersuface of simultaneity $\Sigma_{t'}$, however, their assigned predictions are different simply because they have access to different information on distinct regions of spacetime. With respect to the observer near the $1^{st}$ pointer, her assigned predictions still give equal probabilities to both pointers being to the right or to the left, even though at the same relative time $t'$, the observer near the $2^{nd}$ pointer already associates both pointers being certainly to the right. 

\subsection{Why we can not use this to signal superluminally?}

To understand the common mistakes made in arguments claiming that collapse theories conflict with relativity due to the possibility of superluminal signaling, let us examine in detail the difference between using ontic quantum states defined on any spacelike hypersurface and those defined on the past light cone to specify the local beable. In the first case, for instance, if we had used instead the quantum states on the spacelike hypersurfaces, $\Sigma_{0}, \Sigma_{t}, \Sigma_{t'}, \Sigma_{t''}$ (equations \ref{psihyp0} to \ref{psihypt''}), to calculate the mass density, then, we would have obtained
\begin{align}
	\tilde{m}(\Sigma_0,x)&=\frac{m}{2}[\delta(x-L_1)+\delta(x-R_1)+\delta(x-L_2)+\delta(x-R_2)]\label{tildeMDonSigma0}\\
	\tilde{m}(\Sigma_t,x)&=m[\delta(x-R_1)+\delta(x-R_2)]\label{tildeMDonSigmat}\\
	\tilde{m}(\Sigma_{t'},x)&=m[\delta(x-R_1)+\delta(x-R_2)]\label{tildeMDonSigmat'}\\
	\tilde{m}(\Sigma_{t''},x)&=m[\delta(x-R_1)+\delta(x-R_2)].\label{tildeMDonSigmat''}
\end{align}
That is, the mass density in a region, around a pointer, is affected even \textit{before} it interacts with its nearest detector, so that the outcome is already determined. Suppose that two observers, Alice and Bob, arranged an experiment of this sort to monitor the gravitational field affected by matter. She is going to measure the position of the $1st$ pointer (on the left of the Figure \ref{hypersurfMD}) and he is going to measure, spacelike separated from her, the $2nd$ pointer (to the right of figure \ref{hypersurfMD}). At $t=0$ with respect to $\Sigma_0$, Alice knows according to (\ref{psihyp0}) that the gravitational field of her pointer corresponds to two bumps of mass $m/2$ at $L_1$ and $R_1$, as given by eq. (\ref{tildeMDonSigma0}). However, at $t'$ with respect to $\Sigma_{t'}$, the collapse that Bob induces on his pointer changes the gravitational field in Alice's region, even before she measures her pointer, its mass density changes to just one bump of mass $m$ at $R_1$. This could be used to generate a superluminal channel between them. 

Note that the above possibility could not occur using the recipe to specify the local beables with the state on the $PLC(x)$. In this case, the two bumps in Alice's region would not change until she actually measures (see Equations \ref{MDonSigma0} and \ref{MDonSigmat'}). 
Thus, we obtain a relativistic prescription for describing the matter density, providing the kind of locality required to prevent superluminal signaling. However, we must also account for the non-local quantum correlations observed in Bell-type experiments. As previously mentioned, in relativistic quantum collapse theories, these non-local effects are attributed to the quantum state---an ontic, non-local entity defined on any spacelike hypersurface. Since these non-local states are objective, the Bell-type correlations arise from the conditional probabilities assigned by them. Predictions, which are derived from epistemic states, align statistically with these correlations because they are based on the information accessible to a given observer—information that remains unaffected by events spacelike-separated from them. 

In more general words, the relativistic consistency conditions i) and ii) stated in \ref{Section:relatontictheory}, concerning local operations, state assignment to hypersurfaces, and `microcausality', let us prove a non-signaling theorem for relativistic collapse theories, maintaining the Bell-type correlations \cite{myrvold2003relativistic}. Finally, it is also interesting to note that although in the above example the total mass is conserved on each hypersurface $M_{\Sigma}=\int_{\Sigma}m(x)= 2m$, in general, we will have no mass conservation, as we will see in a further example.

\chapter{Fully self-consistent semiclassical gravity}

In this chapter, we propose a semiclassical framework that couples the expectation value of a quantum field evolving according to the relativistic collapse dynamics described in \S \ref{Section:aRelatQCTheory} with the classical Einstein tensor. The expectation value at any spacetime point is obtained from the state of the field assigned to the past light cone of that point, as in \S \ref{ssection:RelCollocalbeable}. Since the evolution of the field deviates from unitarity at the random collapse locations punctuating spacetime, the expectation value of the energy-momentum tensor is not conserved---neither at the collapse events nor, given the way the expectation value is computed, in their causal future. We therefore take the semiclassical Einstein equations to hold only in patches and provide a well-defined prescription for gluing them together to form what we call a \textit{Semiclassical Collapse Universe}.

\section{General framework: Semiclassical Collapse Universes (SCUs)}\label{Section:SCU}

We propose to fully describe the universe by means of (i) patch-wise constructed spacetimes $(M, g_{ab})$, which in general will be distinct, where $g_{ab}$ is a (continuous) symmetric tensor (with signature $- + + +$) smooth by patches; and (ii) for the matter sector, a quantum field theory following an objective collapse dynamics, living on spacetime $(M, g_{ab})$ and satisfying
\begin{equation}\label{SCU}
	G_{ab}[g](x)=8\pi G \mathcal{T}_{ab}(x),
\end{equation}
everywhere except at the randomly distributed collapse events and their future light cones, where $g_{ab}$ is discontinuous in general, but we require the induced 3-metric to always be continuous. This entire construction constitutes a \textit{Semiclassical Collapse Universe} or SCU.

Thus, a SCU consist of the set $\{ M, \, g_{ab}(x), \, \{ x_i, Z_i \}, \, \hat{L}_i(Z_i), \, \mathcal{S} \}
$ with $M$ a 4-D manifold,  $g_{ab}$  a Lorentzian metric smooth by patches, $\{ x_i, Z_i \}$ a set of points $x_i \in M$,
each associated to a random variable $Z_i$, $\hat{L}_i(Z_i)$ the collapse operators satisfying Eqs. (\ref{Comp}) and (\ref{MC}) and the \textit{follium} $\mathcal{S} : \mathcal{F} \rightarrow \mathcal{H}$ providing an assignment of a quantum state of the matter fields, $ |\Psi_\Sigma\rangle \in \mathcal{H}$, to each spacelike hypersurface $\Sigma \in \mathcal{F}$ (with $\mathcal{F}$ the collection of all hypersurfaces in $\mathcal{M}$). We assume that the evolution of the field is punctuated by the set of collapse events ${x_i}$ which is distributed over $M$ with a constant mean number of elements per unit 4-volume of the metric $g_{ab}(x)$ and the associated outcomes to each collapse is represented by the random variables $Z_i$ consistent with the probability distribution in Eq. (\ref{Prob}). Naturally, we assume that the follium $\mathcal{S}$ is consistent with the dynamics given by a Tomonaga-Schwinger type equation as (\ref{TomonagaSchw}) and, where collapses occur, with the rule of Eq. (\ref{CollRule}). Finally, Eq. (\ref{SCU})
is valid everywhere, except on the collapse points $x_i$ and their future light cones, where we assume that the induced 3-metric is continuous.

In the next subsection, we provide more details about this final gluing condition and, in particular, explain why it is sufficient to fully determine the metric everywhere.

\subsection{Gluing conditions}

As we saw in chapter \ref{Chapter:SemiClass}, there are instances where the semiclassical Einstein equations can not hold in the entire spacetime. In particular, if the quantum state suffers collapses, we will have regions where the expectation value of the energy-momentum tensor is not divergenceless, so the equations (\ref{SemiClass}) become inconsistent with the Bianchi identities. 
However, in our SCUs framework, we will assume that the semiclassical equations (\ref{SCU}) are valid ``before'' and ``after'' the collapse. More precisely, these equations are valid everywhere except over the future light-cone $\partial J^+(x^\mu)$ of the collapse event $x^\mu$. Thus, we shall impose well-defined gluing conditions in those regions, to ensure the ``smallest possible jump'' for the metric and its derivatives, consistent with some constraints to be specified and the collapse theory.  

A motivation for our SCU framework was the approach developed in \cite{juarez2023initial,juarez2024hadamard}, where a collapse is assumed to induce a discontinuous and stochastic change in the expectation value of the energy-momentum tensor defined with respect to the quantum state on an arbitrary Cauchy hypersurface $\Sigma_C$. To construct the entire spacetime, the idea is to look for the ``smallests jump possible'' across $\Sigma_C$. Thus, based on physical considerations, continuity is imposed on the induced $3$-metric, $g_{ij}$, and the transverse traceless term of the extrinsic curvature on $\Sigma_C$. This, together with the constraints uniquely determines the remaining part of the post-collapse extrinsic curvature.

Our analysis, although motivated by \cite{juarez2023initial}, presents a significant conceptual difference, we do not take the matter sector to be described with respect to non-local quantum states associated to Cauchy hypersurfaces which, among other things, could lead to superluminal signaling. Instead, as has been already stated, we consider matter to be described by a relativistic ontology. That is, we consider the local beables of our SCUs to be represented by the energy-momentum density tensor $\mathcal{T}_{ab}(x)$ defined by Eq. (\ref{E-Mdensitytensor2}) and associated to the event $x$ and its past-light-cone.  

In order to explain the demand for the induced 3-metric to be continuous, we focus on the simple case where only one collapse occurs at $x^\mu$. In consequence, this ``affects'' the quantum state associated to any Cauchy hypersurface that intersects the future of $x^\mu$. In other words, a collapse occurring at  $x^\mu$ induces a ``jump'' in the energy-momentum density tensor on $\partial J^+(x^\mu)$. And since we couple the Einstein tensor at every spacetime point with the energy-momentum density tensor via equation (\ref{SCU}), an associated ``jump'' is expected in the geometry of spacetime on $\partial J^+(x^\mu)$. This situation divides the SCU into two patches, $J^+(x^\mu)$ and $M-J^+(x^\mu)$. Therefore, instead of imposing gluing conditions on spacelike hypersurfaces as in \cite{juarez2023initial}, we must impose them on future light cones whose tips coincide with each collapse event, in this case $\partial J^+(x^\mu)$.

With this in mind, we have to address the initial value formulation on null hypersurfaces, which presents important differences compared to the more familiar spacelike case.  A thorough treatment of the situation, where the initial hypersurface is a future null cone, $C_O$, with vertex at $O$, is given in \cite{choquet2011cauchy}, where constraints for the Einstein equations on $C_O$ are derived and solved. Our approach, regarding the gluing conditions, is based on the results and considerations in \cite{choquet2011cauchy}. We take our spacetime example in section \ref{Section:Example}, to be an exemplification of a solution consistent with the results in \cite{choquet2011cauchy}. The key result is that if we prescribe on $C_O$ as initial data a quadratic form, $\bar{g}$, that satisfies a set of four equations known as wave-map gauge constraints, $\mathcal{C}_\alpha=T_{\alpha\beta}l^\beta|_{C_O}$, and $\bar{g}_{\alpha\beta}(O)=\eta_{\alpha\beta}$ in some coordinate system, then the Einstein equation admits a solution, $g$,  in a neighborhood of the vertex such that $g|_{C_O}=\bar{g}$. \footnote{Although in \cite{choquet2011cauchy} uniqueness is demonstrated only for the vacuum case, it is plausible that for reasonable energy-momentum tensors the solution should also be unique.}

Furthermore and of special relevance to our analysis, \cite{choquet2011cauchy} studies a method to construct initial data that satisfies the wave-map constraints given a 3-metric, $\tilde{g}$, in $C_O$. To achieve this, the authors introduce a special coordinate system, $\{x^{\prime\alpha}\}$, where $C_O$ is represented by the equation $x^{\prime 0}=(x^{\prime 1})^2+(x^{\prime 2})^2+(x^{\prime 3})^2$. This coordinate system can be constructed using Riemann normal coordinates at $O$ with an orthogonal frame \cite{choquet2008general}. Based on $\{x^{\prime\alpha}\}$, null adapted coordinates, $\{x^\alpha\}$, are defined in which the data we aim to construct takes the form
\begin{equation}
	\bar{g}:=g|_{C_0}:=\bar{g}_{00}(dx^0)^2+2\nu_0dx^0dx^1+2\nu_Adx^0dx^A+\underbrace{\bar{g}_{AB}dx^Adx^B}_{\tilde{g}}.
\end{equation}
Then, working in this coordinate system, if we give an admissible 3-metric, $\tilde{g}=\bar{g}_{AB}dx^Adx^B$, on $C_O$, where by ``admissible'' we mean that $\tilde{g}_{\alpha^\prime\beta^\prime}(O)=\eta_{\alpha^\prime\beta^\prime}$,\footnote{This condition ensured consistency with the fact that $\{x^{\prime\alpha}\}$ is a Riemann normal coordinate system around $O$. Note that we are demanding conditions in the components of the metric associated with $\{x^{\prime\alpha}\}$, using the $\{x^\alpha\}$ coordinates.	Since $\{x^\alpha\}$ are not defined on $O$, this condition is introduces a limit at the vertex.}
it is shown that the wave-map constraints (for reasonable $T_{ab}$) turn into four hierarchical, ordinary differential equations, for the remaining four coefficients of $\bar{g}$, all linear, once the physical constraint,  $(G_{rr}-T_{rr})|_{C_O}=0$, has been solved. Therefore, if we prescribe $\tilde{g}$ and impose initial conditions for the differential equations such that $\bar{g}_{\alpha^\prime\beta^\prime}(O)=\eta_{\alpha^\prime\beta^\prime}$, we can uniquely determine the remaining data $(\bar{g}_{00}, \nu_0, \nu_A)$ of $\bar{g}$ which guarantees the existence of a solution.

Based on this, our proposal for employing the semiclassical Einstein equations (\ref{SCU}) in a scenario where a collapse occurs at  $x^\mu$ is to impose continuity of the induced 3-metric on $\partial J^+(x^\mu)$. The four wave-map constraints, together with the corresponding jump in the expectation value of the energy-momentum tensor, then uniquely determine the remaining components of $\bar{g}$ that ensure the existence of a solution to the Einstein equations in  $J^+(x^\mu)$. Note that, since the induced 3-metric in our proposal is derived from a physical metric, it is always admissible.

\section{Example}\label{Section:Example}

As we have seen in the previous sections, although the quantum state is a non-local beable associated to hypersurfaces that, after a collapse, actualizes instantaneously and independently of the foliation, the mass-energy density, on the other hand, is a local beable that can only be affected by the quantum state at the speed of light. Therefore, if a relativistic quantum collapse theory is intended to be useful in a semiclassical gravity context, it must be possible to compute the spacetime metric everywhere, which in turn, is influenced by the stress-energy tensor.

Take for instance, situations in which a single collapse changes dramatically the mass density over a region of spacetime. Inside the future light cone of the collapse, we have the set of spacetime events which are already influenced by the change in the mass density. Outside, we might have a very different distribution associated to some initial superposed state. Note that in the kind of semiclassical framework we are proposing, there would be mass non-conservation due to the retarded influence in the local beable. Then, the spacetime metric, as given by the Eqs. (\ref{SCU}), could be abruptly and discontinuously different in both regions. Notwithstanding, we can provide a concrete description in idealized, symmetric, situations.

\begin{figure}[ht]
	\centering
	\includegraphics[width=14cm]{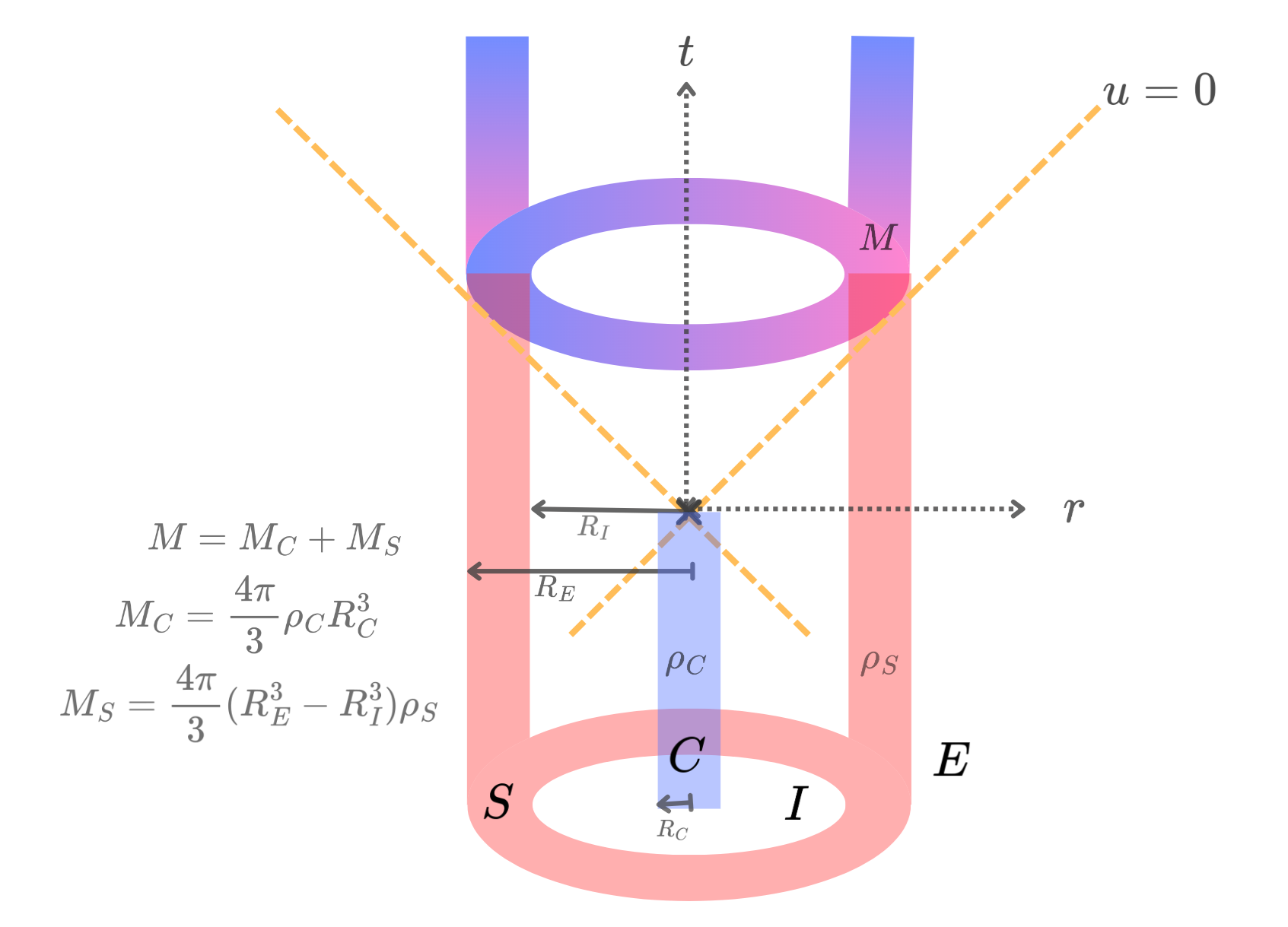}
	\caption{{\small An spherically symmetric superposition of masses suffers a collapse at $r=0$ and $t=0$, localizing the entire mass $M$ at the region of the shell. The central massive core has radius $R_C$ and uniform density $\rho_C$, and the shell has an inner radius $R_I$, an outer radius $R_E$ and uniform density $\rho_S$.}}
	\label{STmetric}
\end{figure}

To this end, let us consider a quantum state giving rise, via Eq. (\ref{E-Mdensitytensor2}), to a system of total mass $M$ in an initially superposed state of being entirely concentrated in a   (three dimensional) spherical shell of mass $M_S$ of inner radius $R_I$, outer radius $R_E$ and uniform
density $\rho_S$, and in a central core of mass $M_C$ with radius $R_C$ and uniform density $\rho_C$. Take, for instance, the particular case where a collapse occurs at the center of the system and produces a localization of the mass to the shell. See figure \ref{STmetric}. 

Before the collapse, we have four distinct regions: the core (C), the interior (I), the shell (S) and the exterior (E). We define
\begin{equation}\label{masses}
	M_C= \frac{4\pi}{3} \rho_C R_C^3,\ \ \ \ M_S= \frac{4\pi}{3} (R_E^3-R_I^3)\rho_S \ \ \textrm{and}\ \ M=M_C+M_S.
\end{equation}
Given that it is a spherically symmetric scenario, we can wirte the metric everywhere as 
\begin{equation}
	ds^2=-e^{2\phi(r)}dt^2+ \big(1-\frac{2m(r)}{r}\big)^{-1}dr^2+ r^2d\Omega^2,
\end{equation}
with
\begin{equation}
	m(r)=4\pi \int_{0}^{r} \rho(r')r'^2dr'
\end{equation}
and
\begin{equation}
	\frac{d\phi(r)}{dr}=\frac{m(r) + 4\pi r^3 P(r)}{r[r-2m(r)]},
\end{equation}
with $P(r)$ the radial pressure. The expressions for $m(r)$ and $\phi(r)$ in every region are given in the Appendix \ref{Appendix:spacetime}.
Using the relations in Eqs. (\ref{masses}), the metric everywhere can be written as a function of $\rho_C$, keeping $M$ constant. Now, we transform into null coordinates with 
\begin{equation}
	u=t-\int_{r_0}^{r} \bigg(\frac{1-2m(s) }{s} \bigg)^{1/2} e^{-\phi(s)} ds,
\end{equation}
so that the metric can be written in general as
\begin{equation}
	ds^2= -e^{2\phi(r,\rho_C)}du^2- 2 e^{2\phi(r,\rho_C)}\bigg( 1-\frac{2m(r,\rho_C)}{r}  \bigg)^{-1/2}dudr  +  r^2 d\Omega^2.
\end{equation}

Now, the occurrence of the collapse at $r=0$ and $t=0$ produces a localization of all the mass on the shell. Note that the effect travels along the causal future of the collapse event (i.e., $u=0$) due to the form of Eq. (\ref{E-Mdensitytensor2}). This produces the mass distribution depicted in Figure \ref{STmetric}, where the metric of the SCU is described by $ds^2(\rho_C)$ for $u<0$ and  $ds^2(0)$ for $u\geq0$. Moreover, on the $u=0$ surface, the metric is going to be discontinuous, as was already expected. However, the induced 3-metric, $r^2d\Omega^2$, on this surface—computed from both below and above---is continuous as required by our gluing condition.

At this point, we note a potential way to avoid discontinuities in the metric: introducing an interpolation between the metrics before and after the collapse over a short time interval. For instance, from $u=0$ to $u=\tau_C$, the metric could be given by $\bar{g}=ds^2(\alpha(u))$, where $\alpha(u)$ continuously interpolates between $\rho_C$ and $0$. It is interesting to note that during the collapse---specifically, in the region between the core and the shell—even though $G_{tt}$ remains zero, $G _{tr}$ is non-zero. This could be interpreted as a radial mass-energy current. In any case, the specific values of the components of the Einstein tensor depend on the details of the interpolation; and without a well-defined prescription for constructing it, this approach to avoiding discontinuities remains, at best, tentative.

\newpage

\section{Gravitationally-Induced-Entanglement experiments}

The SCUs framework outlined in this work, as a semiclassical framework, should lead to empirical predictions manifestly different from approaches associating a quantum nature to spacetime. In line with this, one could look for experimental probes to decide whether spacetime behaves in accordance to one, say as a classical entity, or the other, allowing for quantum characteristics. In section \ref{Chapter:SemiClass} we reviewed some of the thought experiments intended to rule out the semiclassical frameworks and considered by some, as indirect evidence for the quantum gravity program. There, we argued that the arguments can only be put forward for certain (and somewhat naive) semiclassical approaches, or that the experiments can not be carried out even in principle. More recently, more plausible and less restricted tests have been proposed as `direct evidence'. In \cite{MarlettoVedral2017} and \cite{Bose2017}, a general kind of protocol is devised to `witness' the quantum nature of the gravitational field.

In essence, this kind of protocol considers two separated quantum systems $Q_1$ and $Q_2$, and a third system $C$ (e.g., the gravitational field), so that the quantum systems can only interact if mediated by $C$. If the quantum systems are initially in separable states, it is argued that if they end up entangled after interacting only by means of $C$, then, $C$ must also be quantum. The experimental procedures in \cite{MarlettoVedral2017} and \cite{Bose2017} are purportedly technologically feasible. Thus, if after realizing such procedures, the quantum systems are tested and found out to be entangled, we could conclude that spacetime, indeed, has a quantum nature and in consequence, that the semiclassical approaches would be empirically wrong. The advantage of these tests is that no control of the gravitational field is needed, or knowledge of the underlying quantum model. Then, it is interesting to analyze what would actually be the concrete predictions of our SCUs framework in such experiments.

To begin, take for concreteness the case in \cite{Bose2017}. There, the two quantum systems in question are two massive particles, each of which are introduced to a matter-wave interferometer. To this end, particles with embedded spins and Stern-Gerlach apparatuses are used to produce a superposition of two spatial states $\ket{L}_i$ and $\ket{R}_i$ for the $i$-th particle. The interferometers are placed side by side, such that the particles interact only by their mutual gravitational interaction. During the time interval $\tau$ in which both particles are in their respective superposed states, each branch of the joint superposition evolves under a different gravitational potential energy depending on the relative separation between the particles in that branch. Since a quantum state with potential energy $V$ acquires a phase factor $e^{-iV\tau/\hbar}$ under time evolution, each component accumulates a different phase. In the case where we have only the Newtonian gravitational potential $V(r) = -Gm_1m_2/r$, the accumulated phase is $\phi(r) = Gm_1m_2\tau/(\hbar r)$, and since the separations differ between the $\ket{LL}$, $\ket{LR}$, $\ket{RL}$, and $\ket{RR}$ branches, the corresponding phase shifts differ, leading generically to entanglement. Thus, if we recombine the components of their respective superposed states, the amount of entanglement can be tested.

To see this, take state of the particles just after being prepared by their respective Stern-Gerlach apparatuses to be the separable state
\begin{equation}
	\ket{\Phi(t=0)}_{12}=\frac{1}{\sqrt{2}}(\ket{L}_1+\ket{R}_1)\frac{1}{\sqrt{2}}(\ket{L}_2+\ket{R}_2).
\end{equation} 
After a time $\tau$, the gravitational interaction will lead the evolution of the state to
\begin{equation}\label{BoseEntangledState}
	\ket{\Phi(t=\tau)}_{12}=\frac{e^{i\phi}}{\sqrt{2}}\{\ket{L}_1\frac{1}{\sqrt{2}}(\ket{L}_2+e^{i\Delta\phi_{LR}}\ket{R}_2)+\ket{R}_1\frac{1}{\sqrt{2}}(e^{i\Delta\phi_{RL}}\ket{L}_2+\ket{R}_2)\},
\end{equation} 
with $\Delta\phi_{LR}=\phi_{LR}-\phi$ and $\Delta\phi_{RL}=\phi_{RL}-\phi$. If we suppose that the dominant contribution to the gravitational interaction is Newtonian, so that the general relativistic contributions are negligible, then, the expressions for the gravitational potential depend on the distance $d$ between the centers of the superpositions and the distance $\Delta x$ between the $\ket{L}$ and $\ket{R}$ components in the following way:
\begin{equation}
	\phi\sim\frac{Gm_1m_2 \tau}{\hbar d}, \ \ \ \phi_{LR}\sim\frac{Gm_1m_2 \tau}{\hbar (d+\Delta x)}, \ \ \ \phi_{RL}\sim\frac{Gm_1m_2 \tau}{\hbar (d-\Delta x)}.
\end{equation} 
One can readily see that (\ref{BoseEntangledState}) is an entangled state whenever
\begin{equation}
	(\ket{L}_2+e^{i\Delta\phi_{LR}}\ket{R}_2)\neq(e^{i\Delta\phi_{RL}}\ket{L}_2+\ket{R}_2).
\end{equation} 

\subsection{A SCUs treatment of the Marletto-Vedral-Bose experiment}
There are at least two different reasons for why our SCUs framework entails that no entanglement would be corroborated after measuring the aforementioned particles in complementary bases. First, aside from relativistic considerations, collapse models give a precise probabilistic description of how long a superposition would last depending on the `size' of the quantum system. Thus, it could be the case that the empirically adequate parameters of such models prevent the viability of maintaining the required superpositions for long enough time to produce the relevant phase shifts \footnote{In \cite{MarlettoVedral2017} it is actually acknowledged that for realistic realizations of the experiment, say with micro-diamonds, the approximate time interval during which the phase shift accumulates is something like $\tau \sim 3.5 s$, which for the typical parameters of collapse models is longer than the time such a system would last in a superposition. }.

Second,  recall the basic semiclassical equation of SCUs (\ref{SCU}). The right hand side of this equation tells us that the gravitational field produced by quantum matter, say a massive particle in spatial superposition, arises from the local beable defined at every point in spacetime as a `mass density function'. Here, in particular, this is given by the corresponding energy-momentum density $\mathcal{T}_{ab}(x)$. Which in the case of the Marletto-Vedral-Bose experiment where the gravitational interaction is Newtonian, the gravitational field produced by the superposed particles of mass $m_i$, arises as a single classical field produced by the mass density associated to both components of the superposition---which appears as given by particles of mass $\frac{m_i}{2}$ located at the left \textit{and} right of their respective interferometer.

We can see that this proposed experimental protocol would function in a similar fashion, according to our SCUs framework, to the Collela-Overhauser-Werner (COW) experiment \cite{COWexperiment}. There, a quantum test particle in a superposition is subjected to a classical gravitational field. Our framework tells us that the Marletto-Vedral-Bose experiment is just a more complex case of the COW setting, in the sense that we now have two test particles; each particle is subjected to the ``classical'' gravitational field sourced by the mass density associated to the other interferometer\footnote{As is explained on section \S \ref{Section:SGsignaling}, each component of the superposition, associated to the position of a particle, suffers also a self-attraction which tends to join its superposition. This behavior could be negligible compared to the attraction generated by the other component of the superposition, however, since in the experiment the particles are recombined, this effect do not contributes to a further prediction.} (in the case of the COW setting, the source of the classical gravitational field is the Earth).

In more concrete terms, according to our framework, the initial state above would evolve into:
\begin{equation}
	\ket{\Phi(t=\tau)}_{SCU}=\frac{1}{\sqrt{2}}(e^{i\alpha}\ket{L}_1+\ket{R}_1)\frac{1}{\sqrt{2}}(\ket{L}_2+e^{i\alpha}\ket{R}_2).
\end{equation} 
with the phase given by
\begin{equation}
	\alpha=\frac{Gm_1m_2\tau }{\hbar}(\frac{1}{d-\Delta x}- \frac{1}{d+\Delta x}).
\end{equation}
It is clear that no entanglement is generated after the interaction. However, our model provides a concrete prediction---particular relative phases on each side of the eperiment---that could be measured doing an interference experiment like the one performed in \cite{COWexperiment} on any branch of the setting. 
Therefore, if a test like that of the Marletto-Vedral-Bose experiment is carried out in the near future and no entanglement is found, an interference test could be conducted to determine whether the SCU framework proposed in this work provides an empirically adequate description or not.

\section{Conclusions}

Quantum mechanics and general relativity are both immensely successful theories within their respective domains. However, a domain in which both kinds of descriptions intertwine and play a central role has long resisted a unified, consistent, and accurate formulation. The two broad paths taken to address this challenge---either providing a quantum treatment of gravity or embracing its classical nature---have both been intensely studied for decades, without a clear resolution. A semiclassical treatment, in which classical spacetime geometry is described by the Einstein tensor and coupled to the expectation value of the energy-momentum tensor of matter fields, could represent a significant step toward better understanding the interplay between the gravitational and quantum realms—even if such a description is ultimately replaced by a full quantum theory of gravity. For this approach to be successful, it must provide a self-consistent framework and yield testable predictions. However, the semiclassical approach has been severely criticized over the years, often accused of being internally inconsistent and empirically inaccurate.

In this work, we have started by endorsing the view that our physical descriptions should come in the form of proper theories, where by proper it is meant, simply, to be clear about what entities are postulated as existing and how they behave. Given that standard quantum mechanics fails in both requisites, we begin by choosing some of the quantum theories that addresses such problems. In particular, we have taken objective quantum collapse theories and analyzed their proposed relativistic extensions. On the other hand, by independently assessing the arguments against the semiclassical gravity approach, we have realized the natural advantages that such a treatment would carry if properly constructed with the relativistic extensions of the quantum theories under consideration. 

With this in mind, we have proposed a novel, self-consistent, and empirically viable framework for semiclassical gravity that incorporates relativistic quantum collapse theories and addresses the well-known issues associated with traditional semiclassical approaches. In our \textit{Semiclassical Collapse Universe} (SCU) framework, matter fields evolve according to a modified quantum dynamics that includes spontaneous collapse events, providing a clear mechanism for the emergence of definite outcomes. This evolution enables a relativistic prescription for the local beables, defined in terms of the expectation value of the energy-momentum tensor, which is well-defined at each point in spacetime. These local beables are suitable for coupling to the classical Einstein tensor---that is, they are taken to serve as the source of gravity.

We have outlined the general principles of the SCU framework and demonstrated its concrete functioning through an example. Furthermore, we analyzed the implications of our approach for the well-known gravitationally induced entanglement experiments, highlighting potentially testable predictions that could distinguish our model from both standard semiclassical and full quantum gravity approaches. Our paper detailing the SCU framework has already been published in \textit{Physical Review D} and the preprint is available on the arXiv \cite{mucino2025fully}.


\chapter*{General conclusions}\addcontentsline{toc}{chapter}{\protect\numberline{}General Conclusions}
\label{chapter:concl}

This thesis has focused on assessing and extending quantum theories in order to explore their implications across diverse regimes: thermodynamics, semiclassical and quantum-gravity scenarios, as well as the most general measurement framework where all \textit{in principle} possible predictions can be considered. The overarching motivation has been to advance our understanding of the physical \textit{foundations} of the world, and in particular, to shed some light on the nature of time.

Throughout this work, it has been demonstrated that philosophical and foundational reasoning in physics can serve as a primary tool for constructing \textit{new physics}---that is, in developing new theoretical frameworks and generating novel empirical predictions that are amenable to experimental test---rather than being confined to secondary debates over what untestable `interpretations' we like the most.

For instance, it was seen that one of the famous proposals to solve the fundamental problems of quantum mechanics, the objective collapse program, can be extended to a relativistic setting with clear ontological commitments about what actually exist in spacetime and concrete dynamics. This, among other things, allowed us to construct a working semiclassical framework where spacetime can be treated classically but influenced by quantum matter evolving according to a relativistic collapse model. Our approach leads to concrete explanations and verifiable predictions in well-known experiments proposed to test the quantum nature of gravity.

On the other hand, it was also seen that the standard arguments in favor of the complete empirical agreement between pilot-wave theory and standard quantum mechanics rely on a flawed circular reasoning. In contrast, a careful analysis of the pilot-wave dynamics and the general notion of detections, leads to novel possible empirical consequences which should be verified or falsified by actual experiments. The moral, in brief, is that what can or can not be detected according to a certain theory should depend on the dynamics of the given theory and not on external ad hoc presupposes. 

Also, a fundamental open question in physics concerns why we observe macroscopic process evolving asymmetrically in time. Even if time itself has an intrinsic direction, why do we not see milk unmixing from coffee as time passes to the future? A lot has been said about this topic for more than a century, yet if we ever want to have some clarity, we should formulate answers in terms of the dynamics of the stuff we postulate to exist. In this work, considering that the fundamental stuff is quantum in nature, it was investigated whether intrinsically asymmetric quantum dynamics could provide a basis for explaining the typical macroscopic irreversible evolution.




\appendix

\chapter{Spacetime of the spherically symmetric example}\label{Appendix:spacetime}


In general, for static, spherically symmetric spacetimes we can write the metric as
\begin{equation}
	ds^2 = -e^{2 \phi(r)} dt^2 + \left(1-\frac{2m(r)}{r} \right)^{-1} dr^2 + r^2 d\Omega^2 ,
\end{equation}
with 
\begin{equation}
	m(r) = 4 \pi \int_0^r \rho(r') r'^2 dr'
\end{equation}
and
\begin{equation}
	\frac{d\phi}{dr} = \frac{m(r) + 4 \pi r^3 P(r)}{r [r-2 m(r)]} ,
\end{equation}
where $P(r)$ is the radial pressure. In the case of a perfect fluid, the radial pressure is equal to the ``angular pressure''. Here, in contrast, we are not assuming a perfect fluid, so the radial pressure need not be equal to $F(r)$, which is given by
\begin{equation}
	F(r)= \frac{\rho(r) \left(m(r)+4 \pi r^3 P (r)\right)+P (r) \left(4 \pi r^3 P (r)+2 r -3 m(r) \right)}{2 (r-2 m(r))}+\frac{1}{2} r P'(r)
\end{equation}
Below the different regions, illustrated in Figure \ref{STmetric} of the spacetime are considered.

\subsection*{Exterior region (E)}

In the exterior region, the metric is given by
\begin{equation}
	ds^2 = -\left(1 - \frac{2M}{r} \right) dt^2 + \left(1-\frac{2M}{r} \right)^{-1} dr^2 + r^2 d\Omega^2,
\end{equation}
i.e., it is the Schwarzschild metric with mass $M$.

\subsection*{Shell (S)}

Inside the shell we have,
\begin{eqnarray}
	m_S(r) & = & M_C + \frac{4 \pi}{3} \rho_S (r^3 - R_I^3) , \\
	\phi_S(r) & = & A_S + \int \frac{m_S(r) + 4 \pi r^3 P(r)}{r [r-2 m_S(r)]} dr
\end{eqnarray}
with $A_S$ such that 
\begin{equation}
	e^{2 \phi_S(R_E)} = \left(1 - \frac{2M}{R_E} \right) 
\end{equation}
and, to ensure continuity of $P(r)$ and $F(r)$ at $R_I$ and $R_E$, $P(r)$ satisfies
\begin{eqnarray}
	P(R_I) & = & 0, \\
	P(R_E) & = & 0, \\
	\left. \left( \frac{dP}{dr} \right) \right\rvert_{R_I^+} & = & \frac{M_C \rho_S}{\left(2 M_C - R_I \right) R_I} , \\
	\left. \left( \frac{dP}{dr} \right) \right\rvert_{R_E^-} & = & \frac{M \rho_S}{\left(2 M - R_E \right) R_E} .
\end{eqnarray}

\subsection*{Interior region (I)}

In the interior region, between the Core and the Shell, the metric is given by
\begin{equation}
	ds^2 = - e^{2 \phi_S(R_I)} \left(1 - \frac{2M_C}{R_I} \right)^{-1} \left(1 - \frac{2M_C}{r} \right) dt^2 + \left(1-\frac{2M_C}{r} \right)^{-1} dr^2 + r^2 d\Omega^2 .
\end{equation}

\subsection*{Core (C)}
Within the core, 
\begin{eqnarray}
	m_C(r) & = & \frac{4 \pi}{3} \rho_C r^3 , \\
	\phi_C(r) & = & A_C + \int \frac{m_C(r) + 4 \pi r^3 P(r)}{r [r-2 m_C(r)]} dr .
\end{eqnarray}
with $A_C$ such that 
\begin{equation}
	e^{2 \phi_C(R_C)} = e^{2 \phi_S(R_I)} \left(1 - \frac{2M_C}{R_I} \right)^{-1} \left(1 - \frac{2M_C}{R_C} \right) 
\end{equation}
and, to ensure continuity of $P(r)$ and $F(r)$ at $R_C$, $P(r)$ satisfies
\begin{eqnarray}
	P(R_C) & = & 0, \\
	\left. \left( \frac{dP}{dr} \right) \right\rvert_{R_C^-} & = & \frac{M_C \rho_C}{\left(2 M_C - R_C \right) R_C} .
\end{eqnarray}

\setlength\bibitemsep{.1\itemsep}
\printbibliography
\printindex
           
\end{document}